\documentclass[acmsmall]{acmart}

\usepackage{tikz}
\usepackage{pgf-pie}   %
\usepackage{subcaption}   %
\usepackage{xcolor}

\usepackage{booktabs}   %
\usepackage{subcaption} %
\usepackage{algorithm}
\usepackage{algorithmicx}
\usepackage[noend]{algpseudocode}
\usepackage{stmaryrd}
\usepackage{mdframed}
\usepackage{url}
\usepackage{listings}
\usepackage[bottom]{footmisc}
\usepackage{graphicx}
\usepackage{wrapfig}
\usepackage{fancyvrb}
\usepackage{paralist}
\usepackage{caption}
\usepackage{blindtext}
\usepackage{comment}
\usepackage{xspace}
\usepackage{multirow}
\usepackage{multicol}
\usepackage{booktabs}
\usepackage{enumerate}
\usepackage{enumitem}
\usepackage{caption}
\usepackage{graphicx}
\usepackage{listing}
\usepackage{tikz}
\usepackage{ifsym}
\usepackage{circledsteps}
\usepackage{minted}
\usepackage[dvipsnames]{xcolor}
\usepackage{soul}

\usepackage{amsmath}
\usepackage{bbold}

\usepackage{appendix}
\usepackage[nameinlink]{cleveref}
\usepackage{textgreek}

\usepackage{makecell}
\usepackage{tcolorbox}
\usepackage{sidecap}

\usepackage{mathtools, stmaryrd}
\usepackage{siunitx}

\usepackage{subcaption}
\usepackage{minted}
\setminted{
  breaklines,
  breakanywhere,
  autogobble,
  fontsize=\scriptsize,
  numbersep=4pt
}

\usepackage[utf8]{inputenc}
\usepackage{pgfplots}

\newcommand{\todo}[1]{{\color{red}#1}}

\newcommand{\udafinput}{xs}
\newcommand{\xs}{xs}

\newcommand{\Sim}{\Psi}

\newcommand{\udaf}{\mathcal{P}}
\newcommand{\rudaf}{\mathcal{R}}
\newcommand{\textudaf}{insertion-only fold}
\newcommand{\textrudaf}{retractable fold}
\newcommand{\Init}{\mathcal{I}}
\newcommand{\Acc}{\mathsf{acc}}
\DeclareMathOperator{\eval}{\mathsf{eval}}
\DeclareMathOperator{\Retract}{\mathsf{retract}}

\DeclareMathOperator{\Reach}{\mathsf{Reach}}
\DeclareMathOperator{\remove}{remove}

\DeclareMathOperator{\Subseq}{\mathsf{DeletesTo}}

\newcommand{\UDState}{S}
\newcommand{\Elem}{A}
\newcommand{\Out}{O}
\newcommand{\List}[1]{\texttt{[}#1\texttt{]}}
\newcommand{\fold}{\mathsf{fold}}
\newcommand{\sem}[2]{\llbracket #1 \rrbracket_{#2}}
\newcommand{\psem}[2]{\lbrack #1 \rbrack_{#2}}
\newcommand{\Norm}{\mathsf{Norm}}
\newcommand{\invseq}{\omega}
\newcommand{\RetSpec}{\theta}
\newcommand{\Err}{\mathsf{Err}}

\newcommand{\qual}{\chi}

\newcommand{\vals}[1]{\zeta(#1)}

\providecommand{\acc}{\Acc}         %
\providecommand{\SimTwo}{\Sim}  %

\providecommand{\SimBudget}{\Omega}  %

\newcommand{\dsl}[1]{\text{\fontsize{8}{9}\selectfont\ttfamily #1}}

\newcommand{\setof}[1]{\left\{ #1 \right\}}
\newcommand{\lstof}[1]{\left[ #1 \right]}

\newcommand{\prog}{\mathcal{P}}

\crefname{problem}{problem}{problems}
\crefname{lemma}{lemma}{lemmas}
\Crefname{lemma}{Lemma}{Lemmas}
\crefname{claim}{claim}{claims}
\Crefname{claim}{Claim}{Claims}
\crefname{definition}{definition}{definitions}
\Crefname{definition}{Definition}{Definitions}
\crefname{example}{example}{examples}
\Crefname{example}{Example}{Examples}

\newcommand{\idiotbox}[2]{
\fbox{
\begin{minipage}{0.9\linewidth}
{\bf Result for {#1}:}
{#2}
\end{minipage}
}
}

\DeclareMathOperator{\listappend}{+\!+}

\newcommand{\emptylist}{\texttt{[]}}

\makeatletter
\newcommand{\pequiv}[3]{%
    \ifinalign@
        #1 &\equiv \left(#2, #3\right)
    \else
        #1 \equiv \left(#2, #3\right)
    \fi
}
\newcommand{\requiv}[3]{%
    \ifinalign@
        {#1} &\equiv_{#2} {#3}
    \else
        {#1} \equiv_{#2} {#3}
    \fi
}

\newcommand{\irule}[2]%
   {\mkern-2mu\displaystyle\frac{#1}{\vphantom{,}#2}\mkern-2mu}

\newcommand{\bfpara}[1]{\vspace{5pt}\noindent\emph{\textbf{#1}}}

\newcommand{\init}{\mathcal{I}}

\newcommand{\hole}{\Box}
\newcommand{\pp}{\nolinebreak{\bf +}\nolinebreak\hspace{-.10em}{\bf +}}

\newcommand{\Effect}{\mathsf{Effect}}

\newcommand{\tool}{\textsc{Rewind}\xspace}
\newcommand{\ink}{\textsc{Ink}\xspace}

\lstdefinestyle{numbers}
{
  numbers=left,
  numberstyle=\sf,
  xleftmargin=15pt
}

\newmintinline[T]{scala}{fontsize=\footnotesize, escapeinside=||}
\newmintinline[Tnormal]{scala}{escapeinside=||}

\newminted[Scala]{scala}{fontsize=\footnotesize,
escapeinside=||,
numbersep=5pt}

\algnewcommand{\IIf}[1]{\State\algorithmicif\ #1\ \algorithmicthen}

\setcopyright{cc}
\setcctype{by}
\acmDOI{10.1145/3871732}
\acmYear{2027}
\acmJournal{PACMPL}
\acmVolume{11}
\acmNumber{POPL}
\acmArticle{30}
\acmMonth{1}
\acmSubmissionID{popl27main-p237-p}
\received{2026-07-09}
\received[accepted]{2026-10-05}



\begin{document}
\title{Rewinding the Fold: Making Stateful Computations Retractable}

\author{Ziteng Wang}
\orcid{0009-0001-8487-8093}
\affiliation{%
  \institution{University of Texas at Austin}
  \city{Austin}
  \country{USA}
}
\email{ziteng@utexas.edu}

\author{Linus Zheng}
\orcid{0009-0005-9832-4066}
\affiliation{%
  \institution{University of Texas at Austin}
  \city{Austin}
  \country{USA}
}
\email{linusjz@utexas.edu}

\author{Dixin Tang}
\orcid{0000-0002-3316-6651}
\affiliation{%
  \institution{University of Texas at Austin}
  \city{Austin}
  \country{USA}
}
\email{dixin@utexas.edu}

\author{Işıl Dillig}
\orcid{0000-0001-8006-1230}
\affiliation{%
  \institution{University of Texas at Austin}
  \city{Austin}
  \country{USA}
}
\email{isil@cs.utexas.edu}

\begin{abstract}
Incremental computation seeks to maintain outputs efficiently as inputs evolve,
without recomputing from scratch after every change. While much prior work has
focused on updating results when new data arrive, supporting \emph{deletion} is
fundamentally more difficult: once prior inputs have been folded into a compact
summary, the information needed to undo their effect may already have been lost.
This paper studies incremental deletion for fold-based stream computations. This
setting captures a broad class of data-processing tasks, where a program
maintains a running summary of a stream and uses that summary to process future
inputs. Given a fold that supports only additions and a policy describing what type of
deletion support is required, our method synthesizes a new program that supports
deletions by extending the summary state and deriving both the addition and
deletion logic. Our synthesis procedure is backed by a new verification framework
that decomposes correctness into local obligations over relational invariants.
As a result, for every supported sequence of additions and deletions, the
synthesized fold is guaranteed to match what full recomputation would produce.
We implemented this approach in a new tool called \tool{} and evaluated it on
94 benchmark scenarios drawn  from real-world aggregations in the Flink framework. \tool{} synthesizes deletion support for all 94 benchmarks, including cases that contain bugs in the manually-written deletion implementation.
\end{abstract}

\maketitle              %
\section{Introduction}

Incremental computation seeks to efficiently maintain a program's outputs as its inputs evolve, without paying the cost of recomputing from scratch. A standard way to achieve this goal is to derive \emph{update} logic that explains how the output should change as \emph{new} data arrive~\cite{opera,annieliu}. This paradigm is central
to modern data-processing frameworks~\cite{flink,murray2013naiad,differential-dataflow,DBSP:pvldb/BudiuCMRT23}, which commonly support stateful computations by repeatedly applying such update logic to a maintained state as new records arrive. In this way, the system incrementally accumulates the effect of the input stream, using the evolving state as a concise summary of prior computation.

Yet insertion is only part of the story. In many practical workloads, systems must also \emph{undo} the effect of previously incorporated inputs. For example, records may age out of sliding windows, stale outputs may need to be withdrawn, and continuously maintained views must reflect deletions as well as additions. For this reason, modern data-processing frameworks such as Apache Flink~\cite{flink} and Materialize~\cite{materialize-arrangements} also support change-oriented maintenance, making it (in principle) possible to process deletions and modifications incrementally rather than by recomputing from scratch.

In practice, however, interfaces for incremental deletion are harder
to use than their insertion-only counterparts.
For example, while Flink provides a \texttt{retract} interface for
handling deletions,  developers must implement the retraction
logic themselves and maintain any necessary auxiliary state.
This can be challenging  both semantically and algorithmically. Semantically,
retracting a past contribution is harder because the maintained state
may already have discarded the information needed to faithfully
reverse its effect. Algorithmically, deletion support often demands
auxiliary state and carefully orchestrated rollback logic, making
such operators significantly harder to design and implement.
As a result, although modern frameworks facilitate incremental
deletion in principle, exploiting this capability effectively
remains challenging in practice.

Motivated by this observation, this paper addresses the problem of automatically transforming an insertion-only computation into one that also supports \emph{incremental deletions}. We study this problem in the setting of fold-based computations over input streams, which underlie many operators in modern data-processing systems.  Such a setting is compelling for two reasons. First, it is broad: many practically relevant operators can be viewed as repeated updates to a maintained summary. Second, it exposes the core difficulty: because the past influences future behavior only through a \emph{summary}, the ability to retract prior inputs depends entirely on what the computation has chosen to preserve. Hence, this setting isolates the central tension behind incremental deletion.

At a high level, the core contribution of this paper is a method for synthesizing \emph{incremental deletion logic} for fold-based computations. The method takes as input an insertion-only computation together with a \emph{retraction specification} describing which deletions must be supported. Such a specification is necessary because deletion support is inherently partial: a computation may be able to retract some prior inputs incrementally, but not others. Given these inputs, our method determines what additional information, if any, must be retained to support the required deletions, synthesizes the lifted update logic needed to maintain that information, and automatically constructs the corresponding retraction function. The resulting computation is then formally verified to be semantically equivalent to the original insertion-only computation.

From a programming languages perspective, this paper makes two main technical contributions. The first contribution is a semantic account of incremental deletion and a new \emph{verification methodology} for proving equivalence between the insertion-only version and the corresponding retractable one.
This equivalence checking problem is inherently challenging because (1)  the retractable computation may maintain a different internal summary;  (2) executions may contain arbitrary interleavings of insertions and deletions, and  (3) correctness   depends on which deletions must be handled incrementally. Our proposed verification methodology   addresses all of these challenges by reducing a difficult global reasoning problem to local proof obligations over inductive relational invariants. The result is a sound and practically effective method for establishing semantic equivalence despite differences in internal representation and execution history.

Our second technical contribution is a \emph{synthesis framework} for  correct-by-construction retractable computations. The central difficulty is that the space of correct solutions is constrained simultaneously by the choice of retained summary information and by how deletion is performed over that representation. Our approach addresses this tension by jointly searching for auxiliary summary state and a \emph{lifting invariant} that semantically relates the original fold to its enriched counterpart. Once the required auxiliary state is identified, our method derives the retraction logic by inferring a relational invariant as a local specification and then decomposing the implementation into semantic cases, each capturing a different way the removed element can affect the maintained summary.

{
We implemented the proposed approach in a new tool called \tool{} and evaluated it on 94 benchmarks covering  real-world aggregations written in the Flink framework~\cite{flink}. \tool{} successfully synthesized correct retraction logic for all 94 scenarios, including cases where the original manually-written logic is wrong.
We also conducted  ablation studies  demonstrating that all of our proposed  ingredients are crucial for the practicality of \tool{}.} Finally, our performance results show that the synthesized
incremental deletions  substantially outperform recomputation.

To summarize, this paper makes the following contributions:
\begin{itemize}[leftmargin=*]
\item \textbf{Formalization and verification.}
We formalize correctness for retractable folds by requiring that any supported sequence of insertions and deletions produce the same result as re-running the original fold on the remaining elements. We then give a sound and relatively complete verification methodology based on inductive relational invariants.

\item \textbf{Synthesis of retractable folds.}
We develop a synthesis framework that turns an insertion-only fold into a retractable one by jointly discovering the enriched summary state and the operations over it. Rather than assuming a fixed representation, our method determines what information must be retained to make the required deletions possible, synthesizes lifted update logic to maintain that information, and derives a retraction function that is correct by construction.

\item \textbf{Implementation and evaluation.}
We implement the approach in a prototype called \tool{} and evaluate it on 94 benchmarks over 48 real-world aggregations and multiple retraction policies. Our evaluation demonstrates high synthesis success rates, the ability to uncover bugs in manual implementations, and the practical importance of the method's main ingredients.
\end{itemize}

\section{Motivating Example}\label{sec:motivating}

\begin{wrapfigure}{r}{0.48\textwidth}
\small
\vspace{-0.2in}
\begin{minted}[
  fontsize=\footnotesize,
  escapeinside=||,
  numbersep=5pt
]{scala}
case class BidAcc(var maxBid = -Inf, 
                  var highBidCount = 0)
def accumulate(acc: BidAcc, price: Float) = {
  if (price > acc.maxBid) acc.maxBid = price
  if (price > 1000) acc.highBidCount += 1 }
\end{minted}
\vspace{-0.14in}
\caption{Insertion-only aggregation.}
\label{fig:motivating-example}
\vspace{-0.2in}
\end{wrapfigure}
We illustrate our approach using an aggregation function, adapted from~\cite{ink}, that
computes auction statistics, including the maximum bid and the
number of ``high'' bids  exceeding \$1000. This example captures
a common pattern in streaming analytics, where an accumulator
maintains key statistics, such as extremal values or counts. The insertion-only
version of this computation, shown in \Cref{fig:motivating-example}, is
straightforward: the accumulator stores two fields, namely the current maximum
and the high-bid count, and each call to \texttt{accumulate} updates
these fields incrementally.

Now suppose the auction allows bidders to remove previously placed bids, but
only under a specified service policy. One natural policy is a fixed guarantee:
the auction is guaranteed to support up to $N$ bid removals, regardless of which
bids are removed (\textbf{Policy~A}). We express this policy as:
\[
\RetSpec_A(\udafinput{}) =
\min(\mathrm{length}(\udafinput{}), N)
\]
which says that, after bid history $\udafinput$, the system must support
removing $N$ bids if there are at least $N$ bids and all bids otherwise. Another natural policy is to protect the stability of the standing price. Under this policy, if the highest bid seen so far appears $k$ times, the auction guarantees support for $k-1$ arbitrary bid removals, so that the highest bid is not completely removed even in the worst case (\textbf{Policy~B}). We express this policy as:
\[
\RetSpec_B(\udafinput{}) =
\mathrm{count}(\udafinput{}, \max(\udafinput{})) - 1
\]

\begin{figure}[t]
\small
\centering
\begin{subfigure}[t]{0.49\linewidth}
\centering
\begin{minted}[
  fontsize=\footnotesize,
  escapeinside=||,
  numbersep=5pt
]{scala}
case class BidAcc(var maxBid = -Inf, 
    var highBidCount = 0, 
    var pastMaxElems = List(maxlen=N + 1), 
    var rets = 0)
def accumulate(acc: BidAcc, price: Float)  = {
  acc.maxBid = max(acc.maxBid, price)
  if (acc.pastMaxElems.length >= N) {
    acc.pastMaxElems.append(price).dropMin()
  } else {
    acc.pastMaxElems.append(price) }
  if (price > 1000) acc.highBidCount += 1 }
def retract(acc: BidAcc, price: Float) = {
  assert(acc.rets < N, "Cannot retract")
  if (acc.pastMaxElems.membership(price))  
        acc.pastMaxElems.remove(price)
  acc.maxBid = acc.pastMaxElems.maxOr(-Inf)
  acc.rets += 1
  if (price > 1000) acc.highBidCount -= 1 }
\end{minted}
\vspace{-0.7em}
\caption{ Variant supporting up to $N$ withdrawals}
\label{fig:motivating-example-lifted-N}
\end{subfigure}\hfill
\begin{subfigure}[t]{0.49\linewidth}
\centering
\begin{minted}[
  fontsize=\footnotesize,
  escapeinside=||,
  numbersep=5pt
]{scala}
case class BidAcc(var maxBid = -Inf, 
    var highBidCount = 0, 
    var maxBidCount = 0)

def accumulate(acc: BidAcc, price: Float) = {
  if (price > acc.maxBid) {
    acc.maxBid = price; acc.maxBidCount = 1
  } else if (price == acc.maxBid) {
    acc.maxBidCount += 1 }
  if (price > 1000) acc.highBidCount += 1  }

def retract(acc: BidAcc, price: Float) = {
    if (price == acc.maxBid) {
        assert(acc.maxBidCount > 1, 
            "Cannot retract")
        acc.maxBidCount -= 1 }

    if (price > 1000) acc.highBidCount -= 1 }
\end{minted}
\vspace{-0.7em}
\caption{ Variant supporting  multiplicity-based max-bid}
\label{fig:motivating-example-lifted-count}
\end{subfigure}
\caption{Two retractable variants of the auction aggregation: (a) supports a fixed budget of up to $N$ withdrawals; (b) supports the worst-case budget induced by multiplicity of the maximum bid.}
\label{fig:motivating-example-lifted-both}
\vspace{-0.16in}
\end{figure}

These two policies illustrate different kinds of retraction requirements:
Policy~A imposes a fixed removal budget, while Policy~B makes the guarantee
dependent on redundancy in the current auction state.
Furthermore, the original 
aggregation state in \Cref{fig:motivating-example} cannot support either policy as
written, because it discards information that may be needed when bids are later
removed. 

The missing information depends on the policy. To support Policy~A, the
aggregate must retain enough high-valued bids to recover the maximum after a
bounded number of removals, and it must also track how many removals have
already been processed. To support Policy~B, the aggregate only needs to know whether
the current maximum has another copy. Thus, even for the same base computation, different retraction requirements call for different auxiliary
summaries. 

\Cref{fig:motivating-example-lifted-N,fig:motivating-example-lifted-count}
show two corresponding retractable variants of the computation from \Cref{fig:motivating-example} implementing Policies A and B, respectively. Both extend the original
aggregation in three ways: they add auxiliary state, revise the insertion logic
so that this state is maintained as new bids arrive, and implement retraction
logic that updates the enriched state when bids are removed. Crucially, the
details of all three components depend on the policy.

The rest of this paper shows how to automatically generate retractable
computations like those in
\Cref{fig:motivating-example-lifted-N,fig:motivating-example-lifted-count}.
Given an insertion-only computation such as
\Cref{fig:motivating-example} and a user-provided retraction specification,
such as $\RetSpec_A$ or $\RetSpec_B$, our method synthesizes the enriched state,
the lifted insertion logic, and the corresponding retraction procedure. Crucially, the synthesized computation comes with a formal guarantee: it
satisfies the specified retraction requirement, and every supported sequence of
removals yields the same observable result as recomputing the original
aggregation.

\section{Problem Statement}\label{sec:problem}

In this section, we formalize retractable folds and their semantics.

\subsection{Semantics of Retractable Folds}

\begin{definition}[Fold-based computation]\label[definition]{def:fold}
A \emph{fold-based computation} is a triple
$\prog = (\Init,\Acc,\eval)$, where $\Init \in \UDState$ is an initial state,
$\Acc : \UDState \times \Elem \to \UDState$ is an accumulator function, and
$\eval : \UDState \to \Out$ is an evaluation function. {For an input list $\xs=[x_1,\dots,x_n]$, let
$s_0,\dots,s_n$ be a state sequence defined as:}
\[
s_0 \;=\; \Init,\qquad
s_j \;=\; \Acc(s_{j-1},x_j)
\quad\text{for } 1 \le j \le n
\]
The \emph{state semantics} of $\prog$, denoted $\psem{\prog}{\xs}$, is the final state $\psem{\prog}{\xs} \triangleq s_n$, and the \emph{observable semantics} of $\prog$ is
$ 
\sem{\prog}{\xs} \triangleq \eval(\psem{\prog}{\xs})$.
\end{definition}

{Thus, the computation has the standard form
$\eval(\fold(\Acc,\Init,\xs))$, where
$\fold$ builds the internal state, and $\eval$ converts that state
into an observable output. For example, an average computation may 
maintain a sum and count, and $\eval$ returns their quotient.}
A \emph{retractable fold}, which is the central object of study in
this paper, extends \Cref{def:fold} with a \emph{retraction function}
for supporting incremental deletions:

\begin{definition}[Retractable fold]
A \emph{retractable fold} is a quadruple
$ 
\rudaf = (\Init,\Acc,\Retract,\eval)
$, 
where $\Init \in \UDState$ is an initial state, $\Acc : \UDState \times \Elem \to \UDState$ is an accumulator function, $\Retract : \UDState \times \Elem \to \UDState \cup \{\mathsf{Err}\}$ is a retraction function, and $\eval : \UDState \to \Out \cup \{\Err\}$ is an evaluation function.
\end{definition}

At a high level, $\Retract(s,a)$ attempts to remove one prior occurrence of $a$ from the history represented by summary state $s$. Since the maintained state may not preserve enough information to support such a reversal, $\Retract$ is allowed to fail with $\Err$. The function $\eval$ maps the internal state to the observable output, discarding any auxiliary state. To define the semantics of a retractable fold under both insertions and deletions, we represent its input  as an \emph{update sequence}:

\begin{definition}[Update sequence]
An \emph{update sequence} $\invseq$ is a list
$
\invseq \in \List{\{\mathsf{acc}, \mathsf{retract}\} \times \Elem},
$
where each element consists of an operation tag ($\mathsf{acc}$ or $\mathsf{retract}$) 
and an input value $a \in \Elem$.
For $\invseq=\langle(o_1,a_1),\dots,(o_n,a_n)\rangle$, we define:
\[
\invseq^+ = [\, (o_j, a_j) \mid o_j=\mathsf{acc}\,], \quad
\invseq^- = [\, (o_j, a_j) \mid o_j=\mathsf{retract}\,], \quad
\vals{\invseq} = [a_j].
\]
In other words, $\invseq^+$ and $\invseq^-$  yield only the insertions and deletions respectively, whereas 
 $\vals{\invseq}$ drops the operation tags entirely.
\end{definition}

For a pair of input sequences $\invseq, \invseq'$, we write $\invseq \preceq \invseq'$ if $\invseq$ is a prefix of $\invseq'$. 
Next, we define a \emph{valid update sequence} where each deletion contains a matching insertion:

\begin{definition}[Valid sequence]\label[definition]{def:valid-seq}
An update sequence $\invseq = \langle (o_1,a_1), \dots, (o_n,a_n) \rangle$ is
\emph{valid} if
\[
\forall j. (1 \leq j \leq n) \implies  \big(
{
\sum_{1 \leq i \leq j}{\mathbb{1}\{o_i=\mathsf{acc} \land a_i=a_j \}} \geq \sum_{1 \leq i \leq j}{\mathbb{1}\{o_i=
\mathsf{retract} \land a_i=a_j \}}
}
\big)\]
\end{definition}

That is, an input to a retractable fold is valid if every deletion has a matching insertion that precedes it.  From now on, we use the term update sequence to mean one that is valid.

\begin{definition}[Retractable fold semantics]\label[definition]{def:rudaf-semantics}
Consider a retractable fold $\rudaf = (\Init,\Acc,\Retract,\eval)$  and an update sequence $\invseq=\langle(o_1,a_1),\dots,(o_n,a_n)\rangle$.
Let $s_0,\dots,s_n$ be a state sequence defined as:
\[
s_0 \;=\; \Init,\qquad
s_j \;=\;
\begin{cases}
\Acc(s_{j-1},a_j)      & \text{if } {s_{j-1} \neq \mathsf{Err} \ \land\ } o_j=\mathsf{acc} \\[2pt]
\Retract(s_{j-1},a_j)  & \text{if } {s_{j-1} \neq \mathsf{Err} \ \land\ } o_j=\mathsf{retract} \\[2pt]
{\mathsf{Err}} & {\text{otherwise}}
\end{cases}
\]
The \emph{state semantics} of $\rudaf$, denoted $\psem{\rudaf}{\invseq}$,  is the final state
$\psem{\rudaf}{\invseq} \triangleq s_n$.
The \emph{observable semantics} is the corresponding output value:
\[
\sem{\rudaf}{\invseq} \triangleq
( \psem{\rudaf}{\invseq} \neq \mathsf{Err} \ \ ?  \ \   \eval(\psem{\rudaf}{\invseq}) \ \   : \ \  \mathsf{Err} ) 
\]

\end{definition}

\subsection{Retraction Specifications and Equivalence}

To define correctness for a retractable fold, we must first specify
the type of deletion support it is required to provide. As illustrated
in \Cref{sec:motivating}, a computation can admit multiple retractable
variants, each supporting a different class of deletions.
{Requiring unrestricted deletion support may impose unnecessary
storage costs when an application needs only a limited guarantee.
We therefore specify how many arbitrary deletions must be supported,
allowing this number to depend on the insertion history.}
For example, in Policy~A, the budget is bounded by both $N$ and the
number of placed bids, whereas in Policy~B, it depends on the
multiplicity of {the maximum bid in the insertion history}.
{We formalize these requirements as follows:}

\begin{definition}[Retraction specification]\label[definition]{def:retspec}
A \emph{retraction specification} for $\rudaf$ is a function
$
\RetSpec : \List{\Elem} \to \mathbb{N}
$. 
Formally, $\rudaf$ satisfies $\RetSpec$, denoted $\rudaf \models \RetSpec$, iff
$ 
\forall \invseq. \ \big( (  
 \forall \delta \preceq \invseq.\ |\delta^-| \le \RetSpec(\vals{\delta^+}))
\ \Longrightarrow\ 
\sem{\rudaf}{\invseq} \neq \mathsf{Err} \big)
$.

\end{definition}

In other words, $\rudaf$ satisfies $\RetSpec$ if it can process any update
sequence whose prefixes stay within the deletion budget specified by
$\RetSpec$.
{This budget specifies a \emph{minimum} guarantee. In particular,
the fold \emph{may} support additional deletions, but it is not required to.}

\begin{example}
{
Suppose an application must support withdrawing any previously inserted
record. This requirement is expressed by $\RetSpec(\xs)=|\xs|$.
Each new insertion increases the deletion allowance, so a record
that is withdrawn and later reinserted can be withdrawn again.
In contrast, $\RetSpec(\xs)=\min(1,|\xs|)$ guarantees only one
withdrawal over the entire execution, even if more records are
subsequently inserted.
An application may also require a correction allowance proportional
to its workload. For example,
$\RetSpec(\xs)=\lfloor|\xs|/100\rfloor$ guarantees support for a total
of up to one withdrawal per hundred records inserted, regardless
of which records are withdrawn.
}
\end{example}

Next, to define equivalence between the original computation and its retractable version, we introduce the concept of \emph{normalizing} an update sequence:

\begin{definition}[Normalization]\label[definition]{def:normalization}
Given  $\invseq = \langle (o_1,a_1), \dots, (o_n,a_n) \rangle$, the \emph{normalization} of $\invseq$, written $\Norm(\invseq)$,  is the result of repeatedly applying the {following} rewrite rule to a fixpoint:
\[
(o_i,a),\; (o_j,a) \;\leadsto\; \epsilon \quad \quad \textrm{where:}
 \quad
(1) \quad o_i = \mathsf{acc},\quad o_j = \mathsf{retract},\quad i < j,\]
\[
(2) \quad \forall k.\ (1 \le k < i \implies a_k \neq a \ \lor\ o_k \neq \mathsf{acc}), \quad
(3) \quad { \forall k.\ (i < k < j \implies a_k \neq a \ \lor \ o_k \neq \mathsf{retract}) }
\] 
$\Effect(\invseq)$ yields the list of elements in $\Norm(\invseq)$, i.e., 
 $\Effect(\invseq) = \vals{\Norm(\invseq)}$.
\end{definition}
Hence, normalization repeatedly cancels the earliest valid retraction of a value with its earliest preceding accumulation, thereby producing a canonical sequence that reflects the net effect.

\begin{definition}[Equivalence]\label[definition]{def:equivalence}
Let $\udaf = (\Init,\Acc,\eval)$ be a regular fold and 
$\rudaf = (\Init',\Acc',\Retract,\eval')$  a retractable one.
We say that $\rudaf$ is \emph{equivalent to} $\udaf$ modulo $\RetSpec$, denoted $\udaf \equiv_\RetSpec \rudaf$, iff
(1) $\rudaf \models \RetSpec$, and (2) for every valid update sequence $\invseq$ with $xs =  {\Effect(\invseq)}$, 
        if $\sem{\rudaf}{\invseq} \neq \mathsf{Err}$, then $
        \sem{\rudaf}{\invseq} =  \sem{\udaf}{xs}$.
\end{definition}

The definition above captures two requirements: $\rudaf$ must satisfy
the retraction specification, and it must agree semantically with
$\prog$. That is, for any valid update sequence that does not produce
an error, $\rudaf$ returns the same output as $\prog$ run on the
normalized sequence, even if the sequence exceeds the deletion
allowance specified by $\RetSpec$.

\begin{definition}[Problem statement]
Given a regular fold ~$\udaf$ and a retraction specification~$\RetSpec$,
the {synthesis problem} is to construct a retractable fold~$\rudaf$ such that
$\udaf \equiv_\RetSpec \rudaf$.
\end{definition}

For a given $\udaf$ and specification $\RetSpec$, no retractable fold $\rudaf$ over the same internal state $S$ may satisfy $\udaf \equiv_\RetSpec \rudaf$. In such cases, we enrich $\rudaf$'s state to a different type $S'$, a process we call \emph{lifting}.

\section{Verification Methodology}\label{sec:verification}

To solve the synthesis problem, we first need a way to formally verify equivalence between a retractable
computation and its insertion-only counterpart. However, this verification
problem raises several core difficulties that are not directly handled by
standard relational proof techniques:
\begin{itemize}[leftmargin=*]
\item {\bf Different internal states.} First, because the retractable version may maintain
an enriched summary, 
correctness cannot require the two executions to reach the same internal state. 
\item {\bf Arbitrary update interleavings.} Second, the desired guarantee must hold for \emph{all} valid update
sequences, including arbitrarily long executions in which insertions and
deletions are interleaved arbitrarily. This requires an inductive argument showing that each supported
deletion leaves the retractable state consistent with  the normalized history and
able to support subsequent updates. 
\item {\bf Multiple semantic views of execution.} Third, establishing correctness requires tracking  three related but distinct views
of the execution: the mixed sequence executed by the retractable fold, the
normalized insertion-only sequence that determines its observable result, and
the original insertion history that determines which future deletions must still
be supported. This differs from traditional equivalence checking, where it
suffices to relate two executions. 
\item {\bf History-sensitive retraction obligations.} Finally, the retraction
requirement is itself history-sensitive, meaning it depends on which elements
were previously inserted and how many deletions have already been performed.
Thus, the verifier must reason about when a retraction is within the promised
budget, rather than treating deletion as an unconstrained inverse operation.
\end{itemize}
These requirements do not fit naturally into standard relational equivalence checking frameworks. For example, although logics such as RHL~\cite{rhl},  CHL~\cite{chl}, and their variants~\cite{qchl,prhl,apRHL,product,asymmetricProduct}, can relate multiple executions, their specifications are formulated as relational pre/postconditions over a fixed collection of runs. In contrast, correctness for retractable folds requires an inductive invariant that is closed under arbitrary future updates. 

Motivated by these challenges, this section develops a domain-specific relational proof system for verifying retractable computations. The proof system decomposes the global equivalence argument into three local obligations, summarized in \Cref{fig:proof-rules}:

\begin{itemize}[leftmargin=*]
\item {\bf Lifting consistency (Block A).}
The first obligation relates the original  state to the enriched state maintained by the retractable computation. Since the retractable version may store additional information, the two states need not be equal; instead, Block~\textbf{A} shows that they agree observationally on insertion-only histories.

\item {\bf Retraction consistency (Block B).}
The second obligation relates the actual mixed execution of the retractable fold to its own normalized execution. In particular, Block~\textbf{B} proof rules establish that each successful retraction has the same semantic effect as removing a matching insertion from the history, thereby keeping the mixed execution aligned with its normalized counterpart.

\item {\bf Retraction support (Block C).}
The third obligation captures when a retraction is required to succeed. The
Block~\textbf{C}  rules track how much of the deletion budget has already
been consumed and use this information to determine when the current state must
 support another deletion.
\end{itemize}

Finally, Block~\textbf{R} defines the reachability predicate used to characterize states produced by insertion-only histories. These rules collectively reduce the global equivalence theorem to local proof obligations over relational invariants that can be discharged by SMT solvers. We next explain each group of verification conditions in more detail. Throughout, $s$, $\init$, $\acc$, and $\eval$ refer to the original fold, while $s'$, $\init'$, $\acc'$, and $\eval'$ refer to the retractable fold.

\bfpara{Reachability rules.} These are the simplest rules and define the auxiliary predicate $\Reach(\udafinput, s')$, which holds
when $s'$ is the state of the retractable fold that is obtained through insertions 
$\udafinput$ starting at the initial state. 
Rule~\textbf{R1} establishes the base case, and ~\textbf{R2} captures the inductive step.

\begin{figure}[t]
\centering
\small

\vspace{0.6em}

\fbox{%
\begin{minipage}{0.98\linewidth}
\textbf{R.\ Reachability invariant}\medskip

\noindent\textbf{R1 (Base)}\quad
$\Reach(\emptylist{},\,\init')$.

\medskip
\noindent\textbf{R2 (Step)}\quad
$\Reach(\udafinput,\,s') \implies \Reach(\udafinput\listappend{}\List{a},\,\acc'(s',a))$.
\end{minipage}
}

\vspace{0.6em}

\fbox{%
\begin{minipage}{0.98\linewidth}
\textbf{A.\ {Lifting consistency}} \medskip

\noindent\textbf{A1 (Base)}\quad
$\eval(\init) =\eval'(\init')$.

\medskip
\noindent\textbf{A2 (Acc step)}\quad
$\eval(s) =\eval'(s') \implies \eval(\acc(s,a)) = \eval'(\acc'(s',a))$.
\end{minipage}
}

\vspace{0.6em}

\fbox{%
\begin{minipage}{0.98\linewidth}
\textbf{B.\ {Retraction consistency}}\medskip

\noindent\textbf{B1 (Base)}\quad
$\SimTwo(\emptylist,\,\init',\,\init')$.

\medskip
\noindent\textbf{B2 (Acc step)}\quad
$\SimTwo(xs,\,s_1',\,s_2') \implies \SimTwo(xs\listappend{}\List{x},\,\acc'(s_1',\,x),\,\acc'(s_2',\,x))$.

\medskip
\noindent\textbf{B3 (Ret step)}\quad
$\bigl ( \Reach(xs,\,s_1') \wedge\SimTwo(xs,\,s_1',\,s_2')   \wedge x \in xs \wedge  \Retract(s_2',\,x)\neq \Err{}  \\
\phantom{\noindent\textbf{B2 (Acc step)}\quad}
 \land ys = \remove(xs,\,x) \land \Reach(ys,\,s_3') \bigr )
\implies \SimTwo(ys,\,s_3',\,\Retract(s_2',\,x))$.

\medskip
\noindent\textbf{B4 (Eval)}\quad
$\SimTwo(xs,\ s_1',\ s_2') \implies \eval'(s_1') = \eval'(s_2')$.

\end{minipage}
}

\vspace{0.6em}

\fbox{%
\begin{minipage}{0.98\linewidth}
\textbf{C.\ Quantitative retraction support}\medskip

\noindent\textbf{C1 (Base)}\quad
$\SimBudget(\init',\,\init',\,0)$.

\medskip
\noindent\textbf{C2 (Acc step)}\quad
$\SimBudget(s_1',\,s_2',\,n) \implies \SimBudget(\acc'(s_1',\,x),\,\acc'(s_2',\,x),\,n)$.

\medskip
\noindent\textbf{C3 (Ret step)}\quad
$(\SimBudget(s_1',\,s_2',\,n) \wedge \SimTwo(\udafinput{},\,s_3',\,s_2') \wedge x \in \udafinput{} \wedge \Retract(s_2',\,x)\neq \Err{})
\implies \SimBudget(s_1',\,\Retract(s_2',\, x),\,n+1)$.

\medskip
\noindent\textbf{C4 (Capability)}\quad
$(\Reach(ys,\,s_1') \land \SimBudget(s_1',\,s_2',\,n) \land
\SimTwo(xs,\,s_3',\,s_2') \\
\phantom{\noindent\textbf{C4 (Capability)}\quad}
\land \Subseq(ys,\,xs,\,n) \land x \in xs \land n < \RetSpec(ys))
\implies  \Retract(s_2',\,x)\neq \Err{}$.

\end{minipage}
}
\vspace{-0.1in}
\caption{Proof rules  for proving equivalence modulo a retraction specification $\RetSpec$. The primed versions (e.g., $\Acc'$) refer to functions in the retractable fold, and non-primed function names refer to those in the insertion-only one. $\Subseq(ys,xs,n)$ means $xs$ is obtained from $ys$ by deleting exactly $n$ elements.}
\label{fig:proof-rules}
\vspace{-0.2in}
\end{figure}

\bfpara{Lifting consistency rules.}
Next, Block~\textbf{A} rules deal with insertion-only executions and aim to  show that the retractable fold preserves the observable behavior of the original fold even when it contains a richer internal summary. Rule~\textbf{A1} establishes this agreement at the initial states, and Rule~\textbf{A2} shows that the agreement is preserved when both folds process the same insertion.

\bfpara{Retraction consistency rules.}
Block~\textbf{B} defines a crucial relational invariant $\SimTwo$ that connects two different states of the same retractable aggregation: one reached through an insertion-only update sequence, and the other reached by executing a mixed sequence with both insertions and deletions. In particular, $\SimTwo(xs,s_1',s_2')$ means that $s_1'$ is the state obtained by running the retractable fold on the insertion-only history $xs$, while $s_2'$ is the state obtained from some mixed update sequence whose net effect after normalization is exactly $xs$. Thus, $\SimTwo$ says that the mixed state $s_2'$ has the same semantic meaning as the insertion-only state $s_1'$, and its purpose is to show that the semantics of retraction is consistent with the net effect of the  normalized (insertion-only) update sequence.

Concretely, Rule~\textbf{B1} initializes $\SimTwo$ at the empty history, and Rule~\textbf{B2} shows that the relation is preserved when both sides process the same insertion. The key rule is~\textbf{B3}, which captures the effect of a successful retraction. Suppose $s_1'$ is a reachable reference state for an insertion-only history $xs$, and suppose the actual mixed state $s_2'$ is related to $s_1'$ by $\SimTwo(xs,s_1',s_2')$. If $x$ is one of the elements in $xs$ and retracting $x$ from $s_2'$ succeeds, then the rule compares the result against the reference state for the history obtained by removing one occurrence of $x$. That is, letting $ys = \remove(xs,x)$ and letting $s_3'$ be a reachable state for $ys$, the post-retraction state $\Retract(s_2',x)$ must be related to $s_3'$ by $\SimTwo(ys,s_3',\Retract(s_2',x))$. Thus, retracting $x$ from the mixed execution corresponds exactly to removing one matching occurrence of $x$ from the insertion-only history. Finally, rule~\textbf{B4} turns this state relation into observational agreement: whenever two retractable states are related by $\SimTwo$, they must produce the same output under $\eval'$.

\bfpara{Quantitative retraction support rules.}
The previous rules show that retractions preserve the observable semantics of the computation \emph{if} they succeed. But they do not enforce that a retraction actually succeeds when it is required to do so. This is the role of Block~\textbf{C}. Its rules track the gap between two executions: the full insertion-only execution that determines the deletion budget promised by $\RetSpec$, and the actual mixed execution in which some deletions may already have occurred.

To capture this gap, Block~\textbf{C} introduces an indexed relation $\SimBudget(s_1',s_2',n)$, where $s_1'$ is the full insertion-only state, $s_2'$ is the current mixed-execution state, and $n$ counts the deletions already used. Rule~\textbf{C1} initializes the relation with distance $0$; Rule~\textbf{C2} preserves this distance across common insertions; and Rule~\textbf{C3} increments it after a successful retraction.

The key rule is~\textbf{C4}, which turns this bookkeeping into the actual proof obligation. The rule compares two insertion-only histories associated with the same mixed execution. The first, written $ys$, is the full insertion history: it contains every element that has been inserted so far, ignoring deletions. The second,  $xs$, is the effective history that remains after canceling the deletions that have already occurred. Thus, for a mixed update sequence $\invseq$, we can think of $ys$ as $\vals{\invseq^+}$ and $xs$ as $\Effect(\invseq)$. 
Thus, in this rule, the premise $\Reach(ys,s_1')$ identifies the reference state for the full insertion history, which is the history to which $\RetSpec$ applies. The premise $\SimBudget(s_1',s_2',n)$ says that the current state $s_2'$ has already used $n$ deletions relative to that full history. The premise $\SimTwo(xs,s_3',s_2')$ identifies the effective history
represented by the current mixed state, where $s_3'$ is the
corresponding insertion-only reference state.
The premise $\Subseq(ys,xs,n)$ connects the two histories by saying
that $xs$ is obtained from $ys$ by deleting exactly $n$ elements.
With these facts in place, Rule~\textbf{C4} states the promised capability condition. If $n < \RetSpec(ys)$, then the deletion budget promised for the full insertion history $ys$ has not yet been exhausted. Therefore, every element $x$ that still remains in $xs$ must be retractable from the current state $s_2'$.

\begin{wrapfigure}{r}{0.45\textwidth}
\vspace{-0.2in}
\includegraphics[scale=0.55]{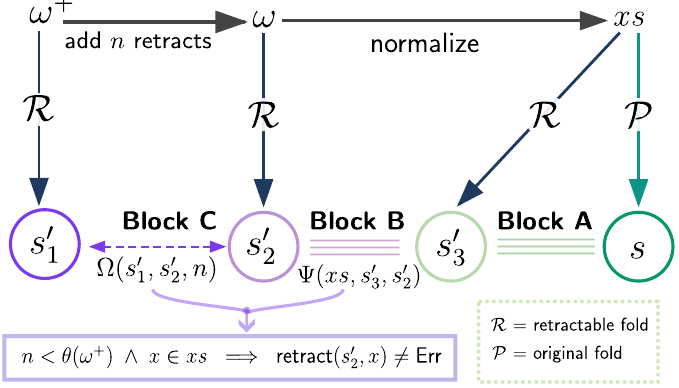}
\vspace{-0.27in}
\caption{Proof methodology}
\label{fig:proof}
\vspace{-0.2in}
\end{wrapfigure}

\bfpara{Putting the pieces together.}
\Cref{fig:proof} illustrates how these  proof rules combine to establish equivalence between $\udaf$ and $\rudaf$ modulo $\RetSpec$. Consider a valid mixed update sequence $\omega$. Let $xs = \Effect(\omega)$ be the  normalized history, and let $n = |\omega^-|$ be the number of retractions in $\omega$. The proof associates four states with this execution. State $s_1'$ is obtained by running $\rudaf$ on the full insertion history $\vals{\omega^+}$, where all insertions are replayed and retractions are ignored. State $s_2'$ is obtained by running $\rudaf$ on the actual mixed sequence $\omega$. State $s_3'$ is obtained by running $\rudaf$ on the effective history $xs$. Finally, state $s$ is obtained by running the original fold $\udaf$ on the same history $xs$.

The right side of the figure establishes observational agreement. Block~\textbf{B} relates the actual mixed state $s_2'$ to the reference retractable state $s_3'$ for the effective history $xs$. This shows that, whenever retractions succeed, executing the mixed sequence has the same effect as the normalized history. Block~\textbf{A} then relates $s_3'$ to the original fold state $s$, showing that the retractable fold agrees with the original fold on insertion-only histories. Hence, Blocks~\textbf{A} and~\textbf{B} together establish that, as long as $\omega$ does not produce $\Err$, the output of $\rudaf$ on $\omega$ agrees with the output of $\udaf$ on $xs$.

The left side of the figure establishes that the execution does not get stuck before the promised deletion budget is exhausted. Block~\textbf{C} relates the full-history state $s_1'$ to the actual mixed state $s_2'$ using the indexed relation $\SimBudget(s_1',s_2',n)$, where $n$ records how many retractions have already been used. Since Block~\textbf{B} also identifies $xs$ as the effective history represented by $s_2'$, Block~\textbf{C} can express the capability obligation imposed by $\RetSpec$: if $n < \RetSpec(\vals{\omega^+})$, then every element still present in $xs$ must be retractable from $s_2'$. Thus, every update sequence whose prefixes remain within the budget specified by $\RetSpec$ avoids $\Err$ and agrees observationally with the original fold.

\begin{example}
Consider the auction example from \Cref{sec:motivating}. Here, \Cref{fig:motivating-example} is the insertion-only fold, and \Cref{fig:motivating-example-lifted-count} is its retractable version under Policy~B. To prove equivalence modulo the specification $\RetSpec_B$ from \Cref{sec:motivating},  we can instantiate the invariants in \Cref{fig:proof-rules} as follows,  abbreviating the state as $(m,h,c)$ for \texttt{maxBid}, \texttt{highBidCount}, and \texttt{maxBidCount} respectively: 
\[
\begin{aligned}
\Reach(xs,(m,h,c)) \triangleq {}&
  h = \#\{y \in xs \mid y > 1000\} \wedge 
(xs = \emptylist \Rightarrow m = -\infty \wedge c = 0)  \\
& \wedge (xs \neq \emptylist \Rightarrow m = \max(xs) \wedge c = \mathrm{count}(xs,m) \wedge c > 0),\\
\SimTwo(xs,(m_1,h_1,c_1),(m_2,h_2,c_2)) \triangleq {}&
  m_1 = m_2 \wedge h_1 = h_2 \wedge c_1 = c_2,\\
\SimBudget((m_1,h_1,c_1),(m_2,h_2,c_2),n) \triangleq {}&
  n \ge 0 \wedge m_1 = m_2 \wedge c_2 + n \geq c_1  \wedge (n = 0 \Rightarrow h_1 = h_2).
\end{aligned}
\]
Here, $\Reach$ gives the insertion-only meaning of the state: $h$ counts high bids, $m$ is the maximum bid, and $c$ is its multiplicity. Since these fields fully characterize the retractable summary, $\SimTwo$ simply requires componentwise equality. $\SimBudget$ tracks the remaining support under Policy~B. If the full insertion history contains $c_1$ copies of the maximum and the current mixed state contains $c_2$ copies, then after $n$ retractions we must have $c_2+n \geq c_1$. This ensures that, as long as the deletion budget is not exhausted, the current state still retains enough copies  to support another  retraction.

\todo{

}
\end{example}

\begin{theorem}[Soundness and Relative Completeness]\label{thm:proof-rules-sound}
Let $\udaf$ be an insertion-only fold, $\rudaf$ a retractable fold, and $\RetSpec$ a retraction specification. Then, we have: \\ \textbf{(1) Soundness:}  If there exist predicates $\SimTwo$, $\SimBudget$,  $\Reach$ satisfying  proof rules in \Cref{fig:proof-rules}, then $\udaf \equiv_{\RetSpec} \rudaf$. \\
\textbf{(2) Relative completeness:} Conversely, if $\udaf \equiv_{\RetSpec} \rudaf$ and the  assumptions stated in \Cref{sec:complete} hold, then there exist predicates $\SimTwo$, $\SimBudget$, and $\Reach$ such that all proof rules in \Cref{fig:proof-rules} are valid.\footnote{Proofs of all theorems are in the appendix.}
\end{theorem}

{Relative completeness relies on three assumptions.
First, the assertion language must be expressive enough to describe
the required invariants.
Second, states with the same observable output must remain
indistinguishable under future insertions.
Third, executions with the same surviving input sequence must agree
on  their outputs and on which future deletions they can support.
A more formal treatment of these assumptions can be found in \Cref{sec:complete}.}

\section{Synthesis Algorithm}\label{sec:synthesis}

Given a standard fold $\udaf$ and retraction requirement $\RetSpec$, the synthesis goal is to generate a retractable fold $\rudaf$ such that $\udaf \equiv_{\RetSpec} \rudaf$. To solve this problem, the algorithm must first determine how to enrich the internal state so that the synthesis task is realizable. It must then update the existing insertion logic to correctly maintain the lifted state and, finally, construct the retraction function from scratch. 
However, the core challenge is that these three tasks are fundamentally entangled: The very existence of a correct retraction function depends on what the lifted state is, yet deciding how to enrich the state   in turn depends on the semantics of retraction.

Our method mitigates this apparent circularity by first reasoning about
\emph{lifting invariants} rather than directly searching over \emph{retractable
folds}. A lifting invariant specifies what information the enriched state is
intended to contain, relative to the original fold state and the insertion
history, without yet specifying how the accumulator maintains that information.
This allows the algorithm to rule out semantically inadequate lifts before
attempting to synthesize either the lifted insertion logic or the retraction
function. Once a suitable lift has been identified and realized, the remaining challenge is
to synthesize a retraction function whose behavior is only \emph{indirectly} specified
by the retraction requirement and is often case-dependent. Our algorithm addresses this challenge by symbolically partitioning the deletion
space into semantically distinct cases, deriving local synthesis constraints for
each case, and solving the resulting subproblems separately. The rest of this section first presents the
top-level synthesis algorithm and then explains its key subprocedures.

\subsection{Synthesis Overview}

\Cref{alg:synthesize-rudaf} gives an overview of our  synthesis procedure.  The top-level search ranges over candidate \emph{lifting specifications}, consisting of a  \emph{lifted state} together with a \emph{lifting invariant} that describes the intended semantics.  
Concretely, a lifting specification has the form:
\[ 
  \{\, (s:\tau,\; a:\tau_{\mathit{aux}}) \mid \qual(s,a, xs) \,\}
\] 
where $\tau$ is the type of the original state, $\tau_{\mathit{aux}}$ is the type of the auxiliary state,
and the logical qualifier $\qual(s, a,xs)$ is the lifting invariant that describes how the auxiliary component $a$ relates to the input sequence $xs$ and existing state $s$.

\begin{example} \label[example]{example:syn-count-lifting}
Consider an insertion-only fold that stores only the current minimum $s=\min(xs)$. A possible lift augments the state $s$ with a counter $c$ tracking the multiplicity of the minimum:
\[
   \{ (\,s:\mathsf{Int},\;c:\mathsf{Int}\,) \;\mid\; 
   \big( s=\infty \Rightarrow c=0 \big) \;\wedge\;
   \big( s \neq \infty \Rightarrow c=\mathrm{count}(xs,s) \big) \}
\]
Here, the qualifier specifies that the auxiliary state $c$ corresponds to the multiplicity of $s$ in $xs$.
\end{example}

Lifting invariants play a key role in the synthesis procedure by enabling \emph{decomposition}.  As
reflected in the proof rules of \Cref{fig:proof-rules}, the lifted accumulator
and the retraction function are tightly coupled: insertions must maintain the
auxiliary state in a form that retractions can use, while the required retractions
constrain what auxiliary information must be retained.
Lifting invariants help break this circularity by fixing the \emph{meaning} of the
auxiliary state before fixing its implementation. Once such an invariant is chosen,
the search can be stratified into two stages. First, the algorithm synthesizes
lifted insertion logic and verifies that it maintains the chosen representation.
Only after this stage succeeds does the algorithm synthesize the retraction function
over that representation. Thus, retraction synthesis can proceed assuming a
well-maintained lifted state; if it fails, the failure is not due to an
incorrectly maintained accumulator.

\begin{algorithm}[t]
\caption{\textsc{RewindFold}}
\label{alg:synthesize-rudaf}
\begin{algorithmic}[1]
\Require Standard fold $\udaf=(\Init, \acc, \eval)$, requirement $\RetSpec(\udafinput)$
\Ensure  Retractable fold $\rudaf=(\Init^\star, \acc^\star,\Retract^\star,\eval^\star)$ or $\bot$
\State $\Sigma \gets \textsc{InferLiftSpec}(\udaf, \RetSpec)$ \label{line:rudaf-syn:infer-types}
\For{\textbf{each} $(\tau_{\mathit{aux}}, \chi, U, G) \in \Sigma$ \textbf{in increasing order}}\label{line:rudaf-syn:for-loop}
  \State $(\mathsf{ok}, \Init^\star, \acc^\star, \eval^\star) \gets \textsc{Lift}(\udaf, \tau_{\mathit{aux}}, \chi, U)$ \label{line:rudaf-syn:lift}
    \If{$ \neg \mathsf{ok}$} \label{line:top-cont}
      \textbf{continue}
  \EndIf
    \State $\Retract^\star \gets \textsc{SynRetract}\big((\Init^\star,\acc^\star,\eval^\star),\udaf,G,\RetSpec\big)$ \label{line:rudaf-syn:synretract}
  \If{$\Retract^\star \neq \bot$} \label{line:rudaf-syn:return-1}
     \Return $(\Init^\star, \acc^\star, \Retract^\star, \eval^\star)$  
  \EndIf
\EndFor
\State \Return $\bot$
\end{algorithmic}
\end{algorithm}

\Cref{alg:synthesize-rudaf} implements this idea. Concretely, it first
invokes \textsc{InferLiftSpec}, which generates candidate lifting specifications
by enumerating auxiliary-state types and candidate invariants from a fixed
component library. These candidates are explored in increasing order of space
cost, reflecting the goal of retaining only the auxiliary information needed to
support the required retractions. This  avoids  implementations that
carry unnecessary state or incur avoidable per-update cost.

For each candidate lifting specification, the algorithm next invokes
\textsc{Lift}, which attempts to synthesize a lifted initializer
$\Init^\star$, accumulator $\acc^\star$, and evaluation function $\eval^\star$.
This step realizes the semantic description given by $\chi$ as concrete
insertion logic over the enriched state. If the proposed auxiliary state cannot
be maintained, the candidate specification is
discarded.

Next, the algorithm invokes \textsc{SynRetract} to synthesize a retraction
function for the realized lift. This is the most subtle part of the algorithm for two reasons. First, the
retraction requirement is indirect: it says when the fold must avoid $\Err$, but
does not by itself specify the local behavior of each call to $\Retract$.
Second, the retraction function is partial and often case-based, as different
deletion scenarios require different updates to the lifted state. As we explain
in \Cref{sec:synth-retract}, \textsc{SynRetract} derives local constraints for
these cases and solves them using symbolic partitioning.

\begin{theorem}[Soundness]
\label{thm:synth-sound}
Let $\udaf = (\Init,\Acc, \eval)$ be an insertion-only fold and let $\RetSpec$ be a retraction specification.
If \Cref{alg:synthesize-rudaf} returns a  retractable fold $\rudaf = (\Init',\Acc',\Retract',\eval')$, then 
$\udaf \equiv_{\RetSpec} \rudaf$.
\end{theorem}

\begin{theorem}[Completeness]
\label{thm:synth-complete}
Let $\udaf$ be an insertion-only fold and  $\RetSpec$  a retraction specification.
Assuming
(i) there exists a correct retractable fold whose lifting is expressible in the DSL from \Cref{fig:lift-dsl},  and
(ii) the oracle, fairness, and search-space assumptions of \Cref{sec:synthesis-proofs} hold, then
\Cref{alg:synthesize-rudaf} will  find a retractable fold
$\rudaf$ such that $\udaf \equiv_{\RetSpec} \rudaf$.
\end{theorem}

{Intuitively, completeness requires a sufficiently expressive
search space, complete solvers, and a fair search strategy that eventually
considers every candidate. Of course, in practice, solver limitations and
timeouts may prevent \tool{} from finding a solution even when
one exists in its search space.}

\subsection{Inference of Lift Specifications}\label{sec:infer-lift-spec}

\begin{algorithm}[t]
\caption{\textsc{InferLiftSpec}}
\label{alg:infer-lifted-types-symbolic}
\begin{algorithmic}[1]
\Require insertion-only fold $\udaf=(\Init, \acc, \eval)$; retraction spec $\theta$; library $L(\tau_a)$ with $(\tau_\mathsf{aux},\mathcal{E}_{\tau})$
\Ensure Candidate liftings $\Sigma$
\State $\Sigma \gets \varnothing;$ \quad $\mathcal{D} \gets \{\, xs \in A^{*} \mid |xs|\le K,\ xs \text{ drawn from test inputs } \mathcal{X}\,\}$
\For{$(\tau_\mathsf{aux}, \mathcal{E}) \in L(\tau_a)$}
  \For{$\chi \in \mathsf{GenQualifierCandidates}(\tau_\mathsf{aux},\acc,\mathcal{E})$}
    \State $(\mathsf{ok},U) \gets \textsc{CheckFunctionalConsistency}(\udaf,\chi,\mathcal{D},\mathcal{X})$ 
    \State $(\mathsf{bad},G) \gets \textsc{CheckRetractability}(\udaf,\theta,\chi,\mathcal{D},\mathcal{X})$
    \If{$ \neg \mathsf{ok} \lor \mathsf{bad}$} \textbf{continue} \EndIf
    \State $\Sigma \gets \Sigma \cup  \{\,(\tau_\mathsf{aux},\chi,U,G)\,\}$
  \EndFor
\EndFor
\State \Return $\textsc{Sort}(\Sigma)$
\end{algorithmic}
\end{algorithm}

\begin{figure}[t]
\centering
\setlength{\fboxsep}{5pt}
{%
\small
\begin{minipage}{0.95\linewidth}
\[
\begin{array}{@{}r@{\;\;}c@{\;\;}l@{}}
T & ::= & \dsl{int} \mid \tau_a \mid (T,\ldots,T)
\\
  & & \mid\; \dsl{sized\_list\_of}(T,n)
        \mid \dsl{sized\_sorted\_bag\_of}(T,n)
        \mid \dsl{sized\_map\_of}(T,T)
\\
  & & \mid\; \dsl{list\_of}(T)
        \mid \dsl{sorted\_bag\_of}(T)
        \mid \dsl{map\_of}(T,T)
\\[0.4em]
E & ::= & s \mid xs \mid \dsl{length}(xs) \mid \dsl{count}(xs,E)
        \mid\; \dsl{nth\_max}(xs,E)
        \mid \dsl{nth\_min}(xs,E)
\\
  & & \mid\; \dsl{multiplicity\_map}(E)
        \mid \dsl{sorted\_bag}(E,\mathrm{Order}, n)
        \mid\; (E,\ldots,E) \mid \dsl{tuple\_select}(E,i)
\\[0.4em]
\chi & ::= & a = E(s, xs)
\end{array}
\]
\end{minipage}%
}
\[
\begin{array}{c}\small
n, i \in \setof{0, 1, 2, \ldots} \quad \mathrm{Order} \in \setof{\dsl{Arrival},\dsl{ReverseArrival}, \dsl{Ascending}, \dsl{Descending}} \\
\end{array}
\]
\vspace{-0.3in}
\caption{
DSL used by \textsc{InferLiftSpec}. The nonterminals $T$ and $E$ range over
candidate auxiliary-state types and summary expressions, respectively.
$\tau_a$ denotes the
input element type of the source fold.
}
\vspace{-0.2in}
\label{fig:lift-dsl}
\end{figure}

The first subprocedure in \Cref{alg:synthesize-rudaf} is
\textsc{InferLiftSpec}, defined in
\Cref{alg:infer-lifted-types-symbolic}. This procedure searches for
candidate ways of enriching the fold state, using the DSL in
\Cref{fig:lift-dsl}. This DSL separates the \emph{shape} of the auxiliary state from its
\emph{meaning}. Concretely, the type language $T$ consists of  possible auxiliary representations,
such as scalars, tuples, fixed-size lists, sorted bags, and maps. In contrast, the expression grammar $E$ describes  generic summaries of the insertion history that
are useful for deletion, such as cardinalities, multiplicities, and order statistics.
{To organize the search, we group these types and expressions
into a component library $L$. Each entry
$(\tau_\mathsf{aux},\mathcal{E})$ pairs an auxiliary-state type
with summary expressions that produce values of that type.
Choosing an expression $E \in \mathcal{E}$ then gives a lifting
invariant $\chi(s,a,xs) \equiv a=E(s,xs)$, which gives
  meaning to the auxiliary state.}

At a high level, {\sc InferLiftSpec} generates candidate specifications by enumerating auxiliary types in increasing order of space cost. Here, space cost is a simple structural measure on auxiliary-state types, with scalar summaries ranked below tuples and bounded containers, and larger containers ranked after smaller ones. For each such type, the procedure then instantiates candidate qualifiers from the DSL and subjects them to two filters: \emph{functional consistency} (\Cref{alg:build-examples}) and \emph{retractability} (\Cref{alg:spec-necessity}).  The first  check asks whether the proposed auxiliary state can actually be maintained by insertion; the second one asks whether it preserves enough information to support the retractions required by $\theta$. Candidates that fail either check are discarded.

\begin{algorithm}[t]
\caption{\textsc{CheckFunctionalConsistency}}
\label{alg:build-examples}
\begin{algorithmic}[1]
\Require $\udaf=(\Init, \acc, \eval)$;\; qualifier $\chi(s,a,xs)$;\; histories $\mathcal{D}$;\; test inputs $\mathcal{X}$
\Ensure $(\mathsf{ok},U)$ with $U:(S\times A_\mathsf{aux}\times A)\rightharpoonup A_\mathsf{aux}$
\State $R \gets \bigl\{\,\bigl((s,a,x),a'\bigr)\ \bigm|\ xs\!\in\!\mathcal{D},\ s=\psem{\udaf}{xs},\ \chi(s,a,xs),\ x\!\in\!\mathcal{X},\ s'=\acc(s,x),\ \chi(s',a',xs\pp[x]) \,\bigr\}$
\State $\mathsf{ok} \gets \left(\forall k,v_1,v_2.\ (k,v_1)\!\in\!R \wedge (k,v_2)\!\in\!R \Rightarrow v_1=v_2\right)$
\State \Return $(\mathsf{ok},\ \{\,k\mapsto v \mid (k,v)\in R\,\})$
\end{algorithmic}
\end{algorithm}

In more detail, the functional consistency check
(\Cref{alg:build-examples}) asks whether a candidate qualifier $\qual$ is \emph{realizable}, which requires that the next auxiliary value  be determined by the current lifted
state and the next input alone. That is, whenever two histories $xs_1$ and
$xs_2$ induce the same lifted state $(s,a)$ under $\qual$, appending the same
input $x$ must also induce the same next auxiliary value $a'$. Otherwise, the
desired auxiliary state depends on information about the past that is not present
in $(s,a)$, and no deterministic lifted accumulator $\acc^\star$ can maintain
$\qual$.

\Cref{alg:build-examples} checks this condition on bounded histories
$\mathcal{D}$ by constructing a relation
$R \subseteq (S\times A_\mathsf{aux}\times A)\times A_\mathsf{aux}$.
For each sampled history $xs$ and input $x$, it computes
$s=\psem{\udaf}{xs}$, chooses an auxiliary value $a$ satisfying
$\qual(s,a,xs)$, and records the value $a'$ required by
$\qual(\acc(s,x),a',xs\pp[x])$. The relation $R$ must be functional, meaning that every
triple $(s,a,x)$ may map to at most one $a'$. If $R$ is non-functional, the
candidate $\qual$ is rejected.
Otherwise, $R$ defines a partial update map $U:(s,a,x)\mapsto a'$ that provides
examples for synthesizing the lifted accumulator.

\begin{example}

Consider again \Cref{example:syn-count-lifting}, where the insertion-only fold stores
$ s=\min(xs)$. 
Suppose we use the candidate lift invariant
\[
\chi \equiv (a=\mathrm{count}(xs,\mathrm{nth\_min}(xs,2)))
\]
 where $\mathrm{nth\_min}(xs,2)$ denotes the second smallest \emph{distinct} element of $xs$. In other words,  $\chi$ says that the auxiliary state corresponds to the multiplicity of the second minimum.  This candidate fails the functional consistency check. To see why, consider the two histories
$
xs_1=\lstof{1,2}$ and $ xs_2=\lstof{1,3}.
$
In both cases, the minimum is $s=1$, and the second minimum occurs exactly once, so both
histories yield the same lifted state
$
(s,a)=(1,1).
$
Now  suppose we append the same new input $x=2$. For
$ 
xs_1\pp\lstof{2}=\lstof{1,2,2}$, 
the second minimum remains $2$ and now occurs twice, so $a'=2$. By contrast, for
$
xs_2\pp\lstof{2}=\lstof{1,3,2}$, 
the second minimum becomes $2$ and occurs only once, so $a'=1$. Thus the same current
lifted state $(s,a)$ and the same next input $x$ lead to two different next auxiliary values.
Hence, $\chi$ does not induce a well-defined update function.
\end{example}

\begin{algorithm}[t]
\caption{\textsc{CheckRetractability}}
\label{alg:spec-necessity}
\begin{algorithmic}[1]
\Require $\udaf=(\Init,\acc,\eval)$;\; spec $\theta$;\; qualifier $\chi(s,a,xs)$;\; histories $\mathcal{D}$;\; test inputs $\mathcal{X}$
\Ensure $(\mathsf{bad},G)$ where $\mathsf{bad}$ iff $\chi$ is insufficient for $\theta$ on $\mathcal{D}$, and each $V\in G$ is a set of histories whose reached states should be related by $\SimTwo$
\State $M \gets \bigl\{\,(ys,xs)\ \bigm|\ ys\!\in\!\mathcal{D},\ \mathsf{Allowed}_\theta(ys,xs)\,\bigr\}$
\State $E_0 \gets \bigl\{\,\bigl((xs,xs),(xs',xs')\bigr)\ \bigm|\ xs,xs'\!\in\!\mathcal{D},\ \exists (s,a).\chi(s,a,xs)\wedge\chi(s,a,xs')\wedge s=\psem{\udaf}{xs}=\psem{\udaf}{xs'}\,\bigr\}$
\State $\mathsf{Acc}(E) \gets \bigl\{\,\bigl((ys\pp[x],xs\pp[x]),(ys'\pp[x],xs'\pp[x])\bigr)\ \bigm|\ \bigl((ys,xs),(ys',xs')\bigr)\!\in\!E,\ x\!\in\!\mathcal{X}\,\bigr\}$
\State $\mathsf{Ret}(E) \gets \bigl\{\,\bigl((ys,\remove(xs,x)),(ys',\remove(xs',x))\bigr)\ \bigm|\ \bigl((ys,xs),(ys',xs')\bigr)\!\in\!E,\ x\!\in\!xs\cap xs'\,\bigr\}$
\State $E^\star \gets \mathsf{EquivClosure}\bigl(\bigl(E_0\ \cup\ \mathsf{Acc}(E^\star)\ \cup\ \mathsf{Ret}(E^\star)\bigr)\cap (M\!\times\!M)\bigr)$
\State $G \gets \bigl\{\, \{\,xs' \mid \exists ys'.\ ((ys,xs),(ys',xs'))\in E^\star \,\}\ \bigm|\ (ys,xs)\in M \,\bigr\}$
\State $\mathsf{bad} \gets \exists \bigl((ys,xs),(ys',xs')\bigr)\!\in\!E^\star.\ \sem{\udaf}{xs}\neq\sem{\udaf}{xs'}$
\State \Return $(\mathsf{bad},G)$
\end{algorithmic}
\end{algorithm}

In addition to checking functional consistency, {\sc InferLiftSpec} performs a
\emph{retractability check} (\Cref{alg:spec-necessity}) to determine whether
the candidate invariant $\chi$ retains enough information to support the
retraction requirement $\theta$. The key idea is simple: histories that lead to the same lifted state will be
indistinguishable to the synthesized retraction function.  Thus, if such
histories can later require different retraction behavior, the candidate lift is
insufficient.

To formalize this idea, \Cref{alg:spec-necessity} considers pairs $(ys,xs)$,
where $ys$ is a full insertion history and $xs$ is an effective history that can
be obtained from $ys$ by deleting at most $\RetSpec(ys)$ elements. These are the
history pairs for which the retraction specification requires correct behavior:
\[
M \triangleq \{\, (ys,xs) \mid ys\in\mathcal D,\ \mathsf{Allowed}_\theta(ys,xs)\,\}, \quad \textrm{where:} 
\]
\[
\mathsf{Allowed}_\theta(ys,xs) \triangleq
\exists zs.~ |zs|\le \RetSpec(ys) \wedge ys=\mathit{merge}(xs,zs),
\]
Here, $\mathit{merge}(xs,zs)$ denotes an order-preserving interleaving, so
$\mathsf{Allowed}_\theta(ys,xs)$ says that $xs$ can result from $ys$ after
deleting an allowed number of elements.

The algorithm then tracks how indistinguishability propagates through future updates. Line~2 initializes a seed relation $E_0$ with insertion-only histories that $\chi$ maps to the same lifted state. Lines~3 and~4 define two closure operations. The operator $\mathsf{Acc}$ says that related history pairs remain related after appending the same new element, while $\mathsf{Ret}$ says that they remain related after deleting the same element from both effective histories. Line~5 computes the least fixed point $E^\star$ containing $E_0$ and closed under these two operations, restricted to the allowed pairs in $M$. Thus, $E^\star$ captures the history pairs that the chosen lift forces the retractable implementation to treat identically.

Finally, lines~6--8 use $E^\star$ to check whether this forced identification is safe. If $E^\star$ relates two pairs whose effective histories $xs$ and $xs'$ produce different outputs under the original fold, then the lift has lost information needed to satisfy $\theta$, and the candidate $\chi$ is rejected. The algorithm also returns groups $G$ of histories that remain indistinguishable under the lift; these groups are later used when synthesizing the retraction function in \Cref{sec:synth-retract}.

\begin{example}\label[example]{example:syn-count-lifting-candidate}
Consider again \Cref{example:syn-count-lifting} that computes $s=\min(xs)$. Suppose $\theta$ requires deleting one element, and suppose the candidate lift augments the minimum with its multiplicity:
\[
\chi(s,a,xs)\equiv a=\mathrm{count}(xs,s),
\]
This candidate fails the retractability check. To see why, observe that the two histories
$xs_1=\lstof{1,2}$ and $xs_2=\lstof{1,3}$
both yield the same lifted state
$
(s,a)=(1,1).
$
So $((xs_1,xs_1),(xs_2,xs_2))\in E_0$, and applying $\mathsf{Ret}$ with $x=1$ propagates this into $((xs_1,\lstof{2}),(xs_2,\lstof{3}))\in E^\star$. However, $\min(\lstof{2})=2$ and $\min(\lstof{3})=3$, meaning that 
 the same lifted state $(1,1)$ is compatible with two different outcomes. This shows that $\chi$ does not preserve enough information to support a correct retract function.
\end{example}

\begin{theorem}[Soundness of \textsc{CheckRetractability}]
\label{thm:check-retract-sound}
If \textsc{CheckRetractability}$(\udaf,\theta,\chi,\mathcal{D},\mathcal{X})$ returns $(\mathsf{true},\_)$, then for every lifted fold that realizes $\chi$, no retract function satisfies the Block B and C rules of \Cref{fig:proof-rules}.
\end{theorem}

\bfpara{Discussion.}
The realizability checks performed by {\sc InferLiftSpec} play an important role in the overall algorithm. In particular, inference of the lifting specification is only the first step of the synthesis pipeline, and \Cref{alg:synthesize-rudaf} must still synthesize both a lifted accumulator and a retraction function. Without these checks, the algorithm could waste effort on lifts that are infeasible from the outset. By filtering such candidates early, our method focuses the later synthesis stages on promising lifts. However, these checks are sound but incomplete: any rejected candidate is infeasible, as stated in \Cref{thm:check-retract-sound}, but some infeasible candidates may still pass {\sc InferLiftSpec} and be eliminated in later stages of the algorithm.

\begin{algorithm}[t]
\caption{\textsc{Lift}}
\label{alg:lift-cegis}
\begin{algorithmic}[1]
\Require $P=(\Init, \acc, \eval)$;\ $T=\{(s:\tau,a:\tau_{\mathsf{aux}})\mid \chi(s,a,xs)\}$;\ update map $U$;\ grammar $\mathcal{G}$
\Ensure $(\mathsf{true},~\Init^\star, \acc^\star,\eval^\star)$ or $(\mathsf{false},~\bot)$
\While{\textbf{true}}
  \State $\mathsf{UpdAux} \gets \textsc{PBE}(\mathcal{G},U)$
  \If{$\mathsf{UpdAux}=\bot$} \Return $(\mathsf{false},~\bot)$ \EndIf
  \State \textbf{let } $\mathsf{Bad}(s,a,x,xs)\!\triangleq\!\chi(s,a,xs)\wedge\Reach(s,xs)\wedge
      \neg\,\chi(\acc(s,x),\mathsf{UpdAux}(s,a,x),xs{++}[x])$
  \State $(\mathsf{sat},\hat s,\hat a,\hat x,\widehat{xs},\hat a') \gets \textsc{SAT}(\mathsf{Bad}(s,a,x,xs)\ \wedge\ \chi(\acc(s,x),a',xs{++}[x]))$
  \If{$\mathsf{sat}$} $U \gets U \cup \{(\hat s,\hat a,\hat x) \mapsto \hat a'\}$ 
    \Else  \  \textbf{let } $\acc^\star((s,a),x) \triangleq (\acc(s,x),\,\mathsf{UpdAux}(s,a,x))$;\ \ \  $\eval^\star(s,a)\triangleq \eval(s);$ \label{line:lift-assign-acc}\ 
    \State \textbf{let } $\Init^\star \triangleq (\Init,a_0)$ where $\models \chi(\init, a_0, [])$\label{line:lift-assign-init}
    \State \Return $(\mathsf{true}, \Init^\star,\acc^\star,\eval^\star)$
  \EndIf
\EndWhile
\end{algorithmic}
\end{algorithm}

\subsection{Synthesizing Lifted Accumulator}

Given a candidate lifting specification
$\{(s:\tau,a:\tau_{\mathsf{aux}})\mid \chi(s,a, xs)\}$, the next step is to
refine the accumulator so that it correctly updates the auxiliary state $a$.
Fortunately, this task is relatively straightforward due to the groundwork laid
by {\sc InferLiftSpec}. First, for each candidate type, we have access to an
update map $U$ consisting of input-output examples $(s,a,x)\mapsto a'$ that
specify how the auxiliary state should evolve. Second, the invariant $\chi$
enables correctness checking by verifying that the synthesized accumulator
preserves $\chi$ inductively. Without this invariant, correctness of the lifted
accumulator would have to be checked indirectly through the proof rules in
\Cref{fig:proof-rules}, which only constrain the behavior needed for
equivalence. By contrast, $\chi$ gives a direct specification of the auxiliary
state, allowing {\sc Lift} to verify that the synthesized insertion logic
realizes the chosen representation before retraction synthesis begins.

Building on the examples and invariant produced by {\sc InferLiftSpec}, the
{\sc Lift} procedure in \Cref{alg:lift-cegis} synthesizes the lifted accumulator
using a standard CEGIS loop. In each iteration, it invokes a PBE solver to synthesize a
candidate update function $\mathsf{UpdAux}$ from the examples in $U$. If
synthesis fails, the procedure returns $\bot$; otherwise, it checks whether
$\mathsf{UpdAux}$ preserves $\chi$. This is done by querying an SMT solver for a
reachable $(s,a)$, input $x$, and sequence $xs$ such that $\chi(s,a,xs)$ holds
but, after applying $\acc$ and $\mathsf{UpdAux}$, the resulting state violates
$\chi$. Any counterexample refines $U$ with the new example $(s,a,x)\mapsto a'$,
and the loop repeats. If no counterexample exists, the lifted accumulator is
constructed by pairing $\acc$ with $\mathsf{UpdAux}$
(line~\ref{line:lift-assign-acc}).

\subsection{Retraction Synthesis with Lifted State}\label{sec:synth-retract}

The most involved component of our synthesis algorithm is the retraction procedure itself.  
At a high level, retraction synthesis must address three key difficulties that make this step substantially more challenging than lifting the insertion logic:

\begin{enumerate}[leftmargin=*]

\item {\bf Lack of local specifications.}
The correctness requirements for retraction, given by Block~B in
\Cref{fig:proof-rules}, do not directly constrain
the behavior of $\Retract$. Instead, they specify retraction behavior \emph{indirectly} through the relational
invariant $\SimTwo$. Furthermore, the lifting
invariant $\chi$ cannot play this role because it describes the meaning of a
lifted state on insertion-only histories, whereas retraction correctness requires
relating two lifted states across a deletion step. Thus, the algorithm must first
infer a suitable $\SimTwo$ relation to  constrain synthesis of
$\Retract$. But this introduces a circular dependency between the synthesis of $\Retract$ and the inference of the $\SimTwo$ relation used to specify its behavior.

\item {\bf Unknown domain.}  
A retraction function need not succeed on every input~$x$; returning \textsf{Err} is permitted whenever~$\RetSpec$ does not require supporting the deletion of~$x$.
However, the key difficulty is that $\RetSpec$ is a function from insertion histories to a numerical deletion budget, so it does not directly translate into a precondition on individual calls to \textsf{Retract}.  Thus, the valid domain of the retraction function is not known in advance and must be \emph{inferred} during synthesis.

\item {\bf Piecewise implementation.}  
The third difficulty is that retraction is inherently piecewise:  the correct behavior of $\Retract$ often depends on how the deleted element interacts with the current summary (e.g., whether it is an extremum or  other copies remain). These distinctions lead to different update rules, causing the  retraction logic to  involve intricate case analysis. 
\end{enumerate}

Our synthesis procedure addresses these difficulties by combining two key ideas.  First, it  infers a \emph{candidate} relational invariant $\SimTwo$ that can be used to impose meaningful constraints on retraction logic by utilizing (a) the examples returned by $\textsc{InferLiftSpec}$ and (b) proof rules that do not require access to the implementation of $\Retract$. 
Second, once such a $\SimTwo$ is given, it performs symbolic execution on bounded input lists to derive a
\emph{symbolic partition} of the input space under that relation. This partition serves two purposes. First, it identifies regions of the input space on which retraction is required to succeed, thereby addressing the \emph{unknown domain} problem. Second, it separates retraction scenarios that are semantically distinct under $\SimTwo$, allowing the algorithm to decompose the synthesis problem into semantically different cases.

\begin{figure}[t]
  \centering
  \small
  \[
  \fbox{$
  \begin{array}{rcll}
\mathsf{ValidCand}(G,\rudaf,\SimTwo)
&\triangleq&
\forall V \in G.\ \forall xs_1,xs_2 \in V.\
\SimTwo(\psem{\rudaf}{xs_1},\psem{\rudaf}{xs_2})
& \text{// compatible states}
\\[1.0ex]
&& {}\wedge\ \SimTwo(\Init^\star,\Init^\star)
& \text{// B1}
\\[1.0ex]
&& {}\wedge\ \forall s_1,s_2,x.\
\SimTwo(s_1,s_2)\Rightarrow
\SimTwo(\acc^\star(s_1,x),\acc^\star(s_2,x))
& \text{// B2}
\\[1.0ex]
&& {}\wedge\ \forall s_1,s_2.\
\SimTwo(s_1,s_2)\Rightarrow
\eval^\star(s_1)=\eval^\star(s_2)
& \text{// B4}
  \end{array}
  $}
  \]
  \vspace{-0.15in}
  \caption{Auxiliary procedure used in \textsc{SynRetract}.}
  \label{fig:synretract-staterel}
  \vspace{-0.2in}
\end{figure}

\bfpara{Algorithm overview.}
\Cref{alg:synretract-partition} implements retraction synthesis as a
nested search. The outer loop searches for a candidate relation $\SimTwo$ that
can serve as a local specification for retraction, whereas the inner loop then attempts
to synthesize a $\Retract$ function that satisfies this specification. A subtle
point here is that, while the relational invariant $\SimTwo$ used in \Cref{fig:proof-rules} is history-indexed (i.e., $\SimTwo(xs,s_1,s_2)$ relates two states relative to a normalized history $xs$), 
the synthesis procedure uses a state-based version of this relation. This is
because $\Retract$ is invoked only with the current lifted state and the element
being deleted and does not have access to the normalized history $xs$. Thus, the
constraints used to synthesize $\Retract$ must ultimately be expressed as local
conditions over states. In the rest of this section, we write $\SimTwo(s_1,s_2)$
for this state-based relation.

The outer loop filters candidate relations using the $\mathsf{ValidCand}$
condition in \Cref{fig:synretract-staterel}.  Intuitively, $\mathsf{ValidCand}$
checks that a candidate relation is compatible with both the examples produced by
\textsc{InferLiftSpec} and the proof obligations from Block~{B}. Here, the
examples correspond to the compatible-history sets $G$ returned by \textsc{InferLiftSpec}: histories in the same set
arise from the same retraction scenario, so the lifted states they reach should be related by any
reasonable $\SimTwo$. Additionally, any suitable candidate for $\SimTwo$ must satisfy the basic closure conditions from Block~B  in \Cref{fig:proof-rules}: it must (1)  hold at the initial state,  (2) be preserved by insertion steps, and (3) imply observational agreement.

\begin{example}\label[example]{ex:synthesis-simtwo}
Continuing \Cref{example:syn-count-lifting}, consider the  lifted
state $(m,c)$ and $\eval^\star(m,c)=m$. At this stage, \textsc{SynRetract}
searches for a binary relation $\SimTwo$ satisfying
$\mathsf{ValidCand}(G,\rudaf,\SimTwo)$.
Suppose $G$ contains a set $V$ with histories $[1,2,1]$ and $[2,1,1]$. Since
both histories yield the same lifted state $(1,2)$, any valid candidate for
$\SimTwo$ should satisfy
$\SimTwo((1,2),(1,2))
$. Now consider the candidate
\[
\SimTwo_{\mathsf{bad}}\big((m_1,c_1),(m_2,c_2)\big)\equiv c_1=c_2.
\]
This candidate is too weak. Although it is compatible with the evidence from
$V$, it also relates $(1,2)$ and $(2,2)$, which violates the observational
agreement requirement in $\mathsf{ValidCand}$, since
$ 
\eval^\star(1,2)=1 \neq 2=\eval^\star(2,2).
$
By contrast,
\begin{equation}\label{eq:sim-good}
\SimTwo\big((m_1,c_1),(m_2,c_2)\big)\equiv (m_1=m_2)\wedge(c_1=c_2)
\end{equation}
is compatible with the histories in $V$ and also satisfies the remaining
conditions of $\mathsf{ValidCand}$. Hence, it is a suitable candidate for the
outer loop.
\end{example}

\begin{algorithm}[t]
\caption{\textsc{SynRetract}}
\label{alg:synretract-partition}
\begin{algorithmic}[1]
\Require Lifted fold $\rudaf=(\Init^\star,\acc^\star,\eval^\star)$, original fold $\udaf$, set of compatible histories $G$, spec $\RetSpec$
\Ensure $\Retract$ or $\bot$
\State $\mathcal{B} \gets \emptyset$
\While{true}
  \State find $\SimTwo \notin \mathcal{B}$ such that $\mathsf{ValidCand}(G,\rudaf,\SimTwo)$
  \If{ $\SimTwo = \bot$}  \Return $\bot$
  \EndIf
  \State $\mathcal{B} \gets \mathcal{B} \cup \{\SimTwo\}$; \quad  $\mathcal{E}\gets\emptyset$;\quad \quad $\Pi\gets[\,]$
  \While{true}
    \State $r \gets \mathsf{SolveSyGuS}(\Pi)$
    \If{$r=\bot$} \textbf{break} \EndIf
    \State $(\mathsf{ok},\ \textsf{cex}) \gets \textsc{Verify}(\udaf,\RetSpec,\rudaf,r,\SimTwo)$
    \If{$\mathsf{ok}$} \Return $r$ \EndIf
    \State \textbf{let } $\textsf{cex}=(xs,\ s^\ast,\ x^\ast)$ \textbf{ and } $\ell \gets |xs|$
    \If{$\ell \notin \mathcal{E}$}
      \State $\Pi \gets \textsc{UpdateSymbolicPartitions}(\Pi,\rudaf,\ell,\SimTwo)$; \ \  $\mathcal{E}\gets\mathcal{E}\cup\{\ell\}$
    \Else
      \State \textbf{let } $(\Gamma_j,\psi_j)\in \Pi$ \textbf{ such that } $(s^\ast,x^\ast)\models \Gamma_j$
      \State $\Pi[\Gamma_j] \gets \psi_j \wedge \SimTwo(\psem{\rudaf}{\remove(xs,x^\ast)},\Retract(s^\ast,x^\ast))$
    \EndIf
  \EndWhile
\EndWhile
\end{algorithmic}
\end{algorithm}

Once a candidate $\SimTwo$ is fixed, the inner loop searches for a retraction function consistent with that relation. This search is organized around a \emph{symbolic partition}
\[
\Pi = \{\, (\gamma_i(s,x),\,\psi_i(s,x)) \mid i=1,\dots,n \,\}
\]
where each guard $\gamma_i(s,x)$ describes a \emph{region} of the  input space, and each formula $\psi_i(s,x)$ gives the \emph{local specification} that $\Retract$ must satisfy on that region. Intuitively, such a   partition decomposes  synthesis into independent subproblems, each corresponding to a  distinct deletion scenario.

\begin{example}\label[example]{example:synretract-symbolic-partition}
Continuing \Cref{example:syn-count-lifting}, consider the lifted state
$(m,c)$, where $m$ is the minimum and $c$ is its multiplicity. The deletion
space splits according to how the deleted element relates to the current
minimum. In particular, symbolic partitioning yields the following two feasible
regions:
\[
\begin{array}{rclrcl}
\gamma_{\mathsf{keep}}((m,c),x) &\equiv& x>m 
& 
\psi_{\mathsf{keep}}((m,c),x) & \equiv &
\Retract((m,c),x)=(m,c) \\[2mm]
\gamma_{\mathsf{dec}}((m,c),x) &\equiv& x=m \wedge c>1 
& 
\psi_{\mathsf{dec}}((m,c),x) & \equiv & 
\Retract((m,c),x)=(m,c-1)
\end{array}
\]
The first region corresponds to deleting a non-minimum element, which leaves the
lifted state unchanged. The second corresponds to deleting one copy of the
minimum while another copy remains, which decrements the multiplicity. The
remaining case, $x=m \wedge c=1$, is excluded: after deleting the unique
minimum, the new minimum cannot be determined from $(m,c)$ alone, so
$\Retract$ may return $\Err$ on that input.
\end{example}

With this partition representation, \Cref{alg:synretract-partition}
proceeds by alternating between synthesis and verification. It starts
from an empty partition and invokes $\mathsf{SolveSyGuS}(\Pi)$ to
synthesize a candidate retraction satisfying the current local
specifications (line~7).
If full verification at line~9 succeeds,
the algorithm terminates successfully. Otherwise, verification produces a counterexample $(xs,s^\ast,x^\ast)$, where $xs$ is an insertion-only history, $s^\ast$ is the current lifted state, and $x^\ast$ is  the deleted element.

The algorithm uses such counterexamples in two different ways, depending on the nature of the counterexample.  If sequence $xs$  has a length $\ell$ that is larger than the ones previously explored,   the counterexample could reveal a new \emph{class} of retraction scenarios, so the algorithm invokes \textsc{UpdateSymbolicPartitions} (explained later) to refine the current partition. By contrast, if $\ell$ has already been explored, then the symbolic execution phase has already exposed the relevant case structure at that length. Hence, failure indicates \emph{not} that a new branch is missing, but that the local specification is too weak. In this case, if $(\gamma_j,\psi_j)$ is the matched partition element, the algorithm replaces $\psi_j$ with the stronger specification:
\[
\psi_j \wedge \SimTwo(\psem{\rudaf}{\remove(xs,x^\ast)},\Retract(s^\ast,x^\ast)).
\]
That is, the local specification of this branch is strengthened to require $\SimTwo$-agreement between the synthesized post-retraction state and the ground truth state.

\begin{algorithm}[t]
\small
\caption{\textsc{UpdateSymbolicPartitions}}
\label{alg:create-symbolic-partitions-map}
\begin{algorithmic}[1]
\Require Partition $\Pi$, lifted fold $\mathcal{P}=(\init, \acc, \eval)$, length $\ell$, relation $\Psi$
\Ensure Updated partition $\Pi$
\State let $xs=[x_1,\ldots,x_\ell]$ be fresh symbols
\For{\textbf{each } $i \in \{1,\ldots,\ell\}$}
  \State $\beta_i(xs,s') \gets \textsc{SymExec}\big(\mathcal{P},\ \remove(xs,x_i)\big)$
  \For{\textbf{each } $(\Phi_{ij}(xs),\ o_{ij}(xs)) \in \textsc{SymExecByPath}(\mathcal{P}, xs)$}
    \State $\varphi_{ij}(s, x, xs) \gets \big(s=o_{ij}(xs) \wedge\ \big(x=x_i\big)\ \wedge\ \Phi_{ij}(xs) \big)$
    \State $\gamma_{ij}(s,x) \gets \mathsf{QE}_{xs}\big(\varphi_{ij}\big)$
    \State $\beta_{ij}(s,x,s') \gets \mathsf{QE}_{xs}\big(\varphi_{ij}\ \wedge\ \beta_i \big)$
    
    \If{$\textsc{unsat}\big(\gamma_{ij}(s,x)\ \wedge\ \exists s'.\, \beta_{ij}(s,x,s')\big)$} 
        \textbf{continue}
    \EndIf
    
    \If{$\textsc{sat}\big(\gamma_{ij}(s,x)\ \wedge\ \exists s'_1,s'_2.\ \beta_{ij}(s,x,s'_1)\ \wedge\ \beta_{ij}(s,x,s'_2)\ \wedge\ \neg \Psi(s'_1,s'_2)\big)$}
        \State $\Pi \gets \textsc{InsertOrMerge}\big(\Pi,\ \gamma_{ij},\ \bot\big)$; \textbf{continue} 
    \EndIf
    
    \State $\psi_{ij}(s,x) \gets \exists s'. \big(\beta_{ij}(s,x,s') \wedge \Psi(s', \Retract(s,x))\big)$
    \State $\Pi \gets \textsc{InsertOrMerge}\big(\Pi,\ \gamma_{ij},\ \psi_{ij}\big)$ 
  \EndFor
\EndFor
\State \Return $\Pi$
\end{algorithmic}
\end{algorithm}

\bfpara{Symbolic partitioning.} A key step in the synthesis process is \Cref{alg:create-symbolic-partitions-map}, which derives local retraction specifications through symbolic execution on a list of size~$\ell$.
To do so, the algorithm first introduces fresh symbols $x_1,\ldots,x_\ell$ to represent each inserted element (line~1) and  considers deleting them one by one.  
For each potential deletion, it re-constructs a ``what-if'' scenario: \emph{What would the accumulator state have been had a particular element~$x_i$ never been added? }
This is captured by the formula~$\beta_i(xs,s')$ at line~3, which serves as the ``ground truth’’ reference for validation.

The inner loop (lines~4--12) then explores the overall execution of~$\udaf$ to reason about the relationship between~$x_i$ and the other elements.  Specifically, it performs 
path-by-path symbolic execution to obtain a condition~$\Phi_{ij}$ that describes when a particular control-flow path is taken and a corresponding symbolic output state~$o_{ij}$.  
We then ask: \emph{“What should happen if~$x_i$ were retracted from this resulting state?”}  
To answer this, the algorithm constructs the formula:
\[
\varphi_{ij}(s,x,xs) \triangleq \Phi_{ij}(xs) \wedge (s = o_{ij}(xs)) \wedge (x = x_i)
\]
representing the situation where the accumulator ends in state~$s$ after processing~$xs$, and the element chosen for deletion is~$x_i$.  
The algorithm then conjoins~$\varphi_{ij}$ with the reference formula~$\beta_i(xs,s')$ and eliminates the history variables~$xs$. This yields two artifacts: a guard~$\gamma_{ij}(s,x)$ (line~6), which identifies the region of the input space covered by this case, and a path-specific deletion relation~$\beta_{ij}(s,x,s')$ (line~7), which characterizes the reference post-deletion state~$s'$ associated with that case.

Next, the algorithm checks whether retraction is even possible for inputs satisfying~$\gamma_{ij}$ by testing the pair $(\gamma_{ij}, \beta_{ij})$ for \emph{realizability} and \emph{ambiguity}:
\begin{itemize}[leftmargin=*]
\item \textbf{Unrealizable case:} Intuitively, an unrealizable case means that the current state~$s$ and element~$x$ 
describe a situation that could never arise from any input history. That is, there is simply no predecessor state~$s'$ that, when~$x$ 
were re-inserted, would yield~$s$. Since retraction is meant to ``undo'' an actual 
insertion, the absence of any such valid predecessor means that no 
implementation could return a correct result.
The satisfiability query at line 8 checks precisely this situation.
\item \textbf{Ambiguous case:} Intuitively, an ambiguous case means that the current state~$s$ 
and element~$x$ could correspond to different predecessors $s_1', s_2'$  that are distinguishable under the relational invariant $\SimTwo$.
{In other words, $\Retract(s, x)$ is ambiguous because we don't have sufficient information to decide whether the output should be $s_1'$ or $s_2'$.}
Thus, under $\SimTwo$, the only $\Retract$ implementation that can satisfy proof rule B3 from \Cref{fig:proof-rules}  is one for which retraction must be  undefined on this region. 
The satisfiability check at lines~9--10 detects precisely this condition. 
\end{itemize}

Thus, if either of these checks fail, $\Retract$ is undefined on $\gamma_{ij}$, and the guard falls outside of the domain of $\Pi$. Otherwise, the algorithm constructs a local specification~$\psi_{ij}$ for~$\gamma_{ij}$ at line 11 enforcing  that ~$\Retract(s,x)$ must be consistent with the effect of deleting~$x$ under $\SimTwo$.
Finally, the pair~$(\gamma_{ij},\psi_{ij})$ is added to the partition~$\Pi$, which accumulates all feasible retraction behaviors.  
Since different paths may yield overlapping guards after quantifier elimination, line~12 calls \textsc{InsertOrMerge} to merge overlapping cases by inserting holes (i.e., unknown expressions) to force generalization. A worked example
showing this refinement process appears in \Cref{app:synretract-trace}.

\section{Implementation}

We implemented our approach in a prototype tool, \tool{},\footnote{{We used Claude Code to assist with the development of \tool{}.}}
written in Python.
\tool{} takes as input a fold and a retraction specification, and outputs a
retractable fold satisfying \Cref{def:equivalence}. The tool communicates with
solvers via the SMT-LIB interface~\cite{smtlib} and uses CVC5~\cite{cvc5} for
validity and satisfiability queries, quantifier elimination, and the SyGuS
queries used by \Cref{alg:lift-cegis,alg:synretract-partition}.

\bfpara{Synthesis.}
The implementation follows \Cref{sec:synthesis}: candidate lifting
specifications are generated from a small summary-component library
and filtered using the checks from \Cref{sec:infer-lift-spec}.
For retraction synthesis, \tool{} builds symbolic partitions via
\Cref{alg:create-symbolic-partitions-map} and uses CVC5's SyGuS engine
to synthesize code for the resulting local specifications.
{Importantly, the grammar supplied to the SyGuS engine excludes
calls to the original fold. This prevents the synthesizer from
simply removing an element from the retained history and rerunning
the original fold. Instead, it must implement retraction using
operations on the chosen data structures.}

\bfpara{Validation and verification.}
Before full verification, \tool{} performs lightweight randomized testing. Concretely, it samples update sequences satisfying the retraction specification and checks that required retractions avoid $\Err$ and that the final result matches the original fold on the normalized history. Candidates that pass are then checked against the verification conditions from \Cref{fig:proof-rules}. The verifier reuses synthesis artifacts when possible: the lifting invariant yields $\Reach$, and the relation chosen by \textsc{SynRetract} yields $\SimTwo$. It then infers $\SimBudget$ using a small portfolio of strategies, including template-based search, and checks all proof obligations with the CVC5 SMT solver.

\bfpara{Framework integration.}
Our approach is intended for data-processing frameworks that support
incremental deletion through user-defined retraction handlers.
In particular, our implementation targets Flink's aggregation model:
the synthesized state serves as the accumulator, and
$\Init$, $\Acc$, $\Retract$, and $\eval$ correspond to
\texttt{\small createAccumulator}, \texttt{\small accumulate}, \texttt{\small retract},
and \texttt{\small getValue}, respectively.

\section{Evaluation}\label{sec:eval}
In this section, we describe the results of an evaluation designed to answer the following  questions:

\begin{enumerate}[label=\textbf{RQ\arabic*.}, leftmargin=*]
    \item How effective is our verifier at checking  candidate retractable folds?
    \item How effective is our synthesizer at producing verified retractable folds?
    \item How efficient are the synthesized  folds compared with recomputation?
    \item How do the main algorithmic components contribute to synthesis success?
\end{enumerate}

 \bfpara{Benchmarks and set-up.}
Our evaluation is based on real-world aggregate computations from Apache Flink,
a widely used stream-processing framework for implementing
stateful computations over continuously arriving data. In Flink, a user-defined
aggregate maintains an accumulator that is incrementally updated via an
\texttt{accumulate} method. Flink  also supports incremental deletion of prior
inputs if the user implements a suitable \texttt{retract} method. This makes Flink a natural source of
benchmarks for our setting because its aggregate functions are stateful folds, and the corresponding \texttt{retract} methods capture real-world
attempts to make such folds retractable.

Our benchmark suite contains 48 Flink aggregations from two sources. The first is prior work on synthesizing homomorphic folds for parallel aggregation~\cite{ink}. Although \ink{} addresses a different problem, its benchmarks are relevant because they contain real-world Flink aggregation code with complex accumulator structures. We include all 23 \ink{} benchmarks and, since they do not provide retraction logic, we evaluate them under three generic specifications: bounded retraction with budgets $k\in \{2,3\}$, modeling small numbers of late or corrected events, and unbounded retraction, modeling long-lived aggregates with arbitrary deletions.
The second source  consists of 25 additional GitHub Flink aggregations with manually written retraction logic, excluding those already in \ink{}.
For each, we extracted the base fold and intended retraction specification. In total, this yields 94 evaluation scenarios, each pairing a source fold with a retraction specification.

\bfpara{Experimental setup.}
All experiments are conducted on a machine with two AMD EPYC 9654 CPUs and
\SI{256}{GB} of memory, running NixOS 25.11. We use a 15-minute timeout for each
benchmark. 

\subsection{RQ1: Verification Effectiveness}

\begin{wraptable}{r}{0.48\linewidth}
\vspace{-0.4in}
\centering
\small
\caption{Statistics for verification benchmarks}
\vspace{-0.1in}
\renewcommand{\arraystretch}{1.1}
\setlength{\tabcolsep}{5pt}
\begin{tabular}{@{}lccc@{}}
\toprule
\textbf{Source} & \textbf{\# Cand.} & \textbf{\# Correct} & \textbf{\# Buggy} \\
\midrule
LLM (\ink{}) & 69 & 53 ($77\%$) & 16 ($23\%$) \\
Human (GitHub) & 25 & 20 ($80\%$)& 5 ($20\%$)\\
\midrule
Overall & 94 & 73 & 21 \\
\bottomrule
\end{tabular}
\label{tab:rq1-candidates}
\vspace{-0.2in}
\end{wraptable}
Our first research question asks whether \tool{} can verify candidate
retractable folds. For this experiment, each benchmark consists of a source
fold, a candidate retractable fold, and a retraction specification. The GitHub
benchmarks already include candidate implementations, namely the manually
written Flink retraction logic. 
For the \ink{} benchmarks, we generate candidate retractable implementations using GPT-5.5 with a standard self-repair loop~\cite{selfrefine}. That is, after each failed validation attempt, we provide structured feedback to the LLM and allow up to three repair rounds. Thus, these LLM-generated
implementations provide plausible
 candidates for evaluating the verifier and also help assess the need for
specialized synthesis methods for retractable folds.

\bfpara{Verification benchmarks.}
{\Cref{tab:rq1-candidates} summarizes statistics for the candidate verification benchmarks. Both sources contain a non-trivial number of incorrect retractable implementations: 16 of the 69 LLM-generated candidates and 5 of the 25 human-written GitHub implementations are buggy.
Thus, retraction logic appears error-prone across both automatically generated and developer-written implementations. Since the LLM candidates were produced with up to three repair rounds, the remaining errors indicate that testing-based repair alone does not reliably catch semantic bugs in retraction logic, motivating techniques that reason directly about retraction semantics.}

{To better understand these failures, we manually inspected the incorrect candidates from both sources. In the human-written implementations, bugs often arise in boundary cases of the retraction semantics, such as incorrectly restoring auxiliary state after a deletion. In the LLM-generated candidates, we observed two recurring patterns. Some fail to guard retractions properly. For example, when the retained summary no longer contains enough information to update the state correctly, they return an updated state instead of $\Err$, causing the result to diverge from full recomputation. Others update bounded auxiliary summaries incorrectly during retraction.  For example, when the same value appears multiple times, they may remove the wrong copy from the retained list, so the auxiliary state no longer corresponds to the elements that remain after deletion.}

{For the verification experiment, we use the 73 source--target pairs that are semantically correct according to our manual analysis. For each pair, we ask whether our verifier can automatically prove equivalence with respect to the source fold and the corresponding retraction specification.}

\bfpara{Baseline.} Since there is no off-the-shelf verifier for checking
equivalence in this setting, we also implement a baseline based on standard
 product-program encodings from the relational verification
literature~\cite{product,covac,asymmetricProduct,selfcomposition,terauchi-aiken-sif}.  Our baseline simulates
 four executions: the retractable fold on the {full
insertion history}, the retractable fold on the actual {mixed sequence}, the original fold on the  normalized history, and the retractable fold on the same normalized history. 
It also maintains relevant ghost state, such as the number
of retractions performed so far.

To make the comparison as strong as possible, we instrument the product program
with assertions that mirror the key ingredients of our proof system. In particular, we ensure that 
(1) the mixed execution  agrees with the normalized retractable execution,  (2) the
retraction budget is respected, (3) in-budget retractions do not return
$\Err$, and (4) the original and retractable folds must produce the same output on
the normalized history. We then encode the resulting product program as a
CHC-based safety problem and use a portfolio solver of Spacer~\cite{spacer} and Eldarica~\cite{eldarica} to infer the invariants
needed to prove these assertions.
Thus, the baseline is given the same semantic ingredients as our verifier, but
must prove the resulting relational safety property monolithically rather than
using the structured proof obligations from \Cref{fig:proof-rules}.

\begin{table}[t]
\centering
\caption{{\bf RQ1:} Verification evaluation results. ``Baseline Auto'' uses
Spacer/Eldarica to infer invariants automatically, while ``Baseline +Inv''
uses invariants inferred by \tool{} and checks the resulting VCs with CVC5.}
\vspace{-0.1in}
\small
\renewcommand{\arraystretch}{1.1}
\setlength{\tabcolsep}{4pt}
\begin{tabular}{@{}lcccccc@{}}
\toprule
\multirow{2}{*}{\textbf{Source}} &
\multirow{2}{*}{\textbf{\# Correct}} &
\multirow{2}{*}{\textbf{\tool{}}} &
\multicolumn{2}{c}{\textbf{Baseline}} &
\multicolumn{2}{c}{\textbf{\tool{} time (s)}} \\
\cmidrule(lr){4-5}
\cmidrule(lr){6-7}
& & & \textbf{Auto} & \textbf{+Inv}
& \textbf{Median} & \textbf{90th pct.} \\
\midrule
LLM (\ink{})    & 53 & 51 (96\%)  & 0 (0\%) & 16 (30\%)   & 12.03 & 84.23 \\
Human (GitHub) & 20 & 20 (100\%) & 0 (0\%) & 5 (25\%) & 5.60  & 33.91 \\
\midrule
Overall        & 73 & 71 (97\%)  & 0 (0\%) & 21 (29\%) & 10.29 & 84.19 \\
\bottomrule
\end{tabular}
\label{tab:rq1-verification}
\vspace{-0.16in}
\end{table}

\bfpara{Results.}
Our main verification results are presented in
\Cref{tab:rq1-verification}.
Overall, \tool{} verifies 71 of the 73 correct candidates (97\%) within the 15-minute timeout. It verifies all 20 correct human-written GitHub candidates and 51 of the 53 correct LLM-generated candidates.
{In contrast, the fully automated baseline is unable to verify any benchmark within the time limit: both Spacer and Eldarica time out while trying to infer the required invariants for the product-program encoding. To separate invariant inference from verification-condition discharge, we also evaluate an oracle-aided variant that is supplied with the invariants inferred by \tool{} and checks the resulting VCs using CVC5. This semi-automated variant verifies 21 of the 73 correct candidates, but the remaining cases still time out during VC discharge. These results suggest that the difficulty is not only finding suitable relational invariants, but also organizing the resulting proof obligations in a form that is tractable for automated solvers. Thus, \tool{}’s structured decomposition is important both for invariant discovery and for VC discharge.}

\bfpara{Failure analysis.}
{In the two unverified cases, \tool{} cannot establish that
deletion produces a state consistent with recomputation. The solver
either times out or returns unknown when asked to strengthen the
relational invariant needed for this proof.}

\medskip
\idiotbox{RQ1}{
{\tool{} verifies 71 of the 73 correct candidate retractable folds. The fully
automated product-program baseline times out on all benchmarks, and even an
oracle-aided variant verifies only 21, showing the benefit of our structured
proof obligations.}
}

\subsection{RQ2: Synthesis Effectiveness}

Our second research question asks whether \tool{} can synthesize a verified retractable computation from a source fold and retraction specification. Unlike in RQ1, the retractable fold is not provided as input; so, \tool{} must discover the auxiliary state, synthesize the retraction logic, and prove equivalence to the source fold.

\begin{table}[t]
\centering
\caption{\textbf{RQ2:} End-to-end synthesis results by lifting requirement.}
\vspace{-0.14in}
\small
\renewcommand{\arraystretch}{1.1}
\setlength{\tabcolsep}{5pt}
\begin{tabular}{@{}lcccc@{}}
\toprule
\textbf{Category} &
\textbf{\# Benchmarks} &
\textbf{Solved} &
\textbf{Median Time (s)} &
\textbf{90th Pct. (s)} \\
\midrule
No lift required & 45       & 45 (100\%)            & 16.58    & 38.89 \\
Lift required    & 49       & 49 (100\%)            & 92.05    & 214.51 \\
\midrule
Overall          & 94       & 94 (100\%)            & 64.62    & 211.85 \\
\bottomrule
\end{tabular}
\label{tab:rq2-synthesis}
\vspace{-0.2in}
\end{table}

{\Cref{tab:rq2-synthesis} reports end-to-end synthesis results for the 94 synthesis scenarios in our benchmark suite. We separate benchmarks according to a key source of difficulty: whether the source fold can be made retractable using its existing accumulator state or whether \tool{} must synthesize additional auxiliary state. Overall, \tool{} synthesizes and verifies a correct retractable computation for \emph{all} 94 scenarios. At first glance, this may seem surprising in light of RQ1, where \tool{} could not verify every correct source--target pair. The key difference is that RQ1 evaluates \tool{} on externally produced implementations, whose state representations and update logic were not chosen with our proof strategy in mind. Such implementations may be correct, but still require invariants that our verifier does not infer automatically. In RQ2, by contrast, \tool{} controls the construction of the retractable computation. In particular, it chooses the auxiliary state through a lifting invariant and synthesizes retraction logic using the same semantic decomposition used during verification. As a result, the synthesized implementations come with proof structure that makes them more amenable to automatic verification.}

{As shown in \Cref{tab:rq2-synthesis}, instances that require lifting are unsurprisingly more challenging, as reflected in synthesis time.  In such cases, \tool{} must both  identify suitable auxiliary state and correctly maintain it during both insertions and deletions. As a result, the synthesis task becomes significantly more difficult.
\Cref{tab:rq2-state-kinds} further characterizes the synthesized auxiliary state. The lifted solutions fall into three categories.  Multiplicity maps are used when the computation needs to distinguish whether an extremal value, such as the current minimum or maximum, has other copies. Tuples and bounded sorted bags retain selected values ordered by magnitude, supporting order statistics such as top-$k$. Finally, a significant number of benchmarks require storing the entire sequence of currently retained inputs. This arises when the result depends on information about the relative order of elements, such as which value was seen first or most recently, that cannot be recovered from the original state after deletion.}

\bfpara{Manual analysis.}
{We manually inspected the benchmarks where \tool{} retains
the full input sequence and found one case where a bounded summary
would suffice. In this case, synthesizing the retraction logic over
the bounded summary times out because it requires complex conditional
updates. \tool{} therefore proceeds to a larger representation that
retains the full sequence and admits simpler retraction logic.
This illustrates that prioritizing smaller representations does not
guarantee a space-minimal result.}

\begin{table}[t]
\centering
\caption{\textbf{RQ2:} Kinds of auxiliary state synthesized by \tool{}.}
\vspace{-0.17in}
\small
\renewcommand{\arraystretch}{1.1}
\setlength{\tabcolsep}{5pt}
\begin{tabular}{@{}lcc@{}}
\toprule
\textbf{Auxiliary state} &
\textbf{\# Instances} &
\textbf{Typical examples} \\
\midrule
Multiplicity map & 2 & min/max with duplicate tracking \\
Tuple or bounded sorted bag & 8 & ordered statistics like top-$k$ \\
Full history & 39 & arrival-order dependent computations \\
\bottomrule
\end{tabular}
\label{tab:rq2-state-kinds}
\vspace{-0.2in}
\end{table}

\vspace{0.1in}
\medskip
\idiotbox{RQ2}{
\tool{} synthesizes verified retractable computations for all 94 benchmarks. Even in the 49 cases where the original accumulator state is insufficient, \tool{} discovers suitable auxiliary state and synthesizes the corresponding update logic.
}

\subsection{RQ3: Efficiency of Synthesized Implementations}
\label{sec:eval-efficiency}

Our third research question asks how efficient the synthesized
folds are compared with recomputation. We first analyze retraction
complexity across all 94 synthesized folds, with particular attention
to those retaining full input history. We then compare runtime
against recomputation on the 45 scenarios supporting unrestricted
retraction.

\begin{wraptable}{r}{0.43\textwidth}
\vspace{-0.3in}
\centering
\caption{Expected amortized complexity.}
\label{tab:rq3-complexity}
\vspace{-0.1in}
\small
\renewcommand{\arraystretch}{1.1}
\setlength{\tabcolsep}{4pt}
\begin{tabular}{@{}lrrrr@{}}
\toprule
\textbf{Scenario} & $O(1)$ & $O(\log n)$ & $O(n)$ & \textbf{Total} \\
\midrule
All     & 72 & 13 & 9 & 94 \\
Full history  & 19 & 11 & 9 & 39 \\
Unrestricted & 30 & 12 & 3 & 45 \\
\bottomrule
\end{tabular}
\vspace{-0.15in}
\end{wraptable}
\bfpara{Retraction complexity.}
{
We manually analyzed the expected amortized complexity of all
synthesized retraction functions and report the results in
\Cref{tab:rq3-complexity}, where $n$ denotes the number of inputs
currently present. As shown in the table, 85 of the 94 folds
(90.4\%) have sublinear retraction cost.
Of particular interest are the 39 folds whose auxiliary state
retains the entire sequence of inputs that have not been deleted.
Of these, 30 (76.9\%) support sublinear retraction,
showing that retaining the input sequence does not necessarily
entail linear-time retraction.
Sublinear retraction is also common among the 45 folds required
to support arbitrary (unrestricted) deletions. Of these, 42 (93.3\%) support $O(1)$ or
$O(\log n)$ retraction, while only three require $O(n)$ time.
}

\begin{wraptable}{r}{0.35\textwidth}
\vspace{-0.35in}
\centering
\caption{Speedup over recomputation.}
\label{tab:rq3-runtime}
\vspace{-0.5em}
\small
\renewcommand{\arraystretch}{1.1}
\setlength{\tabcolsep}{4pt}
\begin{tabular}{@{}lrr@{}}
\toprule
\textbf{Scenario group} & \textbf{\#} & \textbf{Speedup} \\
\midrule
Unrestricted retraction & 45 & $1{,}067.3\times$ \\
Retains input sequence  & 18 & $272.0\times$ \\
\bottomrule
\end{tabular}
\vspace{-2em}
\end{wraptable}
\bfpara{Runtime evaluation.}
We next evaluate whether these complexity differences translate
into lower update times. We consider all 45 scenarios supporting unrestricted retraction,
which allows us to evaluate sustained workloads of insertions
and deletions without exhausting a specified deletion allowance.
Of these, 18 retain the entire sequence
of remaining inputs.
We compare each synthesized fold against a baseline that processes insertions
incrementally but recomputes
the original fold after each deletion.
We evaluate workloads with initial input sizes ranging from
$1{,}000$ to $30{,}000$ records, followed by twice as many updates.
Updates are randomly chosen insertions and deletions with equal
probability, with deletions sampled uniformly from the records
currently present. Both implementations process identical workloads
and produce an output after each update. We measure mean time
per update, including output evaluation but excluding initialization,
using five random seeds and three runs per seed.
We report speedup as the baseline's mean update time divided by
that of the synthesized fold and aggregate speedups across
scenarios using the geometric mean.

\begin{wrapfigure}{r}{0.48\textwidth}
\vspace{-1em}
\centering
\includegraphics[width=\linewidth]{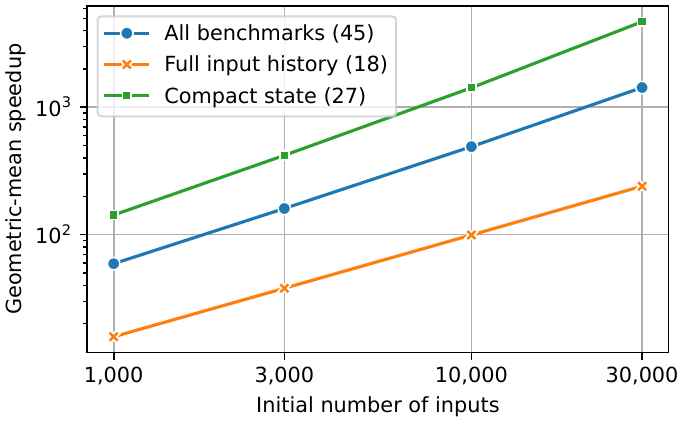}
\vspace{-2em}
\caption{Speedup vs input size.}
\label{fig:rq3-efficiency}
\vspace{-0.6em}
\end{wrapfigure}

\bfpara{Results.}
\Cref{tab:rq3-runtime} summarizes the results for workloads
initialized with $30{,}000$ inputs. All 45 synthesized
implementations outperform recomputation, with speedups ranging
from $1.14\times$ to over $8{,}000\times$ and a geometric mean
of $1{,}067.3\times$.
Notably, even the 18 implementations that retain the entire
input sequence achieve substantial gains, with a geometric-mean
speedup of $272.0\times$ over recomputation.
\Cref{fig:rq3-efficiency} shows that the benefits grow with
the initial number of inputs. Across all 45 scenarios,
geometric-mean speedup increases from $47.9\times$ at
$1{,}000$ initial inputs to $1{,}067.3\times$ at $30{,}000$.
For implementations that retain the input sequence, speedup
increases from $15.2\times$ to $272.0\times$.

\vspace{1em}
\idiotbox{RQ3}{{
85 of 94 synthesized folds support sublinear retraction.
At $30{,}000$ initial inputs, geometric-mean speedup over
recomputation is $1{,}067.3\times$.
}}

\subsection{RQ4: Ablation Studies}

\newcommand{\noliftinv}{\textsc{NoLiftInv}}
\newcommand{\nofeasibility}{\textsc{NoFeasibilityCheck}}
\newcommand{\retractsygus}{\textsc{NoSymbolicPartition}}

Our fourth research question evaluates the importance of \tool{}'s core synthesis
mechanisms. We compare the full system against three ablations that disable one
 component at a time:

\begin{enumerate}[leftmargin=*]
\item \textbf{\noliftinv{}} disables invariant-guided lifting by setting the
logical qualifier to \texttt{true}. Without this guidance, the system must
search for lifted accumulators without a relational explanation of how the
lifted state corresponds to the original fold.

\item \textbf{\nofeasibility{}} disables the feasibility checks (\Cref{alg:build-examples,alg:spec-necessity}) for lifting
specifications. Without these checks, the system may spend
time pursuing lift specifications that cannot be realized by any valid lifted
accumulator or retraction function.

\item \textbf{\retractsygus{}} replaces our symbolic-partitioning-based
retraction synthesis procedure with a direct SyGuS query for the entire
\texttt{retract} function. Hence, this ablation tests whether our symbolic partitioning idea is useful.

\end{enumerate}

\begin{figure}[t]
\centering
\includegraphics[width=0.90\textwidth]{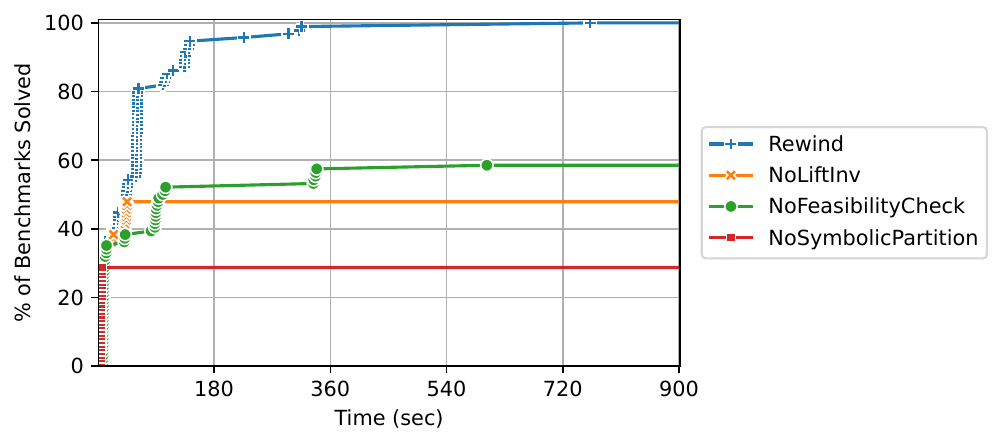}
\vspace{-0.2in}
\caption{\textbf{RQ4:} Comparison between \tool{} and its ablations. The
$x$-axis shows synthesis time, excluding verification time, since all variants
use the same final verifier. The $y$-axis shows the percentage of benchmarks
solved within the corresponding time budget.}
\label{fig:rq3-cdf}
\vspace{-0.2in}
\end{figure}

\Cref{fig:rq3-cdf} reports the results of this ablation study using a
15-minute timeout per benchmark. The running time excludes verification time. 
Because all variants invoke the same verifier after synthesis,  the intention is to isolate
the effect of the synthesis procedure itself. As standard for a cactus plot, the
$y$-axis shows the percentage of benchmarks solved within a given time budget.

The results show that each of our algorithmic ideas contributes to scalability.
Removing lifting invariants (\noliftinv{}) substantially reduces the success
rate, dropping from $100.0\%$ for the full system to $47.9\%$. Disabling the
feasibility checks from \Cref{alg:build-examples,alg:spec-necessity} (\nofeasibility{}) similarly reduces the success rate to $58.5\%$.
The last ablation, \retractsygus{}, has the largest effect: it solves only $28.7\%$ of the benchmarks and plateaus almost immediately. This suggests that direct retraction synthesis is only effective for the simplest cases. For the remaining benchmarks, the retraction logic is too case-dependent to synthesize as a single monolithic SyGuS query, and our symbolic partitioning method is essential for scalability.

\vspace{0.05in}
\medskip
\idiotbox{RQ4}{
Each ablation significantly degrades performance, showing that \tool{} relies on
all three algorithmic ingredients. Disabling symbolic partitioning or lifting
invariants substantially reduces the success rate. Feasibility checks have a
smaller effect on success rate, but a large effect on synthesis time.
}

\section{Related Work}\label{sec:related}

\bfpara{Dynamic and static incremental computation.}
A substantial line of work studies how to maintain outputs efficiently as inputs
change. Self-adjusting computation~\cite{acar_2008,acar_2006,acar_2014}
records dynamic dependencies during execution and propagates input changes to
avoid full recomputation. This idea has led to a number of language and runtime
systems for general-purpose incremental computation, including
\textsc{CEAL}~\cite{ceal} and demand-driven systems such as
\textsc{Adapton}~\cite{adapton,nominal-adapton}. A complementary static line of
work derives incremental programs by transforming existing
programs~\cite{annieliu,ilc,cts}. Unlike these approaches, which
provide general mechanisms for propagating input changes, our setting starts
from a fold with a given forward update function and synthesizes the missing
logic for removing previously processed inputs.
{These approaches are complementary to ours: runtime
systems reuse portions of previous executions, whereas synthesis can
discover auxiliary summaries and update rules not explicit in those
executions. Moreover, general-purpose incremental-computation systems
are not drop-in alternatives for Flink aggregations targeted
by \tool{}.}

\bfpara{Synthesis-based algorithmic transformations.}
Program synthesis has also been used to derive more efficient implementations
from high-level computations. For example, \textsc{Opera} synthesizes online
algorithms from offline ones~\cite{opera}, \textsc{AutoLifter} applies
divide-and-conquer-like algorithmic paradigms by synthesizing lifted
programs~\cite{autolifter}, \textsc{Superfusion} eliminates intermediate data
structures through inductive synthesis~\cite{superfusion}, and Ji et al.
synthesize efficient dynamic-programming algorithms via thinning
transformations~\cite{methyll}.
These systems, like ours, use synthesis to
obtain a more efficient computation. However, they focus on reorganizing how the
output is computed from the input, whereas we synthesize retraction logic  to remove a previously processed input.

\bfpara{Incremental view maintenance and dataflow systems.}
Many database and dataflow systems support efficient maintenance under
insertions and deletions. Systems such as \textsc{DBToaster}, Differential
Dataflow, and \textsc{DBSP} maintain results for queries or dataflows by
exploiting the algebraic structure of the underlying language, where changes to
inputs can be propagated compositionally through operators and deletions can be
represented as negative updates~\cite{koch2014dbtoaster,differential-dataflow,
DBSP:pvldb/BudiuCMRT23}. 
Our setting is different in that the computation is not given as an algebraic query or
dataflow graph. Instead, our setting involves  a stateful fold with an arbitrary accumulator, which 
may not contain enough information to support deletion. \tool{} addresses this
problem by synthesizing both the auxiliary state and the retraction function,
together with a proof of correctness.

\bfpara{Bidirectional transformations.}
Our work is related to bidirectional programming and lenses. In the lens
setting, a program maintains consistency between a source and a derived view via
paired \emph{get} and \emph{put} functions satisfying round-trip
laws~\cite{foster_lenses,relational_lenses,boomerang,quotient_lenses}.
Our problem has a  similar flavor because it also relates a forward
computation to a state update that preserves semantic consistency. However, the
technical problem is different. A lens update is given an edited view and
produces a new source consistent with that view. In contrast, a retraction is
given the current accumulator state and a deleted input element, and must
produce a state equivalent to recomputing the fold on the remaining inputs. Thus,
rather than repairing a source after a view edit, our challenge is to determine what
information the accumulator must retain so that deletions can be processed
incrementally and synthesize the retraction logic.

\bfpara{Program inversion.} Retraction synthesis is closely related to program inversion, which aims to
construct an inverse or partial inverse for a given program. Prior work has
studied inversion for functional programs~\cite{Morita2007Inversion}, symbolic
transducer-based inversion~\cite{genic}, and path-based inductive synthesis of
imperative program inverses~\cite{pins}. Despite this connection, retraction differs from ordinary program inversion in the specification it must satisfy. An inverse of \texttt{accumulate} would recover a pre-state from a post-state and the last inserted element, e.g., an operation satisfying  $\mathsf{unacc}(\acc(s,x),x)=s$.  Such an operation is only meaningful for undoing the most recent update. In contrast, \texttt{retract} must remove the effect of an element from an accumulated history, possibly after many other insertions have occurred.

\bfpara{Synthesis of auxiliary state.}
Prior work has shown that program transformations often become realizable only
after lifting a computation with additional state. In divide-and-conquer
synthesis, Farzan and Nicolet show how to synthesize conditional accumulators
so that partial results can be composed correctly~\cite{Farzan2017DC,
Farzan2019Nested}. Subsequent work on phased synthesis~\cite{parsynt_pldi21}
and SE\textsuperscript{2}GIS~\cite{Farzan2022SE2GIS} makes this process more
compositional and uses symbolic reasoning, including unrealizability witnesses,
to guide synthesis. More recently, \textsc{AutoLifter}~\cite{autolifter}
automatically infers auxiliary variables that make relational specifications
realizable. Our work builds on the same broad insight, but applies it to a
different kind of transformation. In this
setting, the lifted state must support not only the forward computation, but
also a well-defined deletion operation that satisfies a user-specified requirement and whose result agrees with recomputation on
the normalized history.

\bfpara{Solver-aided synthesis.}
Our method builds on standard solver-aided synthesis techniques such as
CEGIS~\cite{sketch,sketch2} and SyGuS~\cite{sygus}, as well as later work that
uses symbolic reasoning to improve synthesis~\cite{Farzan2022SE2GIS,Hu2019Unrealizability,Hu2020Unrealizability}.
These techniques provide some of the solver infrastructure used by \tool{}, but
the main challenge in our setting is formulating retraction as a tractable
synthesis problem. \tool{} addresses this challenge by searching for feasible
lift specifications, inferring relational invariants between mixed and
normalized executions, and using symbolic partitioning to decompose
\texttt{retract} synthesis into simpler subproblems that can be solved using off-the-shelf tools.

\bfpara{Relational verification.}
This paper is also related to the long line of work on relational verification,
which reasons about multiple executions of one or more programs. Relational
Hoare logics and their variants have been used to prove program equivalence,
refinement, security, differential privacy, and quantitative relational
properties~\cite{rhl,chl,qchl,prhl,apRHL}. Another line of work reduces
relational verification to ordinary safety verification using
self-composition, cross-products, and product
programs~\cite{selfcomposition,terauchi-aiken-sif,covac,product,
asymmetricProduct}. We use this latter idea as the basis for our verification
baseline. As discussed in \Cref{sec:verification}, however, verifying
retractable folds raises additional challenges beyond standard 
equivalence checking. Our proof system addresses these challenges by making the
retraction-specific invariants and support obligations explicit.

\bfpara{Retraction support in data processing.}
Modern data-processing systems expose deletion support through explicit
retraction interfaces. For example, Flink allows an \texttt{AggregateFunction} to provide a
\texttt{retract} method for incrementally removing previously processed
records~\cite{flink}. RisingWave provides a similar 
\texttt{retract} interface for user-defined aggregates; without it, the
aggregate cannot be updated incrementally in materialized views~\cite{risingwave}. 
These interfaces motivate our work because they make  it possible for the system to take advantage of  retraction function.
However, they leave the
developer responsible for implementing and validating the corresponding retraction logic.

\section{Conclusion}

This paper studied how to automatically add deletion support to
fold-based incremental computations. We developed a semantic
framework in which retraction specifications describe the required
deletion support, together with a proof system that decomposes
correctness into local relational obligations. Building on this
foundation, our synthesis procedure searches for auxiliary state
through lifting invariants, synthesizes the operations needed to
maintain that state, and derives retraction logic through symbolic
partitioning. We also experimentally evaluated our approach and showed that \tool{}  synthesizes verified
retractable folds for all 94 benchmark scenarios. Furthermore, the resulting
implementations substantially outperform recomputation. 

Our approach also suggests broader opportunities in data
representation synthesis, where changing a representation requires
adapting its operations while preserving correctness.
Retraction adds a further challenge, as the new representation must
support functionality absent from the original program.
Our techniques address this by using lifting invariants to reject
unrealizable representations, leveraging relational reasoning to connect
existing and new operations, and employing symbolic partitioning to derive
smaller synthesis problems from a reference implementation.
We believe these ideas can help automate other program transformations
that introduce new functionality while preserving existing behavior.

\section{Data-Availability Statement}

Our artifact, including the \tool{} implementation, benchmark suite, and evaluation scripts, is available on Zenodo~\cite{wang_2026_23329026}.

\section*{Acknowledgments}
We thank the anonymous reviewers for their thoughtful and constructive feedback.
This work was conducted in a research group supported by NSF awards CCF-1918889, CNS-1908304, CCF-1901376, CNS-2120696, CCF-2210831, CCF-2319471, CCF-2422130, and CCF-2403211, as well as DARPA awards under agreements HR00112590133 and HR0011262E020, and a Renaissance Philanthropy award.

\bibliography{main}

@misc{risingwave, url={https://risingwave.com/}}

@article{DBSP:pvldb/BudiuCMRT23,
author = {Budiu, Mihai and Chajed, Tej and McSherry, Frank and Ryzhyk, Leonid and Tannen, Val},
title = {DBSP: Automatic Incremental View Maintenance for Rich Query Languages},
year = {2023},
issue_date = {March 2023},
publisher = {VLDB Endowment},
volume = {16},
number = {7},
issn = {2150-8097},
url = {https://doi.org/10.14778/3587136.3587137},
doi = {10.14778/3587136.3587137},
journal = {Proc. VLDB Endow.},
month = mar,
pages = {1601–1614},
numpages = {14}
}

@article{opera,
  title={From Batch to Stream: Automatic Generation of Online Algorithms},
  author={Wang, Ziteng and Pailoor, Shankara and Prakash, Aaryan and Wang, Yuepeng and Dillig, I{\c{s}}{\i}l},
  journal={Proceedings of the ACM on Programming Languages},
  volume={8},
  number={PLDI},
  pages={1014--1039},
  year={2024},
  publisher={ACM New York, NY, USA}
}

@article{flink,
  title={Apache flink: Stream and batch processing in a single engine},
  author={Carbone, Paris and Katsifodimos, Asterios and Ewen, Stephan and Markl, Volker and Haridi, Seif and Tzoumas, Kostas},
  journal={The Bulletin of the Technical Committee on Data Engineering},
  volume={38},
  number={4},
  year={2015},
  publisher={Institute of Electrical and Electronics Engineers (IEEE)}
}

@article{DBToaster,
  author    = {Yanif Ahmad and
               Oliver Kennedy and
               Christoph Koch and
               Milos Nikolic},
  title     = {DBToaster: Higher-order Delta Processing for Dynamic, Frequently Fresh
               Views},
  journal   = {{PVLDB}},
  volume    = {5},
  number    = {10},
  pages     = {968--979},
  year      = {2012},
  url       = {http://vldb.org/pvldb/vol5/p968\_yanifahmad\_vldb2012.pdf},
  bibsource = {dblp computer science bibliography, https://dblp.org}
}

@inproceedings{Farzan2022SE2GIS,
  author    = {Azadeh Farzan and Danya Lette and Victor Nicolet},
  title     = {Recursion Synthesis with Unrealizability Witnesses},
  booktitle = {Proceedings of the 43rd {ACM} SIGPLAN International Conference on Programming Language Design and Implementation (PLDI)},
  pages     = {16:1--16:16},
  year      = 2022
}

@inproceedings{Hu2019Unrealizability,
  author    = {Qinheping Hu and Jason Breck and John Cyphert and Loris D'Antoni and Thomas Reps},
  title     = {Proving Unrealizability for Syntax-Guided Synthesis},
  booktitle = {Computer Aided Verification (CAV)},
  pages     = {335--352},
  year      = 2019
}

@inproceedings{Hu2020Unrealizability,
  author    = {Qinheping Hu and John Cyphert and Loris D'Antoni and Thomas Reps},
  title     = {Exact and Approximate Methods for Proving Unrealizability of SyGuS Problems},
  booktitle = {Proceedings of the 41st {ACM} SIGPLAN Conference on Programming Language Design and Implementation (PLDI)},
  pages     = {1128--1142},
  year      = 2020
}

@InProceedings{cvc5,
author="Barbosa, Haniel
and Barrett, Clark
and Brain, Martin
and Kremer, Gereon
and Lachnitt, Hanna
and Mann, Makai
and Mohamed, Abdalrhman
and Mohamed, Mudathir
and Niemetz, Aina
and N{\"o}tzli, Andres
and Ozdemir, Alex
and Preiner, Mathias
and Reynolds, Andrew
and Sheng, Ying
and Tinelli, Cesare
and Zohar, Yoni",
editor="Fisman, Dana
and Rosu, Grigore",
title="cvc5: A Versatile and Industrial-Strength SMT Solver",
booktitle="Tools and Algorithms for the Construction and Analysis of Systems",
year="2022",
publisher="Springer International Publishing",
address="Cham",
pages="415--442",
isbn="978-3-030-99524-9"
}

@inproceedings{sketch,
  author    = {Armando Solar{-}Lezama and
               Liviu Tancau and
               Rastislav Bod{\'{\i}}k and
               Sanjit A. Seshia and
               Vijay A. Saraswat},
  title     = {Combinatorial sketching for finite programs},
  booktitle = {Proceedings of the 12th International Conference on Architectural
               Support for Programming Languages and Operating Systems ({ASPLOS})},
  pages     = {404--415},
  publisher = {{ACM}},
  year      = {2006},
  url       = {https://doi.org/10.1145/1168857.1168907},
  doi       = {10.1145/1168857.1168907},
}

@inproceedings{sketch2,
  author    = {Armando Solar{-}Lezama and
               Christopher Grant Jones and
               Rastislav Bod{\'{\i}}k},
  title     = {Sketching concurrent data structures},
  booktitle = {Proceedings of the {ACM} {SIGPLAN} 2008 Conference on Programming
               Language Design and Implementation ({PLDI})},
  pages     = {136--148},
  publisher = {{ACM}},
  year      = {2008},
  url       = {https://doi.org/10.1145/1375581.1375599},
  doi       = {10.1145/1375581.1375599},
}

@inproceedings{parsynt_pldi21,
author = {Farzan, Azadeh and Nicolet, Victor},
title = {Phased synthesis of divide and conquer programs},
year = {2021},
isbn = {9781450383912},
publisher = {Association for Computing Machinery},
address = {New York, NY, USA},
url = {https://doi.org/10.1145/3453483.3454089},
doi = {10.1145/3453483.3454089},
booktitle = {Proceedings of the 42nd ACM SIGPLAN International Conference on Programming Language Design and Implementation},
pages = {974–986},
numpages = {13},
location = {Virtual, Canada},
series = {PLDI 2021}
}

@inproceedings{sygus,
  author       = {Rajeev Alur and
                  Rastislav Bod{\'{\i}}k and
                  Garvit Juniwal and
                  Milo M. K. Martin and
                  Mukund Raghothaman and
                  Sanjit A. Seshia and
                  Rishabh Singh and
                  Armando Solar{-}Lezama and
                  Emina Torlak and
                  Abhishek Udupa},
  title        = {Syntax-guided synthesis},
  booktitle    = {Formal Methods in Computer-Aided Design, {FMCAD} 2013, Portland, OR,
                  USA, October 20-23, 2013},
  pages        = {1--8},
  publisher    = {{IEEE}},
  year         = {2013},
  url          = {https://ieeexplore.ieee.org/document/6679385/},
  bibsource    = {dblp computer science bibliography, https://dblp.org}
}

@article{adapton,
author = {Hammer, Matthew A. and Phang, Khoo Yit and Hicks, Michael and Foster, Jeffrey S.},
title = {Adapton: Composable, Demand-Driven Incremental Computation},
year = {2014},
issue_date = {June 2014},
publisher = {Association for Computing Machinery},
address = {New York, NY, USA},
volume = {49},
number = {6},
issn = {0362-1340},
url = {https://doi.org/10.1145/2666356.2594324},
doi = {10.1145/2666356.2594324},
journal = {SIGPLAN Not.},
month = {jun},
pages = {156–166},
numpages = {11}
}

@article{annieliu,
author = {Liu, Yanhong A.},
title = {Efficiency by Incrementalization: An Introduction},
year = {2000},
issue_date = {Dec. 1, 2000},
publisher = {Kluwer Academic Publishers},
address = {USA},
volume = {13},
number = {4},
issn = {1388-3690},
url = {https://doi.org/10.1023/A:1026547031739},
doi = {10.1023/A:1026547031739},
journal = {Higher Order Symbol. Comput.},
month = {dec},
pages = {289–313},
numpages = {25}
}

@Article{koch2014dbtoaster,
author={Koch, Christoph
and Ahmad, Yanif
and Kennedy, Oliver
and Nikolic, Milos
and N{\"o}tzli, Andres
and Lupei, Daniel
and Shaikhha, Amir},
title={DBToaster: higher-order delta processing for dynamic, frequently fresh views},
journal={The VLDB Journal},
year={2014},
month={Apr},
day={01},
volume={23},
number={2},
pages={253-278},
issn={0949-877X},
doi={10.1007/s00778-013-0348-4},
url={https://doi.org/10.1007/s00778-013-0348-4}
}

@article{acar_2006,
title = {A Library for Self-Adjusting Computation},
journal = {Electronic Notes in Theoretical Computer Science},
volume = {148},
number = {2},
pages = {127-154},
year = {2006},
note = {Proceedings of the ACM-SIGPLAN Workshop on ML (ML 2005)},
issn = {1571-0661},
doi = {https://doi.org/10.1016/j.entcs.2005.11.043},
url = {https://www.sciencedirect.com/science/article/pii/S1571066106001290},
author = {Umut Acar and Guy Blelloch and Matthias Blume and Robert Harper and Kanat Tangwongsan}
}

@inproceedings{acar_2008,
  author       = {Matthew A. Hammer and
                  Umut A. Acar},
  editor       = {Richard E. Jones and
                  Stephen M. Blackburn},
  title        = {Memory management for self-adjusting computation},
  booktitle    = {Proceedings of the 7th International Symposium on Memory Management,
                  {ISMM} 2008, Tucson, AZ, USA, June 7-8, 2008},
  pages        = {51--60},
  publisher    = {{ACM}},
  year         = {2008},
  url          = {https://doi.org/10.1145/1375634.1375642},
  doi          = {10.1145/1375634.1375642},
  bibsource    = {dblp computer science bibliography, https://dblp.org}
}

@inproceedings{acar_2014,
author = {Acar, Umut A. and Chen, Yan},
title = {Streaming Big Data with Self-Adjusting Computation},
year = {2013},
isbn = {9781450318716},
publisher = {Association for Computing Machinery},
address = {New York, NY, USA},
url = {https://doi.org/10.1145/2429376.2429382},
doi = {10.1145/2429376.2429382},
booktitle = {Proceedings of the 2013 Workshop on Data Driven Functional Programming},
pages = {15–18},
numpages = {4},
location = {Rome, Italy},
series = {DDFP '13}
}

@article{ink,
author = {Wang, Ziteng and Fang, Ruijie and Zheng, Linus and Tang, Dixin and Dillig, I{\c s}{\i}l},
title = {Homomorphism Calculus for User-Defined Aggregations},
year = {2025},
issue_date = {October 2025},
publisher = {Association for Computing Machinery},
address = {New York, NY, USA},
volume = {9},
number = {OOPSLA2},
url = {https://doi.org/10.1145/3763072},
doi = {10.1145/3763072},
journal = {Proc. ACM Program. Lang.},
month = oct,
articleno = {294},
numpages = {27}
}

@article{foster_lenses,
author = {Foster, J. Nathan and Greenwald, Michael B. and Moore, Jonathan T. and Pierce, Benjamin C. and Schmitt, Alan},
title = {Combinators for bidirectional tree transformations: A linguistic approach to the view-update problem},
year = {2007},
issue_date = {May 2007},
publisher = {Association for Computing Machinery},
address = {New York, NY, USA},
volume = {29},
number = {3},
issn = {0164-0925},
url = {https://doi.org/10.1145/1232420.1232424},
doi = {10.1145/1232420.1232424},
journal = {ACM Trans. Program. Lang. Syst.},
month = may,
pages = {17–es},
numpages = {65}
}

@article{genic,
author = {Hu, Qinheping and D'Antoni, Loris},
title = {Automatic program inversion using symbolic transducers},
year = {2017},
issue_date = {June 2017},
publisher = {Association for Computing Machinery},
address = {New York, NY, USA},
volume = {52},
number = {6},
issn = {0362-1340},
url = {https://doi.org/10.1145/3140587.3062345},
doi = {10.1145/3140587.3062345},
journal = {SIGPLAN Not.},
month = jun,
pages = {376–389},
numpages = {14}
}

@article{pins,
author = {Srivastava, Saurabh and Gulwani, Sumit and Chaudhuri, Swarat and Foster, Jeffrey S.},
title = {Path-based inductive synthesis for program inversion},
year = {2011},
issue_date = {June 2011},
publisher = {Association for Computing Machinery},
address = {New York, NY, USA},
volume = {46},
number = {6},
issn = {0362-1340},
url = {https://doi.org/10.1145/1993316.1993557},
doi = {10.1145/1993316.1993557},
journal = {SIGPLAN Not.},
month = jun,
pages = {492–503},
numpages = {12}
}

@article{Morita2007Inversion,
author = {Morita, Kazutaka and Morihata, Akimasa and Matsuzaki, Kiminori and Hu, Zhenjiang and Takeichi, Masato},
title = {Automatic inversion generates divide-and-conquer parallel programs},
year = {2007},
issue_date = {June 2007},
publisher = {Association for Computing Machinery},
address = {New York, NY, USA},
volume = {42},
number = {6},
issn = {0362-1340},
url = {https://doi.org/10.1145/1273442.1250752},
doi = {10.1145/1273442.1250752},
journal = {SIGPLAN Not.},
month = jun,
pages = {146–155},
numpages = {10}
}

@inproceedings{Farzan2017DC,
author = {Farzan, Azadeh and Nicolet, Victor},
title = {Synthesis of divide and conquer parallelism for loops},
year = {2017},
isbn = {9781450349888},
publisher = {Association for Computing Machinery},
address = {New York, NY, USA},
url = {https://doi.org/10.1145/3062341.3062355},
doi = {10.1145/3062341.3062355},
booktitle = {Proceedings of the 38th ACM SIGPLAN Conference on Programming Language Design and Implementation},
pages = {540–555},
numpages = {16},
location = {Barcelona, Spain},
series = {PLDI 2017}
}

@inproceedings{Farzan2019Nested,
author = {Farzan, Azadeh and Nicolet, Victor},
title = {Modular divide-and-conquer parallelization of nested loops},
year = {2019},
isbn = {9781450367127},
publisher = {Association for Computing Machinery},
address = {New York, NY, USA},
url = {https://doi.org/10.1145/3314221.3314612},
doi = {10.1145/3314221.3314612},
booktitle = {Proceedings of the 40th ACM SIGPLAN Conference on Programming Language Design and Implementation},
pages = {610–624},
numpages = {15},
location = {Phoenix, AZ, USA},
series = {PLDI 2019}
}

@MISC{smtlib,
  author =	 {Clark Barrett and Pascal Fontaine and Cesare Tinelli},
  title =	 {{The Satisfiability Modulo Theories Library (SMT-LIB)}},
  howpublished = {{\tt www.SMT-LIB.org}},
  year =	 2016,
}

@inproceedings{murray2013naiad,
  author    = {Derek G. Murray and Frank McSherry and Rebecca Isaacs and Michael Isard and Paul Barham and Mart{\'i}n Abadi},
  title     = {Naiad: A Timely Dataflow System},
  booktitle = {Proceedings of the 24th ACM Symposium on Operating Systems Principles},
  pages     = {439--455},
  year      = {2013},
  doi       = {10.1145/2517349.2522738}
}

@misc{materialize-arrangements,
  author       = {{Materialize}},
  title        = {Arrangements},
  howpublished = {\url{https://materialize.com/docs/get-started/arrangements/}},
  note         = {Accessed 2026-06-12}
}

@inproceedings{rhl,
  author    = {Nick Benton},
  title     = {Simple Relational Correctness Proofs for Static Analyses and Program Transformations},
  booktitle = {Proceedings of the 31st ACM SIGPLAN-SIGACT Symposium on Principles of Programming Languages},
  series    = {POPL 2004},
  pages     = {14--25},
  year      = {2004},
  publisher = {ACM},
  doi       = {10.1145/964001.964003}
}

@inproceedings{chl,
  author    = {Marcelo Sousa and Isil Dillig},
  title     = {Cartesian Hoare Logic for Verifying k-Safety Properties},
  booktitle = {Proceedings of the 37th ACM SIGPLAN Conference on Programming Language Design and Implementation},
  series    = {PLDI 2016},
  pages     = {57--69},
  year      = {2016},
  publisher = {ACM},
  doi       = {10.1145/2908080.2908092}
}

@inproceedings{qchl,
  author    = {Jia Chen and Yu Feng and Isil Dillig},
  title     = {Precise Detection of Side-Channel Vulnerabilities using Quantitative Cartesian Hoare Logic},
  booktitle = {Proceedings of the 2017 ACM SIGSAC Conference on Computer and Communications Security},
  series    = {CCS 2017},
  pages     = {875--890},
  year      = {2017},
  publisher = {ACM},
  doi       = {10.1145/3133956.3134058}
}

@inproceedings{product,
  author    = {Gilles Barthe and Juan Manuel Crespo and C{\'e}sar Kunz},
  title     = {Relational Verification Using Product Programs},
  booktitle = {Proceedings of the 17th International Symposium on Formal Methods},
  series    = {FM 2011},
  pages     = {200--214},
  year      = {2011},
  publisher = {Springer},
  doi       = {10.1007/978-3-642-21437-0_17}
}

@inproceedings{asymmetricProduct,
  author    = {Gilles Barthe and Juan Manuel Crespo and C{\'e}sar Kunz},
  title     = {Beyond 2-Safety: Asymmetric Product Programs for Relational Program Verification},
  booktitle = {Proceedings of the International Symposium on Logical Foundations of Computer Science},
  series    = {LFCS 2013},
  pages     = {29--43},
  year      = {2013},
  publisher = {Springer},
  doi       = {10.1007/978-3-642-35722-0_3}
}

@inproceedings{selfcomposition,
  author    = {Gilles Barthe and Pedro R. D'Argenio and Tamara Rezk},
  title     = {Secure Information Flow by Self-Composition},
  booktitle = {Proceedings of the 17th IEEE Computer Security Foundations Workshop},
  series    = {CSFW 2004},
  pages     = {100--114},
  year      = {2004},
  publisher = {IEEE Computer Society}
}

@inproceedings{terauchi-aiken-sif,
  author    = {Tachio Terauchi and Alex Aiken},
  title     = {Secure Information Flow as a Safety Problem},
  booktitle = {Static Analysis},
  series    = {SAS 2005},
  volume    = {3672},
  pages     = {352--367},
  year      = {2005},
  publisher = {Springer}
}

@inproceedings{covac,
  author    = {Anna Zaks and Amir Pnueli},
  title     = {{CoVaC}: Compiler Validation by Program Analysis of the Cross-Product},
  booktitle = {Formal Methods},
  series    = {FM 2008},
  pages     = {35--51},
  year      = {2008},
  publisher = {Springer}
}

@inproceedings{prhl,
  author    = {Gilles Barthe and Benjamin Gr{\'e}goire and Santiago Zanella B{\'e}guelin},
  title     = {Probabilistic Relational Hoare Logics for Computer-Aided Security Proofs},
  booktitle = {Proceedings of the 11th International Conference on Mathematics of Program Construction},
  series    = {MPC 2012},
  pages     = {1--6},
  year      = {2012},
  publisher = {Springer},
  doi       = {10.1007/978-3-642-31113-0_1}
}

@inproceedings{apRHL,
  author    = {Gilles Barthe and Boris K{\"o}pf and Federico Olmedo and Santiago Zanella B{\'e}guelin},
  title     = {Probabilistic Relational Reasoning for Differential Privacy},
  booktitle = {Proceedings of the 39th Annual ACM SIGPLAN-SIGACT Symposium on Principles of Programming Languages},
  series    = {POPL 2012},
  pages     = {97--110},
  year      = {2012},
  publisher = {ACM},
  doi       = {10.1145/2103656.2103670}
}

@inproceedings{selfrefine,
  title     = {Self-Refine: Iterative Refinement with Self-Feedback},
  author    = {Madaan, Aman and Tandon, Niket and Gupta, Prakhar and Hallinan, Skyler and Gao, Luyu and Wiegreffe, Sarah and Alon, Uri and Dziri, Nouha and Prabhumoye, Shrimai and Yang, Yiming and Gupta, Shashank and Majumder, Bodhisattwa Prasad and Hermann, Katherine and Welleck, Sean and Yazdanbakhsh, Amir and Clark, Peter},
  booktitle = {Advances in Neural Information Processing Systems},
  year      = {2023}
}

@inproceedings{spacer,
  title     = {The Spacer Horn Solver},
  author    = {Komuravelli, Anvesh and Gurfinkel, Arie and Chaki, Sagar},
  booktitle = {Computer Aided Verification},
  series    = {CAV 2016},
  pages     = {181--189},
  year      = {2016},
  publisher = {Springer}
}

@inproceedings{ceal,
  author    = {Hammer, Matthew A. and Acar, Umut A. and Chen, Yan},
  title     = {{CEAL}: A {C}-Based Language for Self-Adjusting Computation},
  booktitle = {Proceedings of the 30th ACM SIGPLAN Conference on Programming Language Design and Implementation},
  series    = {PLDI 2009},
  pages     = {25--37},
  year      = {2009},
  publisher = {ACM},
  doi       = {10.1145/1542476.1542480}
}

@inproceedings{nominal-adapton,
  author    = {Hammer, Matthew A. and Dunfield, Jana and Headley, Kyle and Labich, Nicholas and Foster, Jeffrey S. and Hicks, Michael and Van Horn, David},
  title     = {Incremental Computation with Names},
  booktitle = {Proceedings of the 2015 ACM SIGPLAN International Conference on Object-Oriented Programming, Systems, Languages, and Applications},
  series    = {OOPSLA 2015},
  pages     = {748--766},
  year      = {2015},
  publisher = {ACM},
  doi       = {10.1145/2814270.2814305}
}

@inproceedings{ilc,
author = {Cai, Yufei and Giarrusso, Paolo G. and Rendel, Tillmann and Ostermann, Klaus},
title = {A theory of changes for higher-order languages: incrementalizing λ-calculi by static differentiation},
year = {2014},
isbn = {9781450327848},
publisher = {Association for Computing Machinery},
address = {New York, NY, USA},
url = {https://doi.org/10.1145/2594291.2594304},
doi = {10.1145/2594291.2594304},
booktitle = {Proceedings of the 35th ACM SIGPLAN Conference on Programming Language Design and Implementation},
pages = {145–155},
numpages = {11},
location = {Edinburgh, United Kingdom},
series = {PLDI '14}
}

@InProceedings{cts,
author="Giarrusso, Paolo G.
and R{\'e}gis-Gianas, Yann
and Schuster, Philipp",
editor="Caires, Lu{\'i}s",
title="Incremental $\lambda$-Calculus in Cache-Transfer Style",
booktitle="Programming Languages and Systems",
year="2019",
publisher="Springer International Publishing",
address="Cham",
pages="553--580",
}

@article{autolifter,
  author    = {Ji, Ruyi and Zhao, Yuwei and Xiong, Yingfei and Wang, Di and Zhang, Lu and Hu, Zhenjiang},
  title     = {Decomposition-Based Synthesis for Applying Divide-and-Conquer-Like Algorithmic Paradigms},
  journal   = {ACM Transactions on Programming Languages and Systems},
  year      = {2024},
  publisher = {ACM},
  doi       = {10.1145/3648440}
}

@article{superfusion,
  author    = {Ji, Ruyi and Zhao, Yuwei and Polikarpova, Nadia and Xiong, Yingfei and Hu, Zhenjiang},
  title     = {Superfusion: Eliminating Intermediate Data Structures via Inductive Synthesis},
  journal   = {Proceedings of the ACM on Programming Languages},
  volume    = {8},
  number    = {PLDI},
  year      = {2024},
  publisher = {ACM},
  doi       = {10.1145/3656415}
}

@inproceedings{methyll,
  author    = {Ji, Ruyi and Zhu, Tianran and Xiong, Yingfei and Hu, Zhenjiang},
  title     = {Automating Thinning Theorem: Synthesizing Efficient Dynamic Programming Algorithms},
  booktitle = {Programming Languages and Systems},
  series    = {ESOP 2022},
  year      = {2022},
  publisher = {Springer}
}

@inproceedings{differential-dataflow,
  author    = {McSherry, Frank and Murray, Derek G. and Isaacs, Rebecca and Isard, Michael},
  title     = {Differential Dataflow},
  booktitle = {Proceedings of the Sixth Biennial Conference on Innovative Data Systems Research},
  series    = {CIDR 2013},
  year      = {2013}
}

@inproceedings{relational_lenses,
  author    = {Bohannon, Aaron and Pierce, Benjamin C. and Vaughan, Jeffrey A.},
  title     = {Relational Lenses: A Language for Updatable Views},
  booktitle = {Proceedings of the Twenty-Fifth ACM SIGMOD-SIGACT-SIGART Symposium on Principles of Database Systems},
  series    = {PODS 2006},
  pages     = {338--347},
  year      = {2006},
  publisher = {ACM},
  doi       = {10.1145/1142351.1142399}
}

@inproceedings{boomerang,
  author    = {Bohannon, Aaron and Foster, J. Nathan and Pierce, Benjamin C. and Pilkiewicz, Alexandre and Schmitt, Alan},
  title     = {Boomerang: Resourceful Lenses for String Data},
  booktitle = {Proceedings of the 35th Annual ACM SIGPLAN-SIGACT Symposium on Principles of Programming Languages},
  series    = {POPL 2008},
  pages     = {407--419},
  year      = {2008},
  publisher = {ACM},
  doi       = {10.1145/1328438.1328487}
}

@inproceedings{quotient_lenses,
author = {Foster, J. Nathan and Pilkiewicz, Alexandre and Pierce, Benjamin C.},
title = {Quotient lenses},
year = {2008},
isbn = {9781595939197},
publisher = {Association for Computing Machinery},
address = {New York, NY, USA},
url = {https://doi.org/10.1145/1411204.1411257},
doi = {10.1145/1411204.1411257},
booktitle = {Proceedings of the 13th ACM SIGPLAN International Conference on Functional Programming},
pages = {383–396},
numpages = {14},
location = {Victoria, BC, Canada},
series = {ICFP '08}
}

@INPROCEEDINGS{eldarica,
  author={Hojjat, Hossein and Rümmer, Philipp},
  booktitle={2018 Formal Methods in Computer Aided Design (FMCAD)}, 
  title={The ELDARICA Horn Solver}, 
  year={2018},
  volume={},
  number={},
  pages={1-7},
  doi={10.23919/FMCAD.2018.8603013}}

@software{wang_2026_23329026,
  author       = {Wang, Ziteng and
                  Zheng, Linus and
                  Tang, Dixin and
                  Dillig, Isil},
  title        = {Software Artifact for "Rewinding the Fold: Making
                   Stateful Computations Retractable"
                  },
  month        = oct,
  year         = 2026,
  publisher    = {Zenodo},
  doi          = {10.5281/zenodo.23329026},
  url          = {https://doi.org/10.5281/zenodo.23329026},
}

\newpage
\appendix
\crefalias{section}{appendix}
\crefalias{subsection}{appendix}
\section{Detailed Trace of Symbolic Partition Refinement}
\label{app:synretract-trace}

\begin{example}

Continuing \Cref{example:synretract-symbolic-partition}, we now explain how
\Cref{alg:create-symbolic-partitions-map} derives symbolic partitions as the
explored history length increases. For $\ell=1$, let $xs=[x_1]$. Symbolic
execution yields the single final state $(x_1,1)$, and deleting $x_1$ leads to
the unique predecessor $(\infty,0)$. After quantifier elimination, this yields
the  symbolic partition with just one element:
\[
\gamma_{0}((m,c),x)\equiv (c=1 \wedge x=m),
\qquad
\psi_{0}((m,c),x)\equiv
\Retract((m,c),x) = (\infty,0).
\]
Now, suppose inductive synthesis returns the following solution:
\[
\Retract((m, c), x) = (\infty, 0).
\]
This is consistent with the current local specification, but it fails full verification,  for example, with the  concrete counterexample 
$ 
xs=[1,2],  s^\ast=(1,1),  x^\ast=2
$.
This counterexample has previously unseen length $\ell=2$; hence, 
\Cref{alg:synretract-partition} invokes {\sc UpdateSymbolicPartitions}. For $xs=[x_1,x_2]$, symbolic execution now separates the
possible histories according to whether the deleted element is equal to one of two minima,  larger than the
minimum, or equal to a unique minimum. For the first case, deleting one of two equal minima yields the partition element:
\[
\gamma_{1}((m,c),x) \equiv (c=2 \wedge x=m), \qquad
\psi_{1}((m,c),x) \equiv \Retract((m,c),x) = (m,c-1).
\]
For the second case, we obtain the candidate partitions:
\begin{align*}
\gamma_{2}^a((m,c),x)\equiv (c=1 \wedge x>m),
&\qquad
\psi_{2}^a((m,c),x)\equiv \Retract((m,c),x) = (m,c),\\
\gamma_{2}^b((m,c),x)\equiv (c=2 \wedge x>m),
&\qquad
\psi_{2}^b((m,c),x)\equiv \Retract((m,c),x) = (m,c).
\end{align*}
When performing {\sc InsertOrMerge}, these two get merged into a common partition by generalizing constants into an unknown predicate, yielding the guard $\gamma_{2}((m,c),x)\equiv \square_1 \land x > m$ and its corresponding specification $\Retract((m,c),x) = (m,c)$.

Finally,  the third case where the deleted element is  equal to the unique minimum  has path condition $c=1$ and $x=m$. But this case fails the ambiguity test at lines~9--10 and is pruned. 
Thus, at the end of the second CEGIS iteration, we have the following symbolic partition:
\[
\{\, 
(c=2 \land x=m) \mapsto (m,c-1),\;
(\square_1\{c=1 \lor c=2\} \land x>m) \mapsto (m,c)
\,\}
\]
where the notation $\square\{\phi\}$ indicates that $\square$ must be instantiated with a formula that generalizes $\phi$. 
In the next iteration, suppose \textsc{SolveSyGuS} returns the following plausible solution:
\[
\Retract((m,c),x)=
\text{if } c=2\land x=m \text{ then } (m,1)\;
\text{else if } x>m \text{ then } (m,c)\;
\text{else } \mathsf{Err}
\]
This solution is also incorrect, and \textsf{Verify} can find a new counterexample of length $3$; for example, 
$xs=[1,1,1],  s^\ast=(1,3),  x^\ast=1$.
\Cref{alg:synretract-partition} then invokes {\sc UpdateSymbolicPartitions}, now with $\ell=3$. Symbolic execution discovers a new candidate partition element:
\[
\gamma_4((m,c),x)\equiv (c=3 \wedge x=m), \qquad
\psi_4((m,c),x)\equiv \Retract((m,c),x) = (m,c-1).
\]
This is the same pattern $\gamma_1$ already seen at $\ell=2$, except with $c=3$. Thus, \textsc{InsertOrMerge} merges $\gamma_4$ with $\gamma_1$ and yields:
$\gamma'_1((m,c),x)\equiv (\square_2 \land x=m ),$
{where $\square_2$ is a predicate that generalizes $c = 2 \lor c = 3$}. This gives rise to the new symbolic partition:
\[
\{\, 
(\square_2\{c=2 \lor c=3 \} \land x=m) \mapsto (m, c-1),\;
(\square_1\{c=1 \lor c=2 \lor c=3\} \land x>m) \mapsto (m,c)
\,\}
\]
Instantiating $\square_1$ with $\top$ and $\square_2$ with $c>1$, a plausible synthesized solution is then
\[
\Retract((m,c),x)=
\text{if } x>m \text{ then } (m,c)\;
\text{else if } x=m \land c>1 \text{ then } (m,c-1)\;
\text{else } \mathsf{Err}.
\]
This solution passes {\sf Verify} and the CEGIS loop terminates.
\end{example}

\section{Product-Program Baseline}
\label{app:product-baseline}

This appendix describes the product-program encoding used for the baseline
verifier in \Cref{sec:eval}. The goal of the encoding is to reduce equivalence
modulo a retraction specification to an ordinary safety-verification problem.
The product program simulates the executions that appear in the semantic
correctness condition and asserts, after each update, that these executions
remain appropriately related.

Let $P$ be the original fold, let $R$ be the candidate retractable fold, and let
$\RetSpec$ be the retraction specification. The product program maintains three
ghost variables:
\begin{itemize} 
\item $ys$  for the full insertion history 
\item $xs$ for the effective history 
\item $n$  for the number of retractions so far
\end{itemize}
It also maintains four program states:
\begin{itemize} 
    \item $s_{\mathsf{full}}$, which is the state of $R$ on the full insertion history
$ys$;
\item $s_{\mathsf{mix}}$, which is the state of $R$ on the actual mixed update
sequence;
\item   $s_{\mathsf{eff}}$, which is the state of $R$ on the effective insertion-only
history $xs$; 
\item $s_{\mathsf{orig}}$, which is the state of $P$ on the same effective
history $xs$.
\end{itemize}

\begin{algorithm}[t]
\caption{\textsc{ProductVerify}$(P,R,\RetSpec)$}
\label{alg:product-verify}
\begin{algorithmic}[1]
\State $ys \gets [\,]$; \quad $xs \gets [\,]$; \quad $n \gets 0$
\State $s_{\mathsf{full}} \gets R.\Init$;  $s_{\mathsf{mix}} \gets R.\Init$;  $s_{\mathsf{eff}} \gets R.\Init$; $s_{\mathsf{orig}} \gets P.\Init$
\State \Call{CheckInv}{$P,R,ys,xs,n,s_{\mathsf{full}},s_{\mathsf{mix}},s_{\mathsf{eff}},s_{\mathsf{orig}}$}
\While{$\star$}
  \State choose $op \in \{\mathsf{insert},\mathsf{retract}\}$
  \State choose $a$
  \If{$op=\mathsf{insert}$}
    \State $ys \gets ys \cdot [a]$; $xs \gets xs \cdot [a]$
    \State $s_{\mathsf{full}} \gets R.\acc(s_{\mathsf{full}},a)$; $s_{\mathsf{mix}} \gets R.\acc(s_{\mathsf{mix}},a)$; $s_{\mathsf{eff}} \gets R.\acc(s_{\mathsf{eff}},a)$;  $s_{\mathsf{orig}} \gets P.\acc(s_{\mathsf{orig}},a)$
  \ElsIf{$op=\mathsf{retract}$}
    \State \textbf{assume} $a \in xs$
    \If{$\neg(n < \RetSpec(ys))$} \textbf{assume false}
    \EndIf
    \State $tmp \gets R.\Retract(s_{\mathsf{mix}},a)$
    \State \textbf{assert} $tmp \neq \Err$
    \State $s_{\mathsf{mix}} \gets tmp$
    \State $xs \gets \remove(xs,a)$
    \State $n \gets n+1$
    \State $s_{\mathsf{eff}} \gets R.\Init$;  $s_{\mathsf{orig}} \gets P.\Init$
    \For{\textbf{each} $b \in xs$}
      \State $s_{\mathsf{eff}} \gets R.\acc(s_{\mathsf{eff}},b)$; $s_{\mathsf{orig}} \gets P.\acc(s_{\mathsf{orig}},b)$
    \EndFor
  \EndIf
  \State \Call{CheckInv}{$P,R,ys,xs,n,s_{\mathsf{full}},s_{\mathsf{mix}},s_{\mathsf{eff}},s_{\mathsf{orig}}$}
\EndWhile
\end{algorithmic}
\end{algorithm}

\begin{algorithm}[t]
\caption{\textsc{CheckInv}$(P,R,ys,xs,n,s_{\mathsf{full}},s_{\mathsf{mix}},s_{\mathsf{eff}},s_{\mathsf{orig}})$}
\label{alg:product-check-inv}
\begin{algorithmic}[1]
  \State \textbf{assert} $\Reach(ys,s_{\mathsf{full}})$
  \State \textbf{assert} $\SimBudget(s_{\mathsf{full}},s_{\mathsf{mix}},n)$
  \State \textbf{assert} $\Reach(xs,s_{\mathsf{eff}})$
  \State \textbf{assert} $\SimTwo(xs,s_{\mathsf{eff}},s_{\mathsf{mix}})$
  \State \textbf{assert} $P.\eval(s_{\mathsf{orig}})=R.\eval(s_{\mathsf{eff}})$
\end{algorithmic}
\end{algorithm}

\Cref{alg:product-verify} shows the product-program encoding. Insertion steps
advance all four executions in lockstep (lines~7--9): both ghost histories are
extended with the inserted element, and each simulated state takes the
corresponding accumulator step. Retraction steps are handled only when they are
within the budget prescribed by $\RetSpec$ (lines~11--13): if
$n \ge \RetSpec(ys)$, the path is pruned because the specification imposes no
further support obligation. Otherwise, the product program asserts that the
candidate $\Retract$ succeeds, updates the mixed execution with the returned
state, removes the deleted element from the effective history, and increments
the retraction count (lines~14--18).

After a retraction, the effective-history executions are reconstructed by
replaying the elements of $xs$ from the initial states of $R$ and $P$ in
lockstep (lines~19--22). This makes $s_{\mathsf{eff}}$ and
$s_{\mathsf{orig}}$ explicit reference executions on the normalized history. The
ghost variables are updated  so that $xs$ is obtained from $ys$ by
exactly $n$ deletions.

\Cref{alg:product-check-inv} contains the safety assertions used by the baseline
verifier. The first and third assertions check that $s_{\mathsf{full}}$ and
$s_{\mathsf{eff}}$ are valid insertion-only executions of $R$ on $ys$ and $xs$,
respectively. The $\SimBudget$ assertion relates the full-history execution to
the mixed execution while accounting for the number of retractions already
performed. The $\SimTwo$ assertion relates the mixed execution to the effective
insertion-only execution. Finally, the last assertion checks observational
equivalence between the original fold and the retractable fold on the effective
history.

\newcommand{\canon}{\mathsf{Canon}}

\section{Soundness of Proof Rules}

In this section we prove the soundness direction of
\Cref{thm:proof-rules-sound}: if there exist predicates $\Reach$,
$\SimTwo$, and $\SimBudget$ satisfying all rules in
\Cref{fig:proof-rules}, then the candidate \textrudaf{} $\rudaf$ is
equivalent to $\udaf$ modulo $\RetSpec$ in the sense of
\Cref{def:equivalence}.  Throughout this section we fix an
\textudaf{} $\udaf=(\init,\acc,\eval)$, a \textrudaf{}
$\rudaf=(\init',\acc',\Retract,\eval')$, and a retraction
specification $\RetSpec$.  Up to and including
\Cref{thm:proof-rules-sound-appendix}, we additionally fix predicates
$\Reach$, $\SimTwo$, and $\SimBudget$ satisfying all rules in
\Cref{fig:proof-rules}.

\paragraph{Conventions.}
All update sequences below are valid in the sense of
\Cref{def:valid-seq}; following the convention adopted in
\Cref{sec:problem}, the quantifier in \Cref{def:retspec} also ranges
over valid sequences (clause~(2) of \Cref{def:equivalence} is already
stated for valid sequences).  Validity is closed under taking prefixes, since the
defining condition of \Cref{def:valid-seq} is imposed separately at
every index $j$.  We call an update sequence
\emph{accumulation-only} if every operation tag in it is
$\mathsf{acc}$; accumulation-only sequences are trivially valid.  We use two direct
consequences of \Cref{def:rudaf-semantics}.
First, $\Err$ is absorbing: if $\psem{\rudaf}{\invseq}=\Err$, then
$\psem{\rudaf}{\invseq\pp\invseq''}=\Err$ for every extension of
$\invseq$; contrapositively, if some extension of $\invseq$ has
non-$\Err$ state semantics, then so does $\invseq$ itself.  Second,
if $\psem{\rudaf}{\invseq}\neq\Err$, then
\[
  \psem{\rudaf}{\invseq\pp[(\acc,x)]}
     =\acc'(\psem{\rudaf}{\invseq},x),
  \qquad
  \psem{\rudaf}{\invseq\pp[(\Retract,x)]}
     =\Retract(\psem{\rudaf}{\invseq},x).
\]

For a list $xs$ and an element $x\in xs$, we write $\remove(xs,x)$ for
the list obtained by deleting the first occurrence of $x$.  This convention
makes $\remove$ well defined on lists and matches condition~(2) of
\Cref{def:normalization}, which cancels the earliest accumulation of the
retracted value.  Part~(iii) of \Cref{lem:norm-facts} establishes the
correspondence.

The following normalization facts are used throughout the section.

\begin{lemma}[Elementary properties of normalization]
\label[lemma]{lem:norm-facts}
Let $\invseq$ be a valid update sequence.  Then:
\begin{enumerate}[label=(\roman*),leftmargin=*]
\item $\Norm(\invseq)$ is accumulation-only, and
  $\vals{\Norm(\invseq)}=\Effect(\invseq)$;
\item $\Norm(\invseq\pp[(\acc,x)])=\Norm(\invseq)\pp[(\acc,x)]$, and
  hence \[\Effect(\invseq\pp[(\acc,x)])=\Effect(\invseq)\pp[x];\]
\item if $\invseq\pp[(\Retract,x)]$ is valid, then
  $x\in\Effect(\invseq)$, the sequence
  $\Norm(\invseq\pp[(\Retract,x)])$ is obtained from $\Norm(\invseq)$
  by deleting its first accumulation of $x$, and consequently
  \[\Effect(\invseq\pp[(\Retract,x)])=\remove(\Effect(\invseq),x);\]
\item $\Subseq(\vals{\invseq^+},\Effect(\invseq),|\invseq^-|)$, that
  is, $\Effect(\invseq)$ is obtained from $\vals{\invseq^+}$ by
  deleting exactly $|\invseq^-|$ elements.
\end{enumerate}
\end{lemma}

\begin{proof}
The rewrite rule of \Cref{def:normalization} terminates in a unique
fixpoint on every valid sequence.  Termination follows because every rewrite
strictly shortens the sequence.  For uniqueness, each rewrite preserves
validity.  A rewrite deletes one
accumulation of some value $a$ (at position $i$) and one retraction
of the same value (at position $j>i$), leaving the prefix counts of
every value other than $a$ unchanged.  As for $a$ itself,
condition~(2) makes position $i$ the first accumulation of $a$, so by
validity no retraction of $a$ occurs before $i$, and condition~(3)
excludes retractions of $a$ strictly between $i$ and $j$.  Hence, in
the rewritten sequence, the count condition of \Cref{def:valid-seq}
for $a$ still holds at every index: prefixes ending before $i$ are
unchanged, prefixes ending in $[i,j)$ contain no retraction of $a$ at
all, and prefixes ending at or after $j$ lose one accumulation and
one retraction of $a$.  Next, in a valid sequence, the redex for each
value $a$ is uniquely determined: condition~(2) forces $i$ to be the
position of the first accumulation of $a$, and condition~(3),
together with validity, forces $j$ to be the position of the first
retraction of $a$.  Moreover, rewrites for distinct values neither
enable nor disable one another, because all side conditions of the
rule constrain only positions carrying the value being cancelled.
Consequently, exhaustive rewriting deletes the same set of positions under
every rewrite order, and the fixpoint is unique.

(i)~Suppose, toward a contradiction, that $\Norm(\invseq)$ contains a
retraction, and consider the first retraction in $\Norm(\invseq)$,
say of value $a$ at position $j$.  Since rewriting preserves
validity, $\Norm(\invseq)$ is valid, so some accumulation of $a$
occurs before position $j$; let $i$ be the position of the first
accumulation of $a$.  Then $i<j$, condition~(2) holds by the choice
of $i$, and condition~(3) holds because there is no retraction at all
before position $j$.  Hence the rewrite rule applies to
$\Norm(\invseq)$, contradicting the fact that $\Norm(\invseq)$ is a
fixpoint.  Therefore $\Norm(\invseq)$ is accumulation-only, and
$\vals{\Norm(\invseq)}=\Effect(\invseq)$ holds by the definition of
$\Effect$ in \Cref{def:normalization}.

(ii)~Appending an accumulation preserves validity, so
$\invseq\pp[(\acc,x)]$ is valid.  The trailing accumulation survives every
rewrite because it has no later matching retraction.  Its presence leaves all
other redexes unchanged: conditions (2) and (3) for a redex $(i,j)$ constrain
positions strictly before $j$, whereas the trailing element occupies the last
position.  Exhaustive rewriting therefore follows the same steps on both
sequences, and the resulting fixpoint is
$\Norm(\invseq)\pp[(\acc,x)]$.  Taking value lists yields
\[\Effect(\invseq\pp[(\acc,x)])=\Effect(\invseq)\pp[x].\]

(iii)~The fact $x\in\Effect(\invseq)$ follows by counting.  Every
rewrite deletes exactly one accumulation and one retraction of the
same value, and by (i) the fixpoint $\Norm(\invseq)$ contains no
retraction.  Hence, for every value $a$, exactly as many
accumulations of $a$ are cancelled as there are retractions of $a$ in
$\invseq$, so the number of occurrences of $a$ in $\Effect(\invseq)$
equals the number of accumulations of $a$ in $\invseq$ minus the
number of retractions of $a$ in $\invseq$.  Validity of
$\invseq\pp[(\Retract,x)]$, instantiated at its last index, says that
$\invseq$ contains strictly more accumulations of $x$ than
retractions of $x$; therefore $x$ occurs at least once in
$\Effect(\invseq)$.

For the remaining claims, use the order independence established above and
normalize $\invseq\pp[(\Retract,x)]$ by first rewriting its prefix.
Every redex of $\invseq$ remains a redex of
$\invseq\pp[(\Retract,x)]$, because conditions (2) and (3) constrain
positions strictly before the retraction being cancelled, whereas the appended
element occupies the last position.  Performing the rewrites internal to
$\invseq$ first reaches
$\Norm(\invseq)\pp[(\Retract,x)]$.  Since $\Norm(\invseq)$ is
accumulation-only and contains an accumulation of $x$, one further
rewrite applies: it cancels the trailing retraction against the first
accumulation of $x$ in $\Norm(\invseq)$ (condition~(2) selects the
first accumulation of $x$, and condition~(3) is vacuous because no
other retraction is present).  The result is accumulation-only.  Since every
redex requires a retraction at position $j$, this result is a fixpoint.  Thus
$\Norm(\invseq\pp[(\Retract,x)])$ is $\Norm(\invseq)$ with its first
accumulation of $x$ deleted, and taking value lists gives
\[\Effect(\invseq\pp[(\Retract,x)])=\remove(\Effect(\invseq),x).\]

(iv)~By the counting argument in (iii), exhaustive rewriting of
$\invseq$ performs exactly $|\invseq^-|$ rewrites: each rewrite
deletes exactly one retraction, and the fixpoint contains none.  Each
rewrite also deletes exactly one accumulation, and deletions preserve
the relative order of the surviving entries.  Hence $\Norm(\invseq)$
is obtained from $\invseq^+$ by deleting exactly $|\invseq^-|$
entries.  Taking value lists, $\Effect(\invseq)$ is obtained from
$\vals{\invseq^+}$ by deleting exactly $|\invseq^-|$ elements, which
is precisely $\Subseq(\vals{\invseq^+},\Effect(\invseq),|\invseq^-|)$.
\end{proof}

The next lemma shows a property of the retractable-fold semantics used below.

\begin{lemma}[Accumulation-only executions avoid $\Err$]
\label[lemma]{lem:acc-only-noerr}
For every accumulation-only update sequence $\invseq$, the state
semantics $\psem{\rudaf}{\invseq}$ is an element of the state type of
$\rudaf$; in particular, $\psem{\rudaf}{\invseq}\neq\Err$.
\end{lemma}

\begin{proof}
Let $s_0,\dots,s_n$ be the state sequence of
\Cref{def:rudaf-semantics} for $\rudaf$ on $\invseq$.  We show by
induction on $j$ that every $s_j$ is a proper state.  For $j=0$ we
have $s_0=\init'$, which is a state by the definition of a
retractable fold.  For $j>0$, the induction hypothesis gives
$s_{j-1}\neq\Err$, and $o_j=\mathsf{acc}$ because $\invseq$ is
accumulation-only, so $s_j=\acc'(s_{j-1},a_j)$.  The codomain of
$\acc'$ is the state type itself.  In the definition of a retractable fold,
$\Err$ belongs to the codomains of only $\Retract$ and $\eval'$.  Hence $s_j$ is a
proper state, and in particular
\[\psem{\rudaf}{\invseq}=s_n\neq\Err.\]
\end{proof}

\begin{lemma}[Reachability soundness]
\label[lemma]{lem:sound-R}
For every accumulation-only update sequence $\invseq$,
\[
  \Reach\bigl(\vals{\invseq},\psem{\rudaf}{\invseq}\bigr).
\]
\end{lemma}

\begin{proof}
By induction on $\invseq$.

\emph{Base case.}
For $\invseq=\emptylist$ we have $\vals{\invseq}=\emptylist$ and
$\psem{\rudaf}{\invseq}=\init'$, so rule R1 gives the claim.

\emph{Inductive step (accumulation).}
Let $\invseq'=\invseq\pp[(\acc,x)]$ with $\invseq$
accumulation-only, and let $xs=\vals{\invseq}$ and
$s'=\psem{\rudaf}{\invseq}$.  By \Cref{lem:acc-only-noerr},
$s'\neq\Err$, so
\[
  \vals{\invseq'}=xs\pp[x],
  \qquad
  \psem{\rudaf}{\invseq'}=\acc'(s',x).
\]
The induction hypothesis gives $\Reach(xs,s')$.  Instantiating rule
R2 with this $xs$, this $s'$, and the element $x$ yields
$\Reach(xs\pp[x],\acc'(s',x))$, which is the claim for $\invseq'$.
\end{proof}

\begin{lemma}[Existence of reachable insertion states]
\label[lemma]{lem:sound-R-exists}
For every list $xs$, there exists a state $s'\neq\Err$ such that
$\Reach(xs,s')$.
\end{lemma}

\begin{proof}
Let $\canon(xs)$ denote the accumulation-only update sequence that
inserts the elements of $xs$ in order, so that
$\vals{\canon(xs)}=xs$.  Take $s'=\psem{\rudaf}{\canon(xs)}$.  Then
$s'\neq\Err$ by \Cref{lem:acc-only-noerr}, and \Cref{lem:sound-R}
applied to $\canon(xs)$ gives $\Reach(xs,s')$.
\end{proof}

\begin{lemma}[Lifting preserves semantics]
\label[lemma]{lem:sound-A}
For every accumulation-only update sequence $\invseq$,
\[
  \sem{\udaf}{\vals{\invseq}}
  =
  \sem{\rudaf}{\invseq}.
\]
In particular, $\sem{\rudaf}{\invseq}\neq\Err$ for every
accumulation-only $\invseq$.
\end{lemma}

\begin{proof}
We first show, by induction on $\invseq$, that rules A1 and A2 imply,
for every accumulation-only $\invseq$,
\[
  \eval\bigl(\psem{\udaf}{\vals{\invseq}}\bigr)
  =
  \eval'\bigl(\psem{\rudaf}{\invseq}\bigr).
\]

\emph{Base case.}
For $\invseq=\emptylist$ we have
$\psem{\udaf}{\vals{\emptylist}}=\init$ and
$\psem{\rudaf}{\emptylist}=\init'$, so the desired equality is
exactly rule A1.

\emph{Inductive step (accumulation).}
Let $\invseq'=\invseq\pp[(\acc,x)]$ with $\invseq$
accumulation-only, and let $s=\psem{\udaf}{\vals{\invseq}}$ and
$s'=\psem{\rudaf}{\invseq}$.  By \Cref{def:fold} and, since
$s'\neq\Err$ by \Cref{lem:acc-only-noerr}, by
\Cref{def:rudaf-semantics},
\[
  \psem{\udaf}{\vals{\invseq'}}=\acc(s,x),
  \qquad
  \psem{\rudaf}{\invseq'}=\acc'(s',x).
\]
The induction hypothesis gives $\eval(s)=\eval'(s')$, and rule A2
applied to this equality yields
$\eval(\acc(s,x))=\eval'(\acc'(s',x))$, which is the desired equality
for $\invseq'$.

Now let $\invseq$ be an arbitrary accumulation-only update sequence.
By \Cref{lem:acc-only-noerr}, $\psem{\rudaf}{\invseq}\neq\Err$, so
unfolding the observable semantics of \Cref{def:rudaf-semantics}
gives
$\sem{\rudaf}{\invseq}=\eval'(\psem{\rudaf}{\invseq})$.  Likewise, by
\Cref{def:fold},
$\sem{\udaf}{\vals{\invseq}}=\eval(\psem{\udaf}{\vals{\invseq}})$.
Combining these with the equality proved above yields
$\sem{\udaf}{\vals{\invseq}}=\sem{\rudaf}{\invseq}$.  The codomain of
$\eval$ is the output type $O$, which excludes $\Err$.  Hence
\[\sem{\udaf}{\vals{\invseq}}\neq\Err \text{ and } \sem{\rudaf}{\invseq}\neq\Err.\]
\end{proof}

\begin{lemma}[Retraction consistency invariant]
\label[lemma]{lem:sound-B-reach}
For every valid update sequence $\invseq$ with
$\psem{\rudaf}{\invseq}\neq\Err$, there exists a state $s_1'$ such
that
\[
  \Reach(\Effect(\invseq),s_1')
  \quad\text{and}\quad
  \SimTwo(\Effect(\invseq),s_1',\psem{\rudaf}{\invseq}).
\]
\end{lemma}

\begin{proof}
By induction on $\invseq$.

\emph{Base case.}
For $\invseq=\emptylist$ we have $\Effect(\emptylist)=\emptylist$ and
$\psem{\rudaf}{\emptylist}=\init'$.  Take $s_1'=\init'$: rule R1
gives $\Reach(\emptylist,\init')$, and rule B1 gives
$\SimTwo(\emptylist,\init',\init')$.

\emph{Inductive step (accumulation).}
Let $\invseq'=\invseq\pp[(\acc,x)]$ be valid with
$\psem{\rudaf}{\invseq'}\neq\Err$.  Since $\Err$ is absorbing,
$\psem{\rudaf}{\invseq}\neq\Err$, and $\invseq$ is valid as a prefix
of $\invseq'$.  Let $xs=\Effect(\invseq)$.  The induction hypothesis
yields a state $s_1'$ with $\Reach(xs,s_1')$ and
$\SimTwo(xs,s_1',\psem{\rudaf}{\invseq})$.  By part (ii) of
\Cref{lem:norm-facts}, $\Effect(\invseq')=xs\pp[x]$, and by the
semantics,
$\psem{\rudaf}{\invseq'}=\acc'(\psem{\rudaf}{\invseq},x)$.  Rule R2
gives $\Reach(xs\pp[x],\acc'(s_1',x))$, and rule B2 gives
$\SimTwo(xs\pp[x],\acc'(s_1',x),\acc'(\psem{\rudaf}{\invseq},x))$.
Hence $\acc'(s_1',x)$ witnesses the claim for $\invseq'$.

\emph{Inductive step (retraction).}
Let $\invseq'=\invseq\pp[(\Retract,x)]$ be valid with
$\psem{\rudaf}{\invseq'}\neq\Err$.  As before,
$\psem{\rudaf}{\invseq}\neq\Err$ and $\invseq$ is valid.  Write
$s_2'=\psem{\rudaf}{\invseq}$, $xs=\Effect(\invseq)$, and
$ys=\remove(xs,x)$.  The premises of rule B3 hold:
\begin{itemize}[leftmargin=*]
\item $\Reach(xs,s_1')$ and $\SimTwo(xs,s_1',s_2')$ hold for the
  state $s_1'$ provided by the induction hypothesis applied to
  $\invseq$;
\item $x\in xs$ holds by part (iii) of \Cref{lem:norm-facts}, since
  $\invseq\pp[(\Retract,x)]$ is valid;
\item $\Retract(s_2',x)\neq\Err$ holds because the semantics gives
 \[\psem{\rudaf}{\invseq'}=\Retract(s_2',x)\neq\Err\] by assumption;
\item $ys=\remove(xs,x)$ holds by the definition of $ys$;
\item $\Reach(ys,s_3')$ holds for some state $s_3'$ by
  \Cref{lem:sound-R-exists}.
\end{itemize}
Rule B3 therefore yields $\SimTwo(ys,s_3',\Retract(s_2',x))$.  By
part (iii) of \Cref{lem:norm-facts}, $ys=\Effect(\invseq')$, and
$\Retract(s_2',x)=\psem{\rudaf}{\invseq'}$.  Since $s_3'$ also
satisfies $\Reach(\Effect(\invseq'),s_3')$, it witnesses the claim
for $\invseq'$.
\end{proof}

\begin{lemma}[Normalization preserves semantics]
\label[lemma]{lem:sound-B}
For every valid update sequence $\invseq$ with
$\psem{\rudaf}{\invseq}\neq\Err$,
\[
  \sem{\rudaf}{\Norm(\invseq)}
  =
  \sem{\rudaf}{\invseq}.
\]
\end{lemma}

\begin{proof}
We prove the following invariant by induction on $\invseq$:
\begin{equation}
\tag{$\ast$}
\text{for every valid $\invseq$:}\quad
  \psem{\rudaf}{\invseq}\neq\Err
  \;\Longrightarrow\;
  \SimTwo\bigl(\Effect(\invseq),
    \psem{\rudaf}{\Norm(\invseq)},
    \psem{\rudaf}{\invseq}\bigr).
\end{equation}
The middle argument of $\SimTwo$ in ($\ast$) is a proper state because
$\Norm(\invseq)$ is accumulation-only by part (i) of
\Cref{lem:norm-facts}, so \Cref{lem:acc-only-noerr} applies.

\emph{Base case.}
For $\invseq=\emptylist$ we have $\Norm(\emptylist)=\emptylist$,
$\Effect(\emptylist)=\emptylist$, and
$\psem{\rudaf}{\emptylist}=\init'$, so rule B1 gives ($\ast$).

\emph{Inductive step (accumulation).}
Let $\invseq'=\invseq\pp[(\acc,x)]$ be valid with
$\psem{\rudaf}{\invseq'}\neq\Err$.  Then
$\psem{\rudaf}{\invseq}\neq\Err$ since $\Err$ is absorbing, and
$\invseq$ is valid as a prefix, so the induction hypothesis gives
\[
  \SimTwo\bigl(\Effect(\invseq),
    \psem{\rudaf}{\Norm(\invseq)},
    \psem{\rudaf}{\invseq}\bigr).
\]
By part (ii) of \Cref{lem:norm-facts},
$\Norm(\invseq')=\Norm(\invseq)\pp[(\acc,x)]$ and
$\Effect(\invseq')=\Effect(\invseq)\pp[x]$.  Since both
$\psem{\rudaf}{\Norm(\invseq)}$ and $\psem{\rudaf}{\invseq}$ are
non-$\Err$, the semantics gives
\[
  \psem{\rudaf}{\Norm(\invseq')}
    =\acc'(\psem{\rudaf}{\Norm(\invseq)},x),
  \qquad
  \psem{\rudaf}{\invseq'}
    =\acc'(\psem{\rudaf}{\invseq},x).
\]
Applying rule B2 to the induction hypothesis yields exactly ($\ast$) for $\invseq'$.

\emph{Inductive step (retraction).}
Let $\invseq'=\invseq\pp[(\Retract,x)]$ be valid with
$\psem{\rudaf}{\invseq'}\neq\Err$.  Then
$\psem{\rudaf}{\invseq}\neq\Err$ and $\invseq$ is valid.  Write
$s_2'=\psem{\rudaf}{\invseq}$, $xs=\Effect(\invseq)$, and
$ys=\remove(xs,x)$.  We instantiate rule B3 with
$s_1'=\psem{\rudaf}{\Norm(\invseq)}$ and
$s_3'=\psem{\rudaf}{\Norm(\invseq')}$ and discharge its premises one
by one:
\begin{itemize}[leftmargin=*]
\item $\Reach(xs,s_1')$: by part (i) of \Cref{lem:norm-facts},
  $\Norm(\invseq)$ is accumulation-only with
  $\vals{\Norm(\invseq)}=xs$, so \Cref{lem:sound-R} applies;
\item $\SimTwo(xs,s_1',s_2')$: by the induction hypothesis for $\invseq$;
\item $x\in xs$: by part (iii) of \Cref{lem:norm-facts}, since
  $\invseq'$ is valid;
\item $\Retract(s_2',x)\neq\Err$: the semantics gives
  $\psem{\rudaf}{\invseq'}=\Retract(s_2',x)\neq\Err$ by assumption;
\item $ys=\remove(xs,x)$: by the definition of $ys$;
\item $\Reach(ys,s_3')$: by parts (i) and (iii) of
  \Cref{lem:norm-facts}, $\Norm(\invseq')$ is accumulation-only with
  $\vals{\Norm(\invseq')}=\Effect(\invseq')=ys$, so
  \Cref{lem:sound-R} applies.
\end{itemize}
Rule B3 yields $\SimTwo(ys,s_3',\Retract(s_2',x))$.  Since
$ys=\Effect(\invseq')$, $s_3'=\psem{\rudaf}{\Norm(\invseq')}$, and
$\Retract(s_2',x)=\psem{\rudaf}{\invseq'}$, we obtain ($\ast$) for
$\invseq'$, completing the induction.

Now fix a valid $\invseq$ with $\psem{\rudaf}{\invseq}\neq\Err$.
Applying rule B4 to ($\ast$) gives
\[
  \eval'(\psem{\rudaf}{\Norm(\invseq)})
  =
  \eval'(\psem{\rudaf}{\invseq}).
\]
Both state arguments are non-$\Err$: the right one by assumption, and
the left one by \Cref{lem:acc-only-noerr}, because $\Norm(\invseq)$
is accumulation-only.  Unfolding the observable semantics of
\Cref{def:rudaf-semantics} on both sides gives
$\sem{\rudaf}{\Norm(\invseq)}=\sem{\rudaf}{\invseq}$.
\end{proof}

\begin{lemma}[Budget invariant]
\label[lemma]{lem:sound-C-inv}
For every valid update sequence $\invseq$ with
$\psem{\rudaf}{\invseq}\neq\Err$,
\[
  \SimBudget\bigl(\psem{\rudaf}{\invseq^+},
             \psem{\rudaf}{\invseq},
             |\invseq^-|\bigr).
\]
\end{lemma}

\begin{proof}
We proceed by induction on $\invseq$.  The index of $\SimBudget$ counts the
retractions separating the insertion-only replay $\invseq^+$ from the mixed
execution $\invseq$.  The sequence $\invseq^+$ is accumulation-only, so
\[\psem{\rudaf}{\invseq^+}\neq\Err\] by \Cref{lem:acc-only-noerr}.

\emph{Base case.}
For $\invseq=\emptylist$ we have $\invseq^+=\emptylist$ and
$\invseq^-=\emptylist$, so both state arguments are $\init'$ and
$|\invseq^-|=0$.  Rule C1 shows the claim.

\emph{Inductive step (accumulation).}
Let $\invseq'=\invseq\pp[(\acc,x)]$ be valid with
$\psem{\rudaf}{\invseq'}\neq\Err$.  Then
$\psem{\rudaf}{\invseq}\neq\Err$ since $\Err$ is absorbing, and
$\invseq$ is valid as a prefix, so the induction hypothesis gives
\[
  \SimBudget\bigl(\psem{\rudaf}{\invseq^+},
             \psem{\rudaf}{\invseq},
             |\invseq^-|\bigr).
\]
We have $\invseq'^+=\invseq^+\pp[(\acc,x)]$ and
$\invseq'^-=\invseq^-$.  Since $\psem{\rudaf}{\invseq^+}\neq\Err$ and
$\psem{\rudaf}{\invseq}\neq\Err$, the semantics gives
$\psem{\rudaf}{\invseq'^+}=\acc'(\psem{\rudaf}{\invseq^+},x)$ and
$\psem{\rudaf}{\invseq'}=\acc'(\psem{\rudaf}{\invseq},x)$.  Applying
rule C2 to the induction hypothesis yields
\[
  \SimBudget\bigl(\acc'(\psem{\rudaf}{\invseq^+},x),
             \acc'(\psem{\rudaf}{\invseq},x),
             |\invseq^-|\bigr),
\]
which proves the claim for $\invseq'$.

\emph{Inductive step (retraction).}
Let $\invseq'=\invseq\pp[(\Retract,x)]$ be valid with
$\psem{\rudaf}{\invseq'}\neq\Err$.  Then
$\psem{\rudaf}{\invseq}\neq\Err$ and $\invseq$ is valid.  Write
$s_2'=\psem{\rudaf}{\invseq}$ and $xs=\Effect(\invseq)$.  The
semantics gives $\psem{\rudaf}{\invseq'}=\Retract(s_2',x)\neq\Err$,
and part (iii) of \Cref{lem:norm-facts} gives $x\in xs$.  By
\Cref{lem:sound-B-reach} applied to $\invseq$, there is a state
$s_3'$ with $\SimTwo(xs,s_3',s_2')$.  Together with the induction
hypothesis
$\SimBudget(\psem{\rudaf}{\invseq^+},s_2',|\invseq^-|)$, all premises
of rule C3 are discharged, and the rule yields
\[
  \SimBudget\bigl(\psem{\rudaf}{\invseq^+},
             \Retract(s_2',x),
             |\invseq^-|+1\bigr).
\]
Since $\invseq'^+=\invseq^+$,
$|\invseq'^-|=|\invseq^-|+1$, and
$\psem{\rudaf}{\invseq'}=\Retract(s_2',x)$, this proves the claim for
$\invseq'$.
\end{proof}

\begin{lemma}[Budget support]
\label[lemma]{lem:sound-C-support}
Let $\invseq$ be a valid update sequence such that every prefix
$\delta\preceq\invseq$ satisfies
$|\delta^-|\le\RetSpec(\vals{\delta^+})$.  Then
$\psem{\rudaf}{\invseq}\neq\Err$.
\end{lemma}

\begin{proof}
Suppose, toward a contradiction, that
$\psem{\rudaf}{\invseq}=\Err$, and let $s_0,\dots,s_m$ be the state
sequence of \Cref{def:rudaf-semantics}.  Since $s_0=\init'\neq\Err$,
there is a least index $j$ with $s_j=\Err$, and then
$s_{j-1}\neq\Err$.  If $o_j=\mathsf{acc}$, then
$s_j=\acc'(s_{j-1},a_j)$ is a proper state because $\acc'$ has the state type
as its codomain (as in \Cref{lem:acc-only-noerr}).  Hence
$o_j=\mathsf{retract}$, so
$\invseq$ has a prefix $\invseq_0=\delta\pp[(\Retract,x)]$ such that
$s_2'\triangleq\psem{\rudaf}{\delta}=s_{j-1}\neq\Err$ and
$\Retract(s_2',x)=\Err$.  Both $\delta$ and $\invseq_0$ are valid,
being prefixes of $\invseq$.  Define
\[
  ys\triangleq\vals{\delta^+},\qquad
  xs\triangleq\Effect(\delta),\qquad
  n\triangleq |\delta^-|.
\]
The premises of rule C4 hold with $s_1'=\psem{\rudaf}{\delta^+}$ and the
state $s_3'$ defined below:
\begin{itemize}[leftmargin=*]
\item $\Reach(ys,s_1')$: the sequence $\delta^+$ is
  accumulation-only with $\vals{\delta^+}=ys$, so \Cref{lem:sound-R}
  applies;
\item $\SimBudget(s_1',s_2',n)$: by \Cref{lem:sound-C-inv} applied to
  $\delta$, using $\psem{\rudaf}{\delta}\neq\Err$;
\item $\SimTwo(xs,s_3',s_2')$: by \Cref{lem:sound-B-reach} applied to
  $\delta$, which yields a state $s_3'$ with $\Reach(xs,s_3')$ and
  $\SimTwo(xs,s_3',s_2')$;
\item $\Subseq(ys,xs,n)$: by part (iv) of \Cref{lem:norm-facts}
  applied to $\delta$;
\item $x\in xs$: by part (iii) of \Cref{lem:norm-facts}, since
  $\invseq_0=\delta\pp[(\Retract,x)]$ is valid;
\item $n<\RetSpec(ys)$: applying the budget hypothesis to the prefix
  $\invseq_0$ gives
  $n+1=|\invseq_0^-|\le\RetSpec(\vals{\invseq_0^+})=\RetSpec(ys)$,
  where the last equality holds because $\invseq_0^+=\delta^+$.
\end{itemize}
Rule C4 therefore yields $\Retract(s_2',x)\neq\Err$, contradicting
the choice of $\invseq_0$.  Thus
\[\psem{\rudaf}{\invseq}\neq\Err.\]
\end{proof}

\begin{theorem}[Soundness]
\label{thm:proof-rules-sound-appendix}
If there exist predicates $\Reach$, $\SimTwo$, and $\SimBudget$
satisfying all rules in \Cref{fig:proof-rules}, then
$\udaf\equiv_{\RetSpec}\rudaf$.
\end{theorem}

\begin{proof}
By \Cref{def:equivalence}, we must establish (1)
$\rudaf\models\RetSpec$, and (2) semantic agreement on all valid
update sequences that do not produce $\Err$..

(1) By \Cref{def:retspec}, we must show that
$\sem{\rudaf}{\invseq}\neq\Err$ for every valid update sequence
$\invseq$ whose prefixes respect the budget.  So fix an arbitrary
valid $\invseq$ such that every prefix $\delta\preceq\invseq$
satisfies $|\delta^-|\le\RetSpec(\vals{\delta^+})$.  By
\Cref{lem:sound-C-support}, $\psem{\rudaf}{\invseq}\neq\Err.$
It remains to rule out the possibility that $\eval'$ itself returns
$\Err$ on the final state.  By \Cref{lem:sound-B},
$\sem{\rudaf}{\invseq}=\sem{\rudaf}{\Norm(\invseq)}$.  By part (i) of
\Cref{lem:norm-facts}, $\Norm(\invseq)$ is accumulation-only with
$\vals{\Norm(\invseq)}=\Effect(\invseq)$, so \Cref{lem:sound-A} gives
$\sem{\rudaf}{\Norm(\invseq)}=\sem{\udaf}{\Effect(\invseq)}$.
Finally, by \Cref{def:fold},
$\sem{\udaf}{\Effect(\invseq)}=\eval(\psem{\udaf}{\Effect(\invseq)})$,
and the output type $O$ excludes $\Err$.  Chaining these equalities,
$\sem{\rudaf}{\invseq}=\sem{\udaf}{\Effect(\invseq)}\neq\Err$.  Since
$\invseq$ was arbitrary, $\rudaf\models\RetSpec$.

(2) Let $\invseq$ be an arbitrary valid update sequence with
$xs=\Effect(\invseq)$ and $\sem{\rudaf}{\invseq}\neq\Err$.  By
\Cref{def:rudaf-semantics}, $\psem{\rudaf}{\invseq}\neq\Err$, since
otherwise $\sem{\rudaf}{\invseq}$ would be $\Err$.  By
\Cref{lem:sound-B},
$\sem{\rudaf}{\invseq}=\sem{\rudaf}{\Norm(\invseq)}$, and, as
in part (1), \Cref{lem:sound-A} applied to the accumulation-only
sequence $\Norm(\invseq)$ gives
$\sem{\rudaf}{\Norm(\invseq)}=\sem{\udaf}{xs}$.  Hence
\[
  \sem{\rudaf}{\invseq}
  =
  \sem{\rudaf}{\Norm(\invseq)}
  =
  \sem{\udaf}{xs},
\]
which is precisely clause (2) of \Cref{def:equivalence}.
\end{proof}

\subsection{Completeness proof}
\label{sec:complete}

This subsection elaborates on the relative completeness part of
\Cref{thm:proof-rules-sound}.  We fix an \textudaf{}
$\udaf=(\init,\acc,\eval)$, a \textrudaf{}
$\rudaf=(\init',\acc',\Retract,\eval')$, and a retraction
specification $\RetSpec$ as in \Cref{sec:problem}, and we assume
throughout that $\udaf\equiv_{\RetSpec}\rudaf$ in the sense of
\Cref{def:equivalence}.  We construct predicates $\Reach$, $\SimTwo$, and
$\SimBudget$ under which every rule of
\Cref{fig:proof-rules} is valid, where a rule is \emph{valid} if
every instantiation of its metavariables that satisfies all premises
also satisfies the conclusion.  The construction is relative to the
assumptions stated below.  These assumptions correspond to the two conditions
summarized after \Cref{thm:proof-rules-sound}: (a)~logical
expressiveness (Assumption~A.3) and (b)~content-determined behavior
(Assumption~A.2), together with an accumulator congruence condition
(Assumption~A.1) corresponding to Block~A.

\paragraph{Preliminaries.}
Recall that $\canon(xs)$ denotes the accumulation-only update
sequence that inserts the elements of $xs$ in order, so that
$\vals{\canon(xs)}=xs$.  We collect the elementary facts about update
sequences and normalization used below; they follow directly from
\Cref{def:valid-seq,def:rudaf-semantics,def:normalization}
(cf.\ \Cref{lem:norm-facts,lem:acc-only-noerr}).
\begin{enumerate}[label=(F\arabic*),leftmargin=*]
\item Accumulation-only update sequences are valid and are fixpoints
  of $\Norm$; in particular $\Effect(\canon(xs))=xs$.  Moreover,
  $\psem{\rudaf}{\invseq}\neq\Err$ for every accumulation-only
  $\invseq$, because $\acc'$ maps states to states.  Under
  \Cref{def:rudaf-semantics}, retraction steps and absorption are the sources
  of $\Err$.
\item If $\invseq$ is valid, then $\invseq\pp[(\acc,x)]$ is valid and
  $\Effect(\invseq\pp[(\acc,x)])=\Effect(\invseq)\pp[x]$.
\item If $\invseq$ is valid, then $\invseq\pp[(\Retract,x)]$ is valid
  iff $x\in\Effect(\invseq)$, and in that case
  \[\Effect(\invseq\pp[(\Retract,x)])=\remove(\Effect(\invseq),x).\]
\item $\Err$ is absorbing.  In particular, if
  $\psem{\rudaf}{\invseq}\neq\Err$, then
  $\psem{\rudaf}{\delta}\neq\Err$ for every prefix
  $\delta\preceq\invseq$.
\item We read the side predicate $\Subseq$ of \Cref{fig:proof-rules}
  as follows: $\Subseq(ys,xs,n)$ holds iff there are values
  $z_1,\dots,z_n$ such that
  $xs=\remove(\cdots\remove(\remove(ys,z_1),z_2)\cdots,z_n)$, where
  each intermediate application of $\remove$ is defined, i.e., $z_i$
  occurs in the list it is applied to.  This formalizes the caption
  of \Cref{fig:proof-rules} ($xs$ is obtained from $ys$ by deleting
  exactly $n$ elements, one occurrence per deleted element).  The soundness
  proof uses this form when normalization deletes one accumulated occurrence
  per retraction.
\end{enumerate}

\paragraph{Assumption A.1 (Accumulator congruence).}
The insertion-only components of the two folds are observationally
congruent: $\eval(\init)=\eval'(\init')$, and for all states $s,s'$
and elements $a$,
\[
  \eval(s)=\eval'(s')
  \Longrightarrow
  \eval(\acc(s,a))=\eval'(\acc'(s',a)).
\]
The assumption states that inserting the same element preserves observational
agreement between original and retractable states.  Rules A1 and A2 quantify
over all states and are independent of $\Reach$, $\SimTwo$, and $\SimBudget$.
Their validity is therefore equivalent to Assumption~A.1 for every definition
of the three semantic invariants.

\paragraph{Assumption A.2 (Content-determined behavior).}
For all valid update sequences $\invseq_1,\invseq_2$ with
$\psem{\rudaf}{\invseq_1}\neq\Err$,
$\psem{\rudaf}{\invseq_2}\neq\Err$, and
$\Effect(\invseq_1)=\Effect(\invseq_2)=xs$, the following hold:
\begin{enumerate}[label=(\alph*),leftmargin=*]
\item $\eval'(\psem{\rudaf}{\invseq_1})
      =\eval'(\psem{\rudaf}{\invseq_2})$;
\item for every $x\in xs$,
  $\Retract(\psem{\rudaf}{\invseq_1},x)\neq\Err$
  iff
  $\Retract(\psem{\rudaf}{\invseq_2},x)\neq\Err$.
\end{enumerate}
The assumption requires histories with the same active content to have the
same observable output and one-step retraction support.  Whenever
$\sem{\rudaf}{\invseq_1}\neq\Err$ and
$\sem{\rudaf}{\invseq_2}\neq\Err$, clause~(2) of
\Cref{def:equivalence} already forces both sides of part~(a) to equal
$\sem{\udaf}{xs}$.  Part~(a) additionally requires that whether
$\eval'$ itself fails is content-determined, and part~(b) states the
analogous requirement for the definedness of a single retraction.
This extra strength matters because a valid history may exceed the
retraction budget promised by $\RetSpec$ and yet be successful; on
the future failure behavior of such histories,
\Cref{def:equivalence} is silent.

\paragraph{Assumption A.3 (Logical expressiveness).}
The assertion logic is expressive enough to represent semantic
relations over \textrudaf{} states, lists of elements, and natural
numbers.  More precisely, for each of the three semantic relations
defined in the proof below, there exists a predicate of the assertion
language with the same truth set.  As noted after
\Cref{thm:proof-rules-sound}, automation also requires decidability of the
resulting proof obligations.  The existence result below uses expressiveness.

\begin{theorem}[Relative completeness]
\label{thm:rel-complete-appendix}
Suppose $\udaf\equiv_{\RetSpec}\rudaf$ and Assumptions~A.1--A.3 hold.
Then there exist predicates $\Reach,\SimTwo$, and $\SimBudget$ such
that all proof rules in \Cref{fig:proof-rules} are valid.
\end{theorem}

\begin{proof}
We define $\Reach,\SimTwo$, and $\SimBudget$ as semantic relations
and verify each rule of \Cref{fig:proof-rules}.  By Assumption~A.3,
each relation can then be replaced by an assertion of the proof logic
with the same truth set, which yields the theorem.

\paragraph*{Reachability $\Reach$ (Block R)}
Define
\[
  \Reach(xs,s')
  \Longleftrightarrow
  s'=\psem{\rudaf}{\canon(xs)}.
\]

\emph{R1.}  We have $\canon(\emptylist)=\emptylist$ and
$\psem{\rudaf}{\emptylist}=\init'$, so $\Reach(\emptylist,\init')$
holds.

\emph{R2.}  Suppose $\Reach(xs,s')$, i.e.,
$s'=\psem{\rudaf}{\canon(xs)}$.  Since
$\canon(xs\pp[a])=\canon(xs)\pp[(\acc,a)]$ and
$\psem{\rudaf}{\canon(xs)}\neq\Err$ by~(F1), the semantics unfolds to
\[
  \psem{\rudaf}{\canon(xs\pp[a])}
  =\acc'(\psem{\rudaf}{\canon(xs)},a)
  =\acc'(s',a),
\]
so $\Reach(xs\pp[a],\acc'(s',a))$ holds.

\paragraph*{Lifting consistency (Block A)}
Rule A1 is the first conjunct of Assumption~A.1, and rule A2 is the
second conjunct, instantiated with the states $s,s'$ and the element
$a$ appearing in the rule.

\paragraph*{Retraction consistency $\SimTwo$ (Block B)}
Define
\[
\begin{aligned}
  \SimTwo(xs,s_1',s_2') \Longleftrightarrow {}&
    \Reach(xs,s_1') \;\wedge\;
    \exists\invseq_2.\;
      \invseq_2\text{ valid} \wedge
      \Effect(\invseq_2)=xs \wedge{}\\
    &\psem{\rudaf}{\invseq_2}=s_2' \wedge s_2'\neq\Err.
\end{aligned}
\]
The conjunct $\Reach(xs,s_1')$ pins the first state component to the
canonical insertion-only run of $xs$ for the proofs of B3 and B4.

\emph{B1.}  Take $\invseq_2=\emptylist$: it is valid,
$\Effect(\emptylist)=\emptylist$, and
$\psem{\rudaf}{\emptylist}=\init'\neq\Err$.  Together with
$\Reach(\emptylist,\init')$ (rule R1 above), this gives
$\SimTwo(\emptylist,\init',\init')$.

\emph{B2.}  Suppose $\SimTwo(xs,s_1',s_2')$ with witness $\invseq_2$.
For the first conjunct, the R2 argument gives
\[\Reach(xs\pp[x],\acc'(s_1',x)).\]
For the second conjunct, consider
the extension $\invseq_2\pp[(\acc,x)]$: it is valid with effect
$xs\pp[x]$ by~(F2), and since $s_2'\neq\Err$, its state semantics is
the proper state $\acc'(s_2',x)$.  Thus, we have
\[\SimTwo(xs\pp[x],\acc'(s_1',x),\acc'(s_2',x)).\]

\emph{B3.}  Assume the premises of the rule: $\Reach(xs,s_1')$,
$\SimTwo(xs,s_1',s_2')$ with witness $\invseq_2$, $x\in xs$,
$\Retract(s_2',x)\neq\Err$, $ys=\remove(xs,x)$, and
$\Reach(ys,s_3')$.  Consider the extension
$\invseq_2\pp[(\Retract,x)]$.  Since
$x\in xs=\Effect(\invseq_2)$, the extension is valid by~(F3), and its
effect is $\remove(xs,x)=ys$.  Its state semantics is
\[\Retract(\psem{\rudaf}{\invseq_2},x)=\Retract(s_2',x)\neq\Err\] by the
premise.  Together with $\Reach(ys,s_3')$,
the extended sequence therefore has
\[\SimTwo(ys,s_3',\Retract(s_2',x)).\]

\emph{B4.}  Suppose $\SimTwo(xs,s_1',s_2')$.  The conjunct
$\Reach(xs,s_1')$ forces $s_1'=\psem{\rudaf}{\canon(xs)}$, and the
witness provides a valid $\invseq_2$ with $\Effect(\invseq_2)=xs$ and
$\psem{\rudaf}{\invseq_2}=s_2'\neq\Err$.  By~(F1), $\canon(xs)$ is a
valid update sequence with $\Effect(\canon(xs))=xs$ and
$\psem{\rudaf}{\canon(xs)}\neq\Err$.  Assumption~A.2(a), applied to
$\invseq_2$ and $\canon(xs)$, yields
\[
  \eval'(s_2')
  =\eval'(\psem{\rudaf}{\canon(xs)})
  =\eval'(s_1'),
\]
which is the conclusion of B4.  Clause~(2) of \Cref{def:equivalence} applies
when
$\sem{\rudaf}{\invseq_2}\neq\Err$.  The witness may exceed the budget of
$\RetSpec$, in which case $\eval'(s_2')$ may equal $\Err$ even though $s_2'$ is
a proper state.  Assumption~A.2(a) covers this case.

\paragraph*{Budget relation $\SimBudget$ (Block C)}
Define
\[
\begin{aligned}
  \SimBudget(s_1',s_2',n) \Longleftrightarrow {}&
    \exists\invseq.\;
    \psem{\rudaf}{\invseq}\neq\Err \wedge
    s_1'=\psem{\rudaf}{\invseq^+} \wedge{}s_2'=\psem{\rudaf}{\invseq} \wedge
     n=|\invseq^-|,
\end{aligned}
\]
Here $\invseq$ ranges over arbitrary update sequences.  Their state semantics
is defined by \Cref{def:rudaf-semantics}, and the relation retains the conjunct
$\psem{\rudaf}{\invseq}\neq\Err$.  Allowing arbitrary witnesses makes C3
closed under extension.  The element $x$ in C3 comes from a $\SimTwo$ witness
that may have a different effect from the $\SimBudget$ witness, so appending
its retraction may produce an invalid update sequence.  The sequence
$\invseq^+$ is accumulation-only, so
$\psem{\rudaf}{\invseq^+}\neq\Err$ by~(F1), and the first component
of the relation is always a proper state.

\emph{C1.}  Take $\invseq=\emptylist$: then
$\invseq^+=\invseq^-=\emptylist$, both state components equal
$\init'\neq\Err$, and $n=0$.  Hence \[\SimBudget(\init',\init',0).\]

\emph{C2.}  Suppose $\SimBudget(s_1',s_2',n)$ with witness $\invseq$,
and let $\invseq'=\invseq\pp[(\acc,x)]$.  Then
$\invseq'^+=\invseq^+\pp[(\acc,x)]$, so
$\psem{\rudaf}{\invseq'^+}=\acc'(s_1',x)$ by~(F1), and since
$s_2'=\psem{\rudaf}{\invseq}\neq\Err$, we get
$\psem{\rudaf}{\invseq'}=\acc'(s_2',x)$, which is a proper state.
Also, $|\invseq'^-|=|\invseq^-|=n$.  Hence
\[\SimBudget(\acc'(s_1',x),\acc'(s_2',x),n).\]

\emph{C3.}  Assume the premises: $\SimBudget(s_1',s_2',n)$ with
witness $\invseq$, $\SimTwo(xs,s_3',s_2')$, $x\in xs$, and
$\Retract(s_2',x)\neq\Err$.  Let
$\invseq'=\invseq\pp[(\Retract,x)]$.  Arbitrary update sequences are
admissible witnesses, so $\invseq'$ is admissible.  Then
$\invseq'^+=\invseq^+$, so the first state component is still $s_1'$;
the state semantics is
$\psem{\rudaf}{\invseq'}=\Retract(s_2',x)\neq\Err$ by the premise;
and $|\invseq'^-|=n+1$.  Hence
\[\SimBudget(s_1',\Retract(s_2',x),n+1).\]

\emph{C4.}  Assume the premises:
\[
  \Reach(ys,s_1') \land \SimBudget(s_1',s_2',n) \land
  \SimTwo(xs,s_3',s_2') \land \Subseq(ys,xs,n) \land
  x\in xs \land n<\RetSpec(ys).
\]
Unfolding $\SimTwo(xs,s_3',s_2')$ yields a valid update sequence
$\invseq_2$ with $\Effect(\invseq_2)=xs$ and
$\psem{\rudaf}{\invseq_2}=s_2'\neq\Err$.  Unfolding
$\Subseq(ys,xs,n)$ by~(F5) yields values $z_1,\dots,z_n$ such that
the iterated removal of $z_1,\dots,z_n$ from $ys$ is defined and
equals $xs$.  Define
\[
  \sigma_0\triangleq\canon(ys),\qquad
  \sigma_i\triangleq\sigma_{i-1}\pp[(\Retract,z_i)]
  \;\;\text{for }1\le i\le n,\qquad
  \sigma\triangleq\sigma_n\pp[(\Retract,x)].
\]
We use an induction on $i$ using (F1) and (F3) to show that
each $\sigma_i$ is valid with
$\Effect(\sigma_i)=\remove(\cdots\remove(ys,z_1)\cdots,z_i)$: the
$i$-th extension is valid because definedness of the $i$-th removal
in~(F5) gives $z_i\in\Effect(\sigma_{i-1})$.  In particular,
$\sigma_n$ is valid with $\Effect(\sigma_n)=xs$, and since $x\in xs$,
the full sequence $\sigma$ is valid as well, again by~(F3).

Next, we check that every prefix $\delta\preceq\sigma$ satisfies
$|\delta^-|\le\RetSpec(\vals{\delta^+})$.  If
$\delta\preceq\canon(ys)$, then $|\delta^-|=0$, and
$\RetSpec(\vals{\delta^+})\ge 0$ because $\RetSpec$ maps into
$\mathbb{N}$.  Otherwise, $\delta$ extends $\canon(ys)$ by
retractions only, so $\vals{\delta^+}=ys$ and
$|\delta^-|\le n+1\le\RetSpec(ys)$, using the premise
$n<\RetSpec(ys)$.  Hence clause~(1) of \Cref{def:equivalence}, i.e.,
$\rudaf\models\RetSpec$, gives $\sem{\rudaf}{\sigma}\neq\Err$, and in
particular $\psem{\rudaf}{\sigma}\neq\Err$.  By $\Err$
absorption~(F4), $\psem{\rudaf}{\sigma_n}\neq\Err$, and unfolding the
last step of the semantics,
\[
  \Retract(\psem{\rudaf}{\sigma_n},x)
  =\psem{\rudaf}{\sigma}
  \neq\Err.
\]
Now $\sigma_n$ and $\invseq_2$ are two valid update sequences with
non-$\Err$ state semantics and the same effect $xs$, and $x\in xs$.
Assumption~A.2(b) therefore transfers the defined-ness of retracting
$x$ from $\psem{\rudaf}{\sigma_n}$ to
$\psem{\rudaf}{\invseq_2}=s_2'$, which yields
\[
\Retract(s_2',x)\neq\Err.
\]

The C4 validity argument depends on $\SimTwo$, $\Subseq$, $x\in xs$, and the
bound $n<\RetSpec(ys)$.  The $\Reach$ and $\SimBudget$ premises are used in the
soundness direction (\Cref{lem:sound-C-support}) to supply the corresponding
instantiation of $ys$, $xs$, and $n$.

Thus all rules of \Cref{fig:proof-rules} are valid under the semantic
definitions of $\Reach$, $\SimTwo$, and $\SimBudget$, and by
Assumption~A.3 these relations are expressible in the assertion
logic.  This completes the proof.
\end{proof}

\section{Proofs about the Synthesis Method}\label{sec:synthesis-proofs}

Throughout this section, we make the following assumptions.

\begin{description}[leftmargin=*]
  \item[(A1) Oracles.]
  The SMT, SyGuS, PBE, symbolic-execution, and quantifier-elimination procedures
  used by \tool{} are sound and complete for the queries issued by the
  algorithms.

  \item[(A2) Fairness.]
  All enumeration and CEGIS loops used by \tool{} are fair.

  \item[(A3) Search spaces.]
  Whenever a completeness result assumes the existence of a correct solution,
  its auxiliary type, qualifier, auxiliary update, relation,
  retraction function, and predicates satisfying the proof rules are expressible
  in the corresponding libraries, grammars, and templates. This includes the
  guard expressions used to fill the holes introduced by
  \textsc{InsertOrMerge}. For the corresponding qualifier, the lifted fold
  returned by \textsc{Lift} admits such a retraction and predicates, and its
  relation described in \Cref{thm:validcand-necessary} occurs in the
  \textsc{SynRetract} search space.

\end{description}

The pseudocode and proofs use two interface conventions. First, an invocation
of (A3) fixes one witness for the entire synthesis run: a single retraction,
its proof predicates, and the guard instantiations used by every partition
refinement. Second, a counterexample $(xs,s^\star,x^\star)$ returned by
\textsc{Verify} is a local deletion obligation: $x^\star\in xs$, the rejected
candidate violates
\[
  \Psi(\psem{\rudaf}{\remove(xs,x^\star)},
       \Retract(s^\star,x^\star)),
\]
and, once length $|xs|$ has been explored, $(s^\star,x^\star)$ is covered by
the retained partition element selected at line~12 of
\Cref{alg:synretract-partition}. Its reference post-deletion state is one of
the local obligations associated with that element. Lines~8--13 rely on this
counterexample contract to define the selection at line~12.

\begin{theorem}[Necessity of $\mathsf{ValidCand}$]
\label{thm:validcand-necessary}
Let $\rudaf=(\Init^\star,\acc^\star,\eval^\star)$ be a lifted fold that
realizes a qualifier $\chi$, and let $G$ be the compatible-history sets
returned by
\textsc{CheckRetractability}$(\udaf,\theta,\chi,\mathcal{D},\mathcal{X})$.
Suppose there is a retraction function $\Retract$ and predicates
$\Reach$, $\SimTwo$, and $\SimBudget$ satisfying Blocks R, B, and C of
\Cref{fig:proof-rules}. Let $\Psi$ be the smallest relation that
contains
\[
  \{(s_1,s_2)\mid \exists xs.\ \SimTwo(xs,s_1,s_2)\}
\]
and is closed under reflexivity, symmetry, and transitivity.
Then $\mathsf{ValidCand}(G,\rudaf,\Psi)$ holds.
\end{theorem}

\begin{proof}[Proof sketch]
We establish the three proof-rule clauses in $\mathsf{ValidCand}$. Rule B1 gives
$\SimTwo(\emptylist,\Init^\star,\Init^\star)$, so
$\Psi(\Init^\star,\Init^\star)$ by the definition of $\Psi$.

For B2, suppose $\Psi(s_1,s_2)$. By the definition of $\Psi$, there is a
finite chain $s_1=u_0,\ldots,u_n=s_2$ in which each adjacent pair is
related by $\SimTwo(xs_i,\cdot,\cdot)$ for some $xs_i$, possibly in the reverse
direction. Applying B2 to every such edge gives a chain from
$\acc^\star(s_1,x)$ to $\acc^\star(s_2,x)$ whose adjacent states are related by
$\SimTwo(xs_i\pp[x],\cdot,\cdot)$. Hence
$\Psi(\acc^\star(s_1,x),\acc^\star(s_2,x))$. For B4, rule B4 gives equal
evaluations across every edge in the original chain. Symmetry of equality and
transitivity along the chain yield $\eval^\star(s_1)=\eval^\star(s_2)$.

We next consider the compatible-history sets in $G$. The seed relation $E_0$ in
\Cref{alg:spec-necessity} relates histories that reach the same lifted state
under $\chi$. Reflexivity of $\Psi$ relates the states reached by a reflexive
pair, and both $\mathsf{Acc}$ and $\mathsf{Ret}$ preserve reflexive pairs. For a
non-reflexive pair
$((ys,xs),(ys',xs'))$ generated while computing $E^\star$, we maintain the
following facts. The full histories $ys$ and $ys'$ reach the same lifted state
$q$. There is a common sequence of $n$ deletions that transforms $ys$ into
$xs$ and $ys'$ into $xs'$. Applying those deletions to $q$ produces a state
$t$ such that
\[
  \SimBudget(q,t,n),
  \qquad
  \SimTwo(xs,\psem{\rudaf}{xs},t)
  \quad\text{and}\quad
  \SimTwo(xs',\psem{\rudaf}{xs'},t).
\]
The deletion sequence also witnesses $\Subseq(ys,xs,n)$ and
$\Subseq(ys',xs',n)$.

For a seed pair, $n=0$ and $q=t$ is the shared lifted state. Repeated
applications of B2 from B1 establish the two $\SimTwo$ facts, while repeated
applications of C2 from C1 establish $\SimBudget(q,q,0)$. For a common
$\mathsf{Acc}$ step with input $x$, take the new full-history state to be
$\acc^\star(q,x)$ and the new mixed state to be $\acc^\star(t,x)$. Rules B2
and C2 preserve the displayed facts, and the same deletion sequence still
connects the two extended histories.

For a common $\mathsf{Ret}$ step with input $x$, Rules R1 and R2 provide
reachability of the two insertion-only reference states and of the full-history
state $q$. The
successor pair remains in $M$, so the extended deletion sequence has length at
most $\theta(ys)$; hence the current count satisfies $n<\theta(ys)$. Rule C4,
using $\SimBudget(q,t,n)$, the first $\SimTwo$ fact, and
$\Subseq(ys,xs,n)$, gives $\Retract(t,x)\neq\Err$. Rule B3 then relates the
unprimed insertion-only reference state after deleting $x$ to
$\Retract(t,x)$. A second
application of B3 uses the primed $\SimTwo$ and reachability facts and gives the
same result on the primed side. Finally, C3 establishes
$\SimBudget(q,\Retract(t,x),n+1)$.

Symmetry preserves these facts. For transitivity, if either premise is
reflexive, the composed pair is the other premise. Otherwise, two pairs sharing
a middle component have the same full-history state, deletion count, and
multiset of deleted values, including multiplicity, since these are determined
by the shared full and effective histories. Choose the deletion order from the
first pair. The third history removes the same values with the same
multiplicities, so this order is valid there as well; removals of different
values commute, and the chosen order reaches the same effective history.
Replaying this common sequence from the shared full-history state establishes
the invariant for the composed pair. Thus every non-reflexive pair in
$E^\star$ has the displayed witnesses, while reflexive pairs satisfy the desired
conclusion directly. In either case, the effective histories reach states
related by $\Psi$.

Finally, each $V\in G$ collects histories paired with the same anchor in
$E^\star$. Thus, for $xs_1,xs_2\in V$, both reached states are
$\Psi$-related to the anchor state, and symmetry and transitivity give
$\Psi(\psem{\rudaf}{xs_1},\psem{\rudaf}{xs_2})$.
\end{proof}

\begin{proof}[Proof of \Cref{thm:synth-sound}]
When \Cref{alg:synthesize-rudaf} returns a retractable fold,
\textsc{SynRetract} has produced a retraction accepted by \textsc{Verify}.
The assembled fold therefore satisfies every rule in
\Cref{fig:proof-rules}. By the soundness direction of
\Cref{thm:proof-rules-sound}, the returned fold is equivalent to the original
fold modulo $\RetSpec$.
\end{proof}

\begin{theorem}[Completeness of \textsc{Lift}]
\label{thm:lift-complete}
Let $P=(\Init,\acc,\eval)$, let
$T=\{(s:\tau,a:\tau_{\mathsf{aux}})\mid\chi(s,a,xs)\}$ be a lifting
specification, let $U_0$ be the initial update map, and let $\mathcal{G}$ be
the grammar given to \textsc{Lift}. Suppose there are $a_0$ and
$\mathsf{UpdAux}^\star\in\mathcal{G}$ such that
\begin{enumerate}[leftmargin=*]
  \item $\chi(\Init,a_0,[])$ holds;
  \item $\mathsf{UpdAux}^\star$ is consistent with $U_0$; and
  \item for every history $xs$, auxiliary value $a$, and input $x$, with
  $s=\psem{P}{xs}$,
  \[
    \chi(s,a,xs)\Rightarrow
    \chi(\acc(s,x),\mathsf{UpdAux}^\star(s,a,x),xs\pp[x]).
  \]
\end{enumerate}
Under (A1) and (A2), \textsc{Lift} eventually returns
$(\mathsf{true},\Init^\star,\acc^\star,\eval^\star)$, where
\[
  \Init^\star=(\Init,a_0'),\qquad
  \acc^\star((s,a),x)=(\acc(s,x),\mathsf{UpdAux}(s,a,x)),\qquad
  \eval^\star(s,a)=\eval(s),
\]
and $\acc^\star$ preserves $\chi$ on all reachable states.
\end{theorem}

\begin{proof}[Proof sketch]
The update map is initially consistent with $\mathsf{UpdAux}^\star$. Whenever
the current candidate fails the inductiveness query in
\Cref{alg:lift-cegis}, the counterexample supplies the unique auxiliary value
required by $\chi$ after that step. The qualifier grammar in
\Cref{fig:lift-dsl} has the form $a=E(s,xs)$, so this value is also the value
returned by $\mathsf{UpdAux}^\star$. The example map therefore remains
consistent with $\mathsf{UpdAux}^\star$ after each refinement.

By (A1), \textsc{PBE} returns a candidate after each refinement. Each
counterexample eliminates the current update function while preserving
$\mathsf{UpdAux}^\star$, so fairness (A2) eventually produces an update
function for which the inductiveness query is unsatisfiable. The definitions at
lines~\ref{line:lift-assign-acc} and~\ref{line:lift-assign-init} then give the
stated initializer, accumulator, and evaluation function. For every reachable
post-state, the functional qualifier supplies the unique auxiliary value
$a'=E(\acc(s,x),xs\pp[x])$. Every model of $\mathsf{Bad}$ therefore extends
to a model of the query in \Cref{alg:lift-cegis}. Unsatisfiability of that query
establishes preservation of $\chi$ on reachable states.
\end{proof}

\begin{lemma}[Functional consistency is necessary for lifting]
\label[lemma]{lem:fc-necessary}
Let $R$ be the relation constructed by \Cref{alg:build-examples} for a
qualifier $\chi$. If there is a deterministic update
$\mathsf{UpdAux}^\star$ that preserves $\chi$ on every sampled history and
test input, then $R$ is functional.
\end{lemma}

\begin{proof}[Proof sketch]
Each $((s,a,x),a')\in R$ records the auxiliary value required by $\chi$ after
updating $(s,a)$ with $x$. Preservation by $\mathsf{UpdAux}^\star$ shows that
this value must be $\mathsf{UpdAux}^\star(s,a,x)$. Since $\chi$ has the form
$a=E(s,xs)$, the required value is unique. Hence two entries with the same key
$(s,a,x)$ have the same value.
\end{proof}

\begin{lemma}[Valid lifts pass the retractability check]
\label[lemma]{lem:retractability-necessary}
Suppose a lifted fold realizing $\chi$ satisfies
$\eval^\star(s,a)=\eval(s)$ and admits a retraction function and predicates
satisfying Blocks R, B, and C of \Cref{fig:proof-rules} for $\theta$. Then
\textsc{CheckRetractability}$(\udaf,\theta,\chi,\mathcal{D},\mathcal{X})$
returns $\mathsf{bad}=\mathsf{false}$.
\end{lemma}

\begin{proof}[Proof sketch]
Suppose to the contrary that the check reports a conflict. Then $E^\star$ relates two
allowed pairs whose effective histories $xs$ and $xs'$ have different
observable results under $\udaf$. As in the proof of
\Cref{thm:validcand-necessary}, induction over the construction of $E^\star$
uses B2 for $\mathsf{Acc}$ steps, B3 for $\mathsf{Ret}$ steps, and Block C to
show that the required retractions are defined. The lifted states reached by
$xs$ and $xs'$ must therefore be related by the relation $\Psi$ from
\Cref{thm:validcand-necessary}. The B4 argument in that proof gives equal
observations along every $\Psi$ chain. By the assumed equation
$\eval^\star(s,a)=\eval(s)$, this contradicts
$\sem{\udaf}{xs}\neq\sem{\udaf}{xs'}$.
\end{proof}

\begin{proof}[Proof of \Cref{thm:check-retract-sound}]
The result is the contrapositive of
\Cref{lem:retractability-necessary}. For a lift covered by the theorem, the
existence of predicates and a retraction function satisfying Blocks R, B, and C
would force the check to return $\mathsf{bad}=\mathsf{false}$.
\end{proof}

\begin{theorem}[Completeness of \textsc{InferLiftSpec}]
\label{thm:inferliftspec-complete}
Let $\udaf=(\Init,\acc,\eval)$, $\theta$, $L(\tau)$, $\mathcal{X}$, and
$K$ be the inputs to \textsc{InferLiftSpec}. Suppose the library and qualifier
enumeration contain a pair $(\tau_{\mathsf{aux}}^\star,\chi^\star)$ for which
there is a deterministic lifted retractable fold that realizes
$\chi^\star$, uses $\eval^\star(s,a)=\eval(s)$, and admits predicates satisfying
Blocks R, B, and C of \Cref{fig:proof-rules}. Under (A1) and (A2),
\textsc{InferLiftSpec} eventually adds
$(\tau_{\mathsf{aux}}^\star,\chi^\star,U^\star,G^\star)$ to $\Sigma$.
\end{theorem}

\begin{proof}[Proof sketch]
Fair enumeration eventually reaches
$(\tau_{\mathsf{aux}}^\star,\chi^\star)$. The auxiliary update of the assumed
fold preserves $\chi^\star$, so \Cref{lem:fc-necessary} shows that
\textsc{CheckFunctionalConsistency} returns $\mathsf{ok}=\mathsf{true}$ and
an update map $U^\star$. By \Cref{lem:retractability-necessary},
\textsc{CheckRetractability} returns $\mathsf{bad}=\mathsf{false}$ and evidence
$G^\star$. The candidate therefore passes the guard in
\Cref{alg:infer-lifted-types-symbolic} and is added to $\Sigma$.
\end{proof}

\begin{theorem}[Properties of the partition checks]
\label{thm:functionality-sound}
Let $\Psi$ be the relation constructed in
\Cref{thm:validcand-necessary}, and let
$(\gamma_{ij},\beta_{ij})$ be a case constructed by
\Cref{alg:create-symbolic-partitions-map}.
\begin{enumerate}[leftmargin=*]
  \item If the check at line~8 succeeds, no input satisfying
  $\gamma_{ij}$ has a reference post-deletion state satisfying
  $\beta_{ij}$.
  \item If the ambiguity check succeeds with witness
  $s,x,s'_1,s'_2$, no single state can be $\Psi$-related to both $s'_1$ and
  $s'_2$.
\end{enumerate}
\end{theorem}

\begin{proof}[Proof sketch]
For the first claim, suppose $(s,x)$ satisfies $\gamma_{ij}$. If some $s'$
satisfied $\beta_{ij}(s,x,s')$, then $(s,x,s')$ would satisfy the conjunction
declared unsatisfiable by the first check. Hence no such reference state exists.

For the second claim, let $s,x,s'_1,s'_2$ be the values returned by the
ambiguity query. If a successful result $t$ satisfied the local requirement for
both reference states, then $\Psi(s'_1,t)$ and $\Psi(s'_2,t)$ would hold.
The construction in \Cref{thm:validcand-necessary} is closed under symmetry and
transitivity, so these two facts would imply $\Psi(s'_1,s'_2)$, contradicting
the ambiguity query. The two B3 obligations are therefore incompatible for
this input.
\end{proof}

\begin{lemma}[Partition completeness for a fixed length]
\label[lemma]{lem:partition-complete-fixed-l}
Fix a lifted fold $\rudaf$ and a length $\ell$. Suppose a retraction function
$\Retract^\star$ and predicates satisfy the proof rules, and let $\Psi$ be the
relation constructed from their history-indexed $\SimTwo$ as in
\Cref{thm:validcand-necessary}.
Under (A1) and (A3), the resulting partition is satisfied by a fixed correct
retraction in the grammar and retains every length-$\ell$ case on which the
specification requires success.
\end{lemma}

\begin{proof}[Proof sketch]
By completeness of symbolic execution, every history of length $\ell$ follows
one of the paths enumerated at line~4 of
\Cref{alg:create-symbolic-partitions-map}. For each deletion position,
$\beta_i$ characterizes the reference post-deletion state for the history with
that element removed. Quantifier elimination therefore produces a guard containing
the current state and deleted element, together with a relation containing the
corresponding reference state.

Consider a concrete history $xs$ of length $\ell$, let
$s=\psem{\rudaf}{xs}$, and delete $x=x_i$. The reference post-deletion state
$s'=\psem{\rudaf}{\remove(xs,x)}$ witnesses $\beta_{ij}(s,x,s')$, so the
unrealizability check retains this input. If the specification requires
the deletion, R1 and R2 give $\Reach(xs,s)$ and
$\Reach(\remove(xs,x),s')$. Repeated applications of B2 and C2 from B1 and C1
give $\SimTwo(xs,s,s)$ and $\SimBudget(s,s,0)$. Since
$0<\RetSpec(xs)$, C4 gives $\Retract^\star(s,x)\neq\Err$, and B3 gives
\[
  \SimTwo(\remove(xs,x),s',\Retract^\star(s,x)).
\]
Thus $\Psi(s',\Retract^\star(s,x))$, which is the local specification.
Applying the same argument to every reference state for $(s,x)$ shows that all
of them are $\Psi$-related to the same result. They are therefore related to
each other, so this required input passes the ambiguity check.

The constraints generated at different lengths share a common model. Let
$\mathsf{Refs}(s,x)$ contain every reference post-deletion state contributed
for $(s,x)$ across all symbolic lengths and by the concrete-refinement branch
of \Cref{alg:synretract-partition}.

Fix the single correct witness supplied by (A3), and call its retraction
$\Retract^\dagger$. Its behavior on optional inputs is fixed for the entire
run. On a retained input, it returns a representative of the single
$\Psi$-class containing $\mathsf{Refs}(s,x)$; on an input whose
obligations contain two distinct $\Psi$-classes, it returns $\Err$. On inputs
where the proof rules require success, it agrees with the successful behavior
of the fixed proof-rule witness. The C4 and B3 argument above shows that every
reference state on such an input belongs to the class of that successful
result. Thus, it returns a successful result on every required input.

The representative choices are shared across all lengths to account for mixed
histories. Assumption (A3) supplies proof-rule correctness and predicates for
the fixed witness. The representative construction above establishes that the
local constraints generated by the algorithm are consistent with this witness.

For overlapping cases, instantiate the guard holes introduced by
\textsc{InsertOrMerge} with the condition that all reference states contributed
on the overlap are $\Psi$-related. On the satisfying region,
$\Retract^\dagger$ meets every incoming local specification. On the remaining
region, the obligations are incompatible, so that overlap is assigned $\bot$.
Required inputs are always in the first region by the C4 and B3 argument above.
By (A3), $\Retract^\dagger$, its proof predicates, and these guard expressions
occur in their search spaces. Hence every finite partition generated by
symbolic refinement has the same fixed correct model and retains every required
case.
\end{proof}

\begin{theorem}[Completeness of retraction synthesis]
\label{thm:synretract-complete}
Fix a lifted fold $\rudaf=(\Init^\star,\acc^\star,\eval^\star)$ that realizes
$\chi$, and let $G$ be the evidence returned by
\textsc{CheckRetractability}. Suppose there is a deterministic retraction
$\Retract^\star$ in the SyGuS grammar and predicates satisfying the proof rules
for $(\Init^\star,\acc^\star,\Retract^\star,\eval^\star)$. Suppose also that
the relation constructed in \Cref{thm:validcand-necessary} is in the
state-relation search space. Under (A1) to (A3), \textsc{SynRetract} eventually
returns a retraction accepted by \textsc{Verify}.
\end{theorem}

\begin{proof}[Proof sketch]
By \Cref{thm:validcand-necessary}, the state relation obtained from the proof
of $\Retract^\star$ satisfies $\mathsf{ValidCand}$. Fairness therefore makes
the outer loop of \Cref{alg:synretract-partition} consider that relation,
unless an earlier relation already produces a verified retraction.

Fix the global correct retraction $\Retract^\dagger$ from
\Cref{lem:partition-complete-fixed-l}. Our induction invariant is that this
retraction satisfies the current partition. The invariant holds initially
because the partition is empty. When verification reports a previously unseen
length,
\Cref{lem:partition-complete-fixed-l} shows that symbolic partitioning preserves
satisfaction by this same retraction and every required case. When the length
has already been seen,
write the counterexample as $(xs,s^\star,x^\star)$. The refinement adds
\[
  \Psi(\psem{\rudaf}{\remove(xs,x^\star)},
       \Retract(s^\star,x^\star)).
\]
By the counterexample contract stated at the beginning of this appendix, the
reference post-deletion state in this conjunct is one of the obligations in
$\mathsf{Refs}(s^\star,x^\star)$ for the retained partition element. If this is
a required-success obligation, the C4 and B3 instantiation from
\Cref{lem:partition-complete-fixed-l} relates it to
$\Retract^\dagger(s^\star,x^\star)$. Otherwise,
$\Retract^\dagger$ uses the representative of the common $\Psi$-class for that
retained element. In both cases it satisfies the new conjunct. The rejected
candidate violates the conjunct by the same counterexample contract, so the
refinement eliminates that candidate while preserving the induction invariant.

Thus the same correct retraction remains a model after every refinement. By
(A1), the refined SyGuS problem remains feasible. Fair enumeration
(A2) eventually proposes the fixed retraction, even if other candidates are
considered again after later partition changes. Its predicates satisfy the proof
rules by (A3), so \textsc{Verify} accepts it. Hence
\textsc{SynRetract} returns a verified retraction.
\end{proof}

\begin{proof}[Proof of \Cref{thm:synth-complete}]
Let the correct retractable fold from hypothesis~(i) use the lifting
specification $(\tau_{\mathsf{aux}}^\star,\chi^\star)$. By (A3), its retraction
and predicates satisfying the proof rules occur in the corresponding search
spaces. By (A2),
\textsc{InferLiftSpec} eventually considers this pair. Its auxiliary update
makes the functional-consistency check succeed by
\Cref{lem:fc-necessary}, while its retraction together with those predicates
makes the retractability check succeed by
\Cref{lem:retractability-necessary}. Thus
\Cref{thm:inferliftspec-complete} places the candidate in $\Sigma$.

The candidates in $\Sigma$ are explored in increasing order. Fairness ensures
that the loop reaches this candidate unless an earlier one already returns a
verified fold, in which case \Cref{thm:synth-sound} already gives the desired
conclusion.

When the top-level loop reaches this candidate, its auxiliary update witnesses
the hypotheses of \Cref{thm:lift-complete}, so \textsc{Lift} returns a lifted
fold preserving $\chi^\star$. By (A3), the required retraction and predicates
satisfying the proof rules are available for this lifted fold. Therefore
\Cref{thm:synretract-complete} makes
\textsc{SynRetract} return a retraction accepted by \textsc{Verify}.

The top-level algorithm then returns the assembled fold. By
\Cref{thm:synth-sound}, it is equivalent to $\udaf$ modulo $\RetSpec$.
\end{proof}

\end{document}